\documentstyle[a4p,12pt,cite,epsfig]{article}
\newcommand {\tcdot} {\mbox{$\ \cdot\cdot\cdot\ $}}
\newcommand {\kbf} {{\bf k}}
\newcommand {\pbf} {{\bf p}}
\newcommand {\rbf} {{\bf r}}                     
\newcommand {\rzbf} {r$_{\rm {\bf z}}({\rm {\bf m}}_{\rm {\bf T}})$}

\newcommand {\SL} {{\it S }}

\newcommand {\epsilonl} {\mbox{\large\bf $\epsilon$}}

\newcommand {\epsilo} {\mbox{\large\bf $\epsilon$}}

\newcommand {\fl} {$\Lambda \Lambda$}

\newcommand {\sfl} {$\Lambda$}
\newcommand {\aflafl} {$\bar \Lambda \bar \Lambda$}

\newcommand {\alphale} {\alpha_{\Lambda}}

\def\beq{\begin{equation}}  
\def\eeq{\end{equation}}
\def\bea{\begin{eqnarray}}
\def\eea{\end{eqnarray}}
\def\eq#1{{Eq.~(\ref{#1})}}

\begin{document} 
\bibliographystyle{unsrt}
\begin{titlepage}
\begin{center}
\vspace{1.9cm}

{\Large\bf Bose-Einstein and Fermi-Dirac Interferometry\\ 
\vspace*{3mm}

in Particle Physics}

\end{center}
\vspace{0.9cm}
 
\begin{center}
{\large G}{\small IDEON} {\large A}{\small LEXANDER}\\
\vspace*{6mm}

School of Physics and Astronomy\\
Raymond and Beverly Sackler Faculty of Exact Sciences\\
Tel-Aviv University,
Tel-Aviv 69978, Israel\\
\end{center}
\vspace{1.2cm}

\begin{abstract}
The application of the Bose-Einstein and Fermi-Dirac interferometry 
to multi-hadron final states
of particle reactions is reviewed.
The underlying theoretical concepts of particle interferometry
is presented where a special emphasis is given to the recently
proposed Fermi-Dirac correlation analysis.
The experimental tools  
used for the interferometry analyses and the interpretation of their
results are discussed in some details. In particular the 
interpretation of the
dimension $r$, as measured from the interferometry analysis, is 
investigated and compared to that measured in heavy-ion collisions.
Finally the similarity between the dependence of $r$ 
on the hadron mass and the interatomic separation on the atomic mass
in Bose condensates is outlined.
\end{abstract}

\vspace{3cm}
\end{titlepage}

\newpage
\tableofcontents
\newpage

\section{Introduction}
The two-particle {\it intensity} interferometry ({\it intensity}
correlation) method has been worked out
and exploited for the first time 
in the 1950's by Hanbury-Brown and Twiss (HBT) 
\cite{hbt1,hbt2,hbt3} to correlate intensity
of electro-magnetic radiation, arriving from extraterrestrial
radio-wave sources, 
and thus measure 
the angular diameter of stars and other 
astronomical objects.  
This method rests on the fact that bosons obey the Bose-Einstein
statistics so that the symmetrisation of the multi-particle wave
function
affects the measured many-particle coincident 
spectra to lead to an 
enhancement, relative to the single particle spectra, whenever they
are emitted close by in time-phase space.
Thus the interest in the Bose-Einstein correlation properties was not 
only restricted to their fundamental quantum theoretical
aspects but above all as a tool in the study of the
properties and spatial extent of sources emitting radiation and
particles
in the fields of astronomy, and non-perturbative QCD aspects of
particles and nuclei interactions.\\  

\begin{figure}[ht]
\centering{\epsfig{file=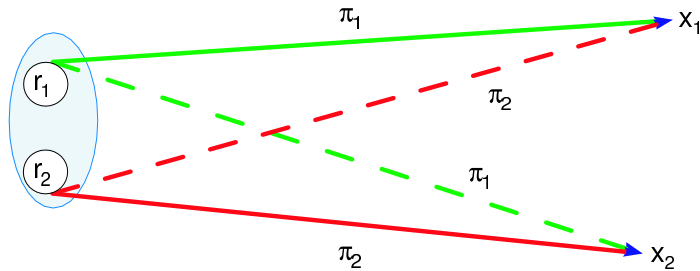,height=3.5cm}}
\caption{\small The two indistinguishable diagrams that 
describe the
emission of two identical bosons, $\pi_1$ and $\pi_2$,
emerging from the two points, $r_1$ and
$r_2$, which lie within an emitter volume, and are detected 
at the positions $x_1$ and $x_2$. 
}
\label{bose_dia}
\end{figure} 
The difference between the {\it intensity} interferometry and the
conventional {\it amplitude} interferometry can be illustrated 
\cite{boal,baym} with the help of
Fig. \ref{bose_dia}. Consider a finite 
source which emits two
indistinguishable particles from the positions $r_1$ and $r_2$
which are later observed at positions $x_1$ and $x_2$.
In an {\it amplitude}
interferometry the positions $x_1$ and $x_2$ could be a slit through
which the emitted particles pass. The particles could then produce an
interference pattern which will depend on the relative phase of the
particle's amplitude as measured at $x_1$ and $x_2$.
In an {\it intensity} interferometry a normalised correlation function
$\stackrel{\sim}K(1,2)$ 
of the particles {\bf 1} and {\bf 2} is formed from the 
average number $\langle n_{1,2} \rangle$
of counts where $n_1$ and $n_2$ are detected simultaneously at 
$x_1$ and $x_2$:
\begin{equation}
\stackrel{\sim}K(1,2)\ =\ \frac{\langle  n_{1,2}\rangle} 
{\langle n_1 \rangle\langle n_2 \rangle}\ -\ 1\ .
\end{equation} 
The correlation function is thus
proportional to the intensity of the particles at $x_1$ and $x_2$.
Because of the symmetrisation of their wave function, identical
particles can have a nonzero correlation function even if the
particles are otherwise noninteracting.\\ 

As can be expected {\it intensity} interferometry is closely related to the
{\it amplitude} interferometry which essentially measures the square of the
amplitudes $A_1$ and $A_2$ falling on the detectors 
$x_1$ and $x_2$:
$$|A_1+A_2|^2\ =\ |A_1|^2+|A_2|^2+(A_1^{\star}A_2 + A_1
A_2^{\star})\ ,$$  
where the last term is the {\it fringe visibility} denoted by $V$, 
which is
the part of the signal which is sensitive to the separation between
the emission points. 
Averaged over random variations its square is given by the product
of the intensities landing on the two detectors: 
$$\langle V^2\rangle\ =\ 2\langle |A_1|^2|A_2|^2 \rangle\ + \
\langle A_1^{{\star}2}A^2_2\rangle\ +\ \langle
A_1^2A_2^{\star 2}\rangle \to 2\langle I_1I_2 \rangle\ .$$
Since  the last two terms of the $\langle V^2 \rangle$ expression
vary rapidly and average to zero, $\langle V^2\rangle$
is proportional to the time averaged correlation of the 
product of the two intensities 
(see e.g. Ref. \cite{Wiedemann}).\\

The Bose-Einstein correlation (BEC) analysis method 
was for the first time introduced in 1959/60 by Goldhaber, Goldhaber, Lee
and Pais (GGLP) \cite{goldhaber,goldhaber1} 
to the hadron sector in order to
study the time-space structure of identical pions produced in particle
interactions. 
Their observation of a BEC enhancement
of pion-pairs emerging from $\overline{p}p$ annihilations
at small relative momenta 
was parametrised in terms of the finite spatial extension of the
$\overline{p}p$ source and the finite localisation
of the decay pions. The momentum range of the enhancement 
could then be related to the size of the particle source in space 
coordinates. Since then the boson interferometry has been
used extensively in variety of particle interactions and
over a wide range of energies.\\

In the mid 1990's the boson intensity interferometry measurement 
method has been
extended to the spin 1/2 baryon sector utilising the Fermi-Dirac
statistical properties and in particular the Pauli exclusion
principle \cite{alex_lipkin}.  
Already the results from the first attempt to study the Fermi-Dirac 
correlation
(FDC) of the $\Lambda \Lambda$ and $\overline{\Lambda}
\overline{\Lambda}$ 
pairs, emerging from the
decay of the gauge $Z^0$ boson, indicated
that the spatial extension of the hadron source is strongly dependent 
on its mass.\\
\begin{figure}[ht]
\centering{\epsfig{file=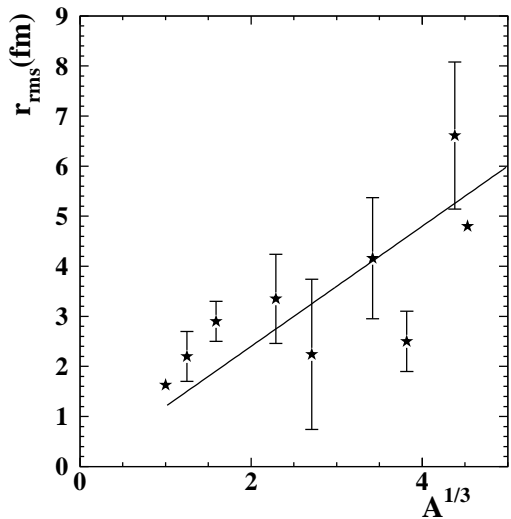,height=7cm,
bbllx=32pt,bblly=156pt,bburx=532pt,bbury=664pt,clip=}}
\caption{\small
An early compilation of values obtained for the emitter dimension $r_{rms}$,  
extracted from interferometry analyses of identical charged pion-pairs
produced in heavy ion collisions are displayed as a function of A$^{1/3}$
where A is the atomic number of the projectile. 
The values are taken from Ref. \cite{chacon1}
and whenever more than one value is quoted for a 
given projectile nucleus, the plotted data 
points represent the average weighted values. The solid line
represents the relation $r_{rms}\ =\ 1.2\times A^{1/3}$ fm.
}
\label{fig_nunu}
\end{figure}

In parallel to the BEC studies of hadronic final states produced in particle
reactions, extensive pion-pair correlation analyses were and are
applied to heavy ion
collisions in order to explore the characteristics of these 
rather complex reactions
\cite{heinz,heinz_bec,Wiedemann}. 
A compilation, taken from Ref. \cite{chacon1},
of the spatial dimension, $r_{rms}$, extracted from
the BEC of two identical charged pions emerging from
nucleus-nucleus collisions, is shown
in Fig. \ref{fig_nunu} as a function of $A^{1/3}$ where $A$ is
the atomic number of the projectile nucleus. As seen in 
Fig. \ref{fig_nunu}, $r_{rms}$
increases more or less linearly with $A^{1/3}$. Moreover, as 
seen from the figure,
the line $r_{rms}$ = 1.2$\times A^{1/3}$ fm, which corresponds to the
dependence of the nucleus radius on its atomic number $A$,
follows rather well the BEC derived spatial dimensions. This
observation, as well as the results from the HBT analyses in
astronomy, led to the expectation that BEC should
also render information on the geometrical extension of the hadron
emitter source in particle collisions.\\

The main purpose of the present report is to cover the field of BEC analyses
applied to hadrons produced in particle collisions 
and only refer to heavy ion reactions whenever
a comparison between the two classes of reactions   
is of a particular interest. A special emphasis is given to recent
developments in the particle interferometry such as the
inclusion of identical fermion correlations and the new
results concerning the spatial dimension dependence on the 
hadron mass. Furthermore, the report is aimed  
mainly to cover the experimental
aspects of the BEC and FDC interferometry and their direct physics
implication and achievements   
while trying to restrict the theoretical discussions to essentials
so as not to overburden the non-expert reader.\\

The organisation of this report is as follows. The basic concepts of
bosons and fermions interferometry are discussed in sections \ref{sect_basic} 
and \ref{sect_fdc}.
In section \ref{sect_exp} details concerning the experimental aspects of the 
correlation measurements are described. Results obtained from the
1-dimensional correlation analyses are summarised
in section \ref{sect_oned}. The observation that 
the spatial dimension deduced from
the correlation measurements depends on the hadron
mass is evaluated in section \ref{sect_rdep} and conceivable grounds
for
this behaviour are discussed. Results obtained from the multi-dimensional  
correlation analyses are dealt with in section \ref{sect_multi} and a 
relation between interatomic separation in Bose condensates and
the spatial dimension of hadrons emerging in high energy reaction
is illustrated. In section \ref{inter} the question concerning the
interpretation of the BEC and FDC deduced geometrical dimension
is discussed.
Finally a summary is presented in section \ref{sect_conclusion}.\\

\section{Basic concepts in particle interferometry}
\label{sect_basic}
\subsection{The Bose-Einstein correlation of two hadrons}
\label{basic}
In deriving the main expressions used for the boson interferometry we follow
closely Ref. \cite{weiner}. As it is well known
in quantum mechanics the interchange of two out of $N$ indistinguishable
bosons does not change the wave function describing this multi-boson
state.
This feature of the Bose-Einstein statistics means that the state $\Psi$
has the property that

$$ \Psi(1,2,\cdots, N)\ =\ \Psi(2,1,\cdots, N)\ ,$$ 
which leads to an interference term in $|\Psi|^2$ that enhances 
near in time-phase space the
production of indistinguishable bosons. Let us first consider a source
of discrete emission points, $\rho_i$, each characterised by a probability
amplitude $F_i(\rbf)$ in the 3-vector $\rbf_i$ phase space
$$F_i(\rbf)\ =\ \rho_i\delta^3(\rbf - \rbf_i)\ .$$
Next we introduce the central assumption pertaining to the BEC effect 
namely, the 
{\it chaotic} or the total {\it incoherence} limit, 
which corresponds to 
the situation where the phases of the production amplitudes 
wildly fluctuate 
in every point of space. In this limit all the phases can be set
to zero.
If $\psi_{\kbf}(\rbf)$ is the wave function of the 
emitted particle,
then the total probability $P(\kbf)$ to observe the emission of one
particle with a 3-momentum vector $\kbf$ is given by summing up the
contributions from all the $i$ points, that is 
$$P(\kbf)\ =\ \sum_i |\rho_i\psi(\rbf_i)|^2\ .$$
For simplicity we will further use plane wave functions
$\psi_{\kbf} \propto e^{i (\kbf \rbf +\phi)}$  
where in the incoherent case 
we can set $\phi = 0$. Next we replace the sum by an integral
so that
\begin{equation}
P(\kbf)\ =\ \int |\rho(\rbf)|^2 d^3r\ .
\label{eq_single}
\end{equation}
The probability to observe two particles with momenta $\kbf_1$ 
and $\kbf_2$ is
\begin{equation}
P(\kbf_1, \kbf_2)\ =\ \int |\psi_{1,2}|^2|\rho(\rbf_1)|^2|\rho(\rbf_2)|^2
d^3r_1d^3r_2\ ,
\label{eq_double}
\end{equation}
where $\psi_{1,2}\ =\ \psi_{1,2}(\kbf_1,\kbf_2,\rbf_1,\rbf_2)$ is the
two-particle wave function.\\

Taking incoherent plane waves, then for two identical bosons
the symmetrised $\psi_{1,2}$ is of the form
\begin{equation}
\psi^s_{1,2}\ =\ \frac{1}{\sqrt{2}}\left [e^{i(\kbf_1\rbf_1+\kbf_2\rbf_2)}
+ e^{i(\kbf_1\rbf_2+\kbf_2\rbf_1)} \right]
\label{eq_psi12}
\end{equation}
so that
\begin{equation}
|\psi^s_{1,2}|^2\ =\ 
1 + \cos[(\kbf_1-\kbf_2)(\rbf_1-\rbf_2)]\ =\ 1\ +\
\cos[\Delta \kbf(\rbf_1-\rbf_2)]\ .
\label{eq_psi12a}
\end{equation}

\noindent
Using Eqs. (\ref{eq_single}), (\ref{eq_double}) and (\ref{eq_psi12a})
one can define a second order correlation function 
\begin{equation}
C_2(\kbf_1,\kbf_2) \equiv \frac{P(\kbf_1,\kbf_2)}{P(\kbf_1)P(\kbf_2)}
\ =\ 1\ +\ \frac{
\int \cos[\Delta \kbf(\rbf_1-\rbf_2)]|\rho(\rbf_1)|^2|\rho(\rbf_2)|^2
d^3r_1d^3r_2}{P(\kbf_1)P(\kbf_2)}\ ,
\label{eq_c2}
\end{equation}
where $\Delta \kbf\ =\ \kbf_1-\kbf_2$.  
Assuming that the emitter extension $\rho(\rbf)$ is localised 
in space and time then it follows that when $\Delta \kbf\ =\ 0$
the last term of  Eq. (\ref{eq_c2}) can vary between the values 0 
to 1.
From Eq. (\ref{eq_c2}) one obtains after integration 
(Fourier transformation)
$$C_2(\Delta \kbf) =\ 1\  +\ |\rho(\Delta \kbf)|^2\ .$$
In many of the two-boson one dimensional BEC analyses one uses the
Lorentz invariant parameter $Q$, 
defined as
$Q^2\ =\ Q^2_2\ =\ -(q_1-q_2)^2\ \equiv\ M^2_2\ -\ 4\mu^2$. Here
$q_1$, $q_2$ and $M^2_2$ are respectively
the 4-momentum vectors and the  
invariant mass squared of the two identical bosons of
mass $\mu$. 

\begin{figure}[ht]
\centering{\epsfig{file=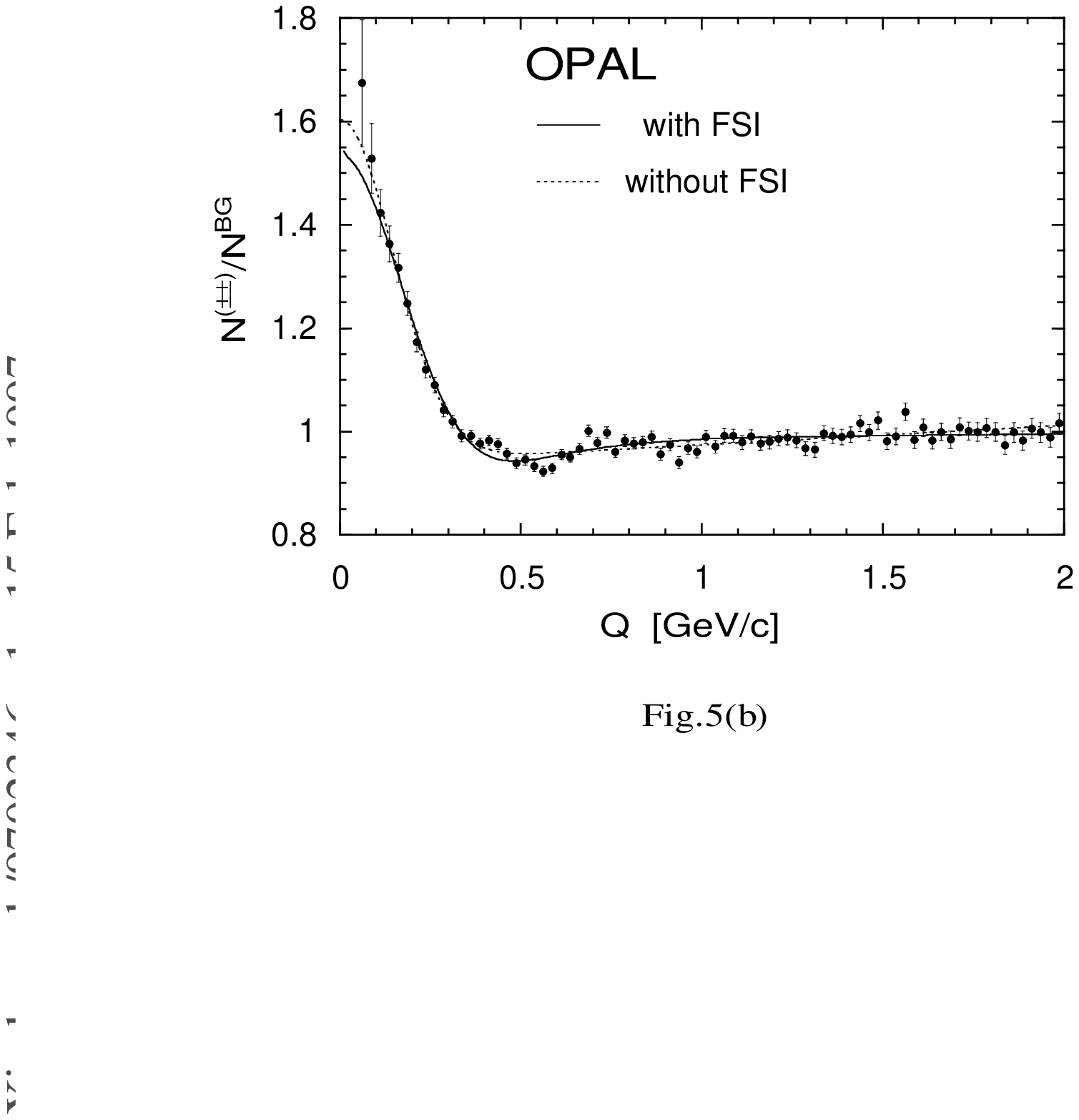,height=6.5cm,bbllx=95pt,bblly=457,
bburx=466pt,bbury=732pt,clip=}
\ \  
\ \epsfig{file=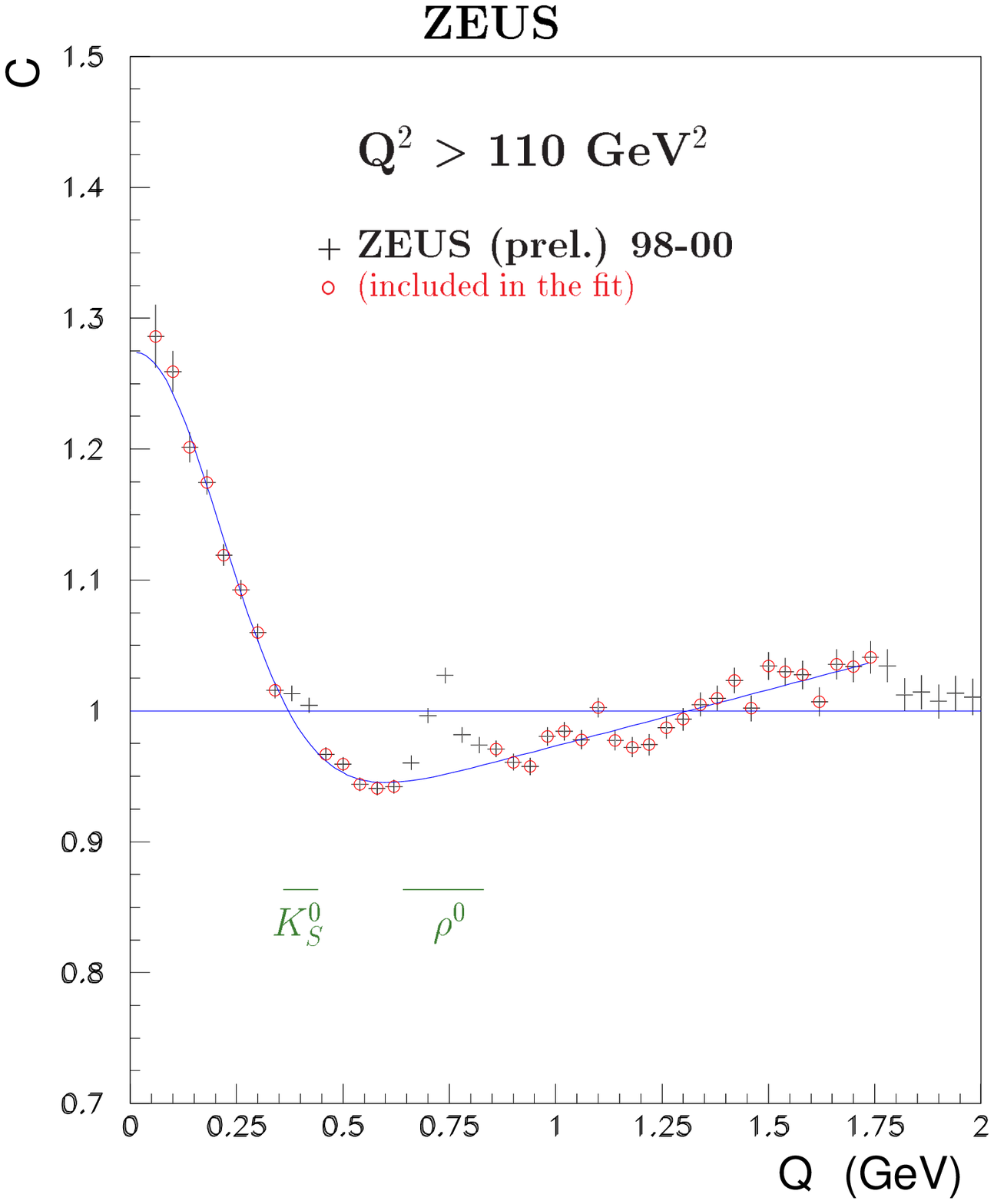,height=8.5cm}}
\caption{\small The $\pi^{\pm}\pi^{\pm}$ BEC as a function of $Q$.
Left: OPAL results \cite{acton} obtained 
in the hadronic $Z^0$ decays.
The solid and dashed lines represent the fit results of
Ref. \cite{final-state} respectively with and  
without the inclusion of final state interactions (see Sec. 
\ref{sec_final}). 
Right: ZEUS results obtained in deep inelastic $ep$ scattering
with momentum transfer of $Q^2 > 110$ GeV$^2$ \cite{zeus_pipi}.
The regions of $K^0_S$ and $\rho^0$ which were omitted from the fit
of the correlation function $C_2(Q)$ are indicated.    
}
\label{fig_example}
\end{figure}

Thus one obtains
\begin{equation}
C_2(Q) =\ 1\  +\ |\rho(Q)|^2\ .
\label{eq_main}
\end{equation}
Assuming further that the source 
is described by a spherical symmetric 
Gaussian density distribution of emitting centres 
$$\rho(r)\ =\ \rho(0)e^{-{\frac{r^2}{2r_0^2}}}\ ,$$ 
then the Bose-Einstein correlation function assumes the form
$$C_2(\Delta \kbf)\ =\ 1\ +\ e^{-r_0^2\Delta \kbf^2}\ .$$
In terms of the variable $Q$ and the dimension $r_G$, 
introduced by Goldhaber et al. 
\cite{goldhaber,goldhaber1},
the correlation function in the completely chaotic limit is equal to    
\begin{equation}
C_2(Q)\ =\ 1\ +\ e^{-r_G^2Q^2}\ .
\label{eq_qr}
\end{equation}
In the completely coherent case it can be shown 
\cite{ringner} that
$C_2(Q)\ =\ 1$. In order to accommodate those cases where the source 
is not completely
chaotic one introduces a chaoticity 
parameter $\lambda_2$ which can vary 
between the value 0, corresponding to a complete coherent case, to
the value 1 at the total chaotic limit. Thus Eq. (\ref{eq_qr}) is transformed
to the GGLP form  
\begin{equation}
C_2(Q)\ =\ N(1\ +\ \lambda_2 e^{-r_G^2Q^2})\ ,
\label{c_two}
\end{equation}
where $N$ is added as a normalisation factor.
Since the strength of the BEC effect depends also on the experimental
data quality, like the purity of the identical boson sample, 
$\lambda_2$ is often also 
referred to as the BEC strength parameter. In the following,
unless otherwise stated, 
we will denote by $r$ the dimension values obtained from 
BEC analyses which used 
the GGLP parametrisation, that is  \mbox{$r$ $\equiv r_G$}.
Two examples of a typical behaviour of the correlation function 
$C_2(Q)$ of identical charged pion-pairs are
shown in Fig. \ref{fig_example}. The first are the results of OPAL
\cite{acton}
where the pion-pairs were taken from the hadronic $Z^0$ decays and the
second reported by the ZEUS collaboration 
\cite{zeus_pipi} in their study of the 
deep inelastic $ep$ scattering produced at the HERA collider.

\subsection{The Kopylov-Podgoretskii parametrisation}
In addition to the Goldhaber parametrisation, given by Eq. (\ref{c_two}),
another favourite parametrisation is that proposed by 
Kopylov and Podgoretskii (KP) \cite{kopylov1,kopylov2,kopylov3} 
which corresponds to a radiating sphere surface of
radius $r_{\small KP}$ with incoherent point-like oscillators of
lifetime $\tau$, namely
\begin{equation}
C(q_t,q_0) = 1 + 
\lambda [4J^2_1(q_tr_{\small KP})/(q_tr_{\small KP}^2]/[1+(q_0\tau)^2]\ ,
\label{eq_kopylov}
\end{equation}
where $J_1$ is the first-order Bessel function. Here $q_0$ is equal to
$|E_1 - E_2|$ and $q_t = |{\bf q}\times\kbf|/|\kbf|$ where
$\kbf = \kbf_1 +\kbf_2$, the sum of the two hadron momenta and
${\bf q} =   \kbf_1 - \kbf_2$ is the difference between them. 
This correlation expression is not Lorentz invariant and its variables are
calculated in the centre of mass system (CMS) 
of the final state hadrons. The
correlation function $C(q_t,q_0)$  
can be written in terms of the Goldhaber geometrical parameter
$r_G$ as \cite{hofmann}
\begin{equation}
C(q_t,q_0) = 1 + \lambda \exp[-r^2_Gq_t^2-r^2_Gq_0^2/(\gamma^2-1)]\ ,
\label{eq_kopylov1}
\end{equation}
where $\gamma$ is the $\gamma -$factor of the identical boson pair.
From a comparison between Eqs. (\ref{eq_kopylov1}) and (\ref{c_two}) 
it is clear that the
parameters $r_G$ and $r_{\small KP}$ have different interpretations. At small
$q_t$ and $q_0$, however, the parametrisation (\ref{eq_kopylov})
can be approximated to be  
\begin{equation}
C(q_t,q_0) = 1 + \lambda \exp[ -(r_{\small KP}/2)^2q_t^2 - (q_0\tau)^2]\ .
\label{eq_kopylov2}
\end{equation}
From  Eqs. (\ref{eq_kopylov1}) and (\ref{eq_kopylov2}) one finds  
the approximate relation $r_{\small KP} \simeq 2r_G$
which is verified experimentally. For example, in a study
\cite{opal_pipi} of the BEC of the $\pi^{\pm}\pi^{\pm}$ system
emerging from the decay of the $Z^0$ boson one obtained
$r_G=0.955\pm 0.012(stat.)\pm 0.015(syst.)$ fm as compared to
$r_{\small KP}=1.778\pm 0.023(stat.)\pm 0.036(syst.)$ fm.\\ 

In contrast to the rather simple and straightforward BEC analysis
offered by the GGLP parametrisation, the KP method expects 
a correlation enhancement at low, near zero,  $q_0$ values so that the analysis
is carried out in several $q_0$ energy slices.
In addition the KP formalism  has an extra free parameter,
$(q_0\tau)^2$, which has to be 
determined through a fit to the data. For these reasons the 
KP correlation investigations require higher statistical data samples 
than those needed for the BEC analysis in the GGLP method. 

\subsection{Higher order Bose-Einstein correlations}
\label{higher}
A Bose-Einstein correlation enhancement is also expected to be present 
in identical boson systems of more than two particles
when they emerge from the interaction within a small 
time-space region.
In the search for these so called, {\it higher order} BEC enhancements, 
one has to differentiate between those produced
from the lower BEC order(s)
and those who are {\it genuine} correlations. 
The normalised
over-all inclusive correlations of $n$ identical bosons is given by
\cite{dremin}
\begin{equation}
R_n\ =\ \frac{\rho_n(p_1,p_2\tcdot p_n)}{\rho_1(p_1)\rho_1(p_2)\tcdot\rho_1(p_n)}\
=\ \sigma^{n-1}\frac{d^n\sigma}{dp_1dp_2\tcdot dp_n} 
\Biggm / 
\left \lgroup \frac{d\sigma}{dp_1}\frac{d\sigma}{dp_2}\tcdot\frac{d\sigma}{dp_n}\right \rgroup\ ,
\label{eq_r}
\end{equation}
where $\sigma$ is the total boson production cross section,
$\rho_1(p_i)$ and $d\sigma/dp_i$ are the single-boson density in
momentum space and the inclusive cross section, respectively.
Similarly $\rho_n(p_1,p_2\tcdot p_n)$ and $d^n\sigma/(dp_1dp_2\tcdot dp_n)$ 
are respectively
the density of the $n$-boson system and its inclusive cross section.
The product of the independent one-particle densities
$\rho_1(p_1)\rho_1(p_2)\tcdot\rho_1(p_n)$ is referred to as the 
reference density
distribution, or reference sample, to which the measured correlations
are compared to. Specifically the inclusive two-boson density 
$\rho_2(p_1,p_2)$ can be written as:
\begin{equation}
\label{eq_rho2}
\rho_2(p_1,p_2) \ = \ \rho_1(p_1)\rho_1(p_2) + K_2(p_1,p_2)\ ,
\end{equation}
where $\rho_1(p_1)\rho_1(p_2)$ represents the two independent boson
momentum spectra and  
$K_2(p_1,p_2)$ describes the two-body correlations. In this simple
case of two identical bosons the normalised density function $R_2$,
defined by Eq. (\ref{eq_r}), already measures the genuine two-body
correlations which here (see Sec. \ref{basic})  
is referred to as the $C_2$ correlation function.
Thus one has
\begin{equation}
C_2 \equiv R_2 \ = \ 1 + \stackrel{\sim}{K}_2(p_1,p_2) \ ,
\end{equation}
where $\stackrel{\sim}{K}_2(p_1,p_2) =
K_2(p_1,p_2)/[\rho_1(p_1)\rho_1(p_2)]$ is the normalised two-body
correlation term which in the GGLP parametrisation 
defined in Eq. (\ref{c_two}), is equal to $\lambda_2 \exp(-Q^2_2 r^2_2).$\\

The inclusive correlation of three identical bosons,
$\rho_3(p_1,p_2,p_3)$, includes
the three independent boson momentum spectra, the two-particle
correlation $K_2$ and the genuine three-particle correlation $K_3$,
namely:
\begin{equation}
\label{eq_rho3}
\rho_3(p_1,p_2,p_3) \ = \ \rho_1(p_1)\rho_1(p_2)\rho_1(p_3) +
\sum_{(3)} \rho_1(p_i)K_2(p_j,p_k) + K_3(p_1,p_2,p_3) \ ,
\end{equation}
where the summation is taken over all the three possible permutations.
The normalised inclusive three-body density, is then given by
\begin{equation}
\label{eq_r3}
R_3 \ = \ \frac{\rho_3(p_1,p_2,p_3)}{\rho_1(p_1)\rho_1(p_2)\rho_1(p_3)} \ = \
1 + R_{1,2} + \stackrel{\sim}{K}_3(p_1,p_2,p_3) \ .
\end{equation}
Here
$$R_{1,2} \ = \ \frac{\sum_{(3)} \rho_1(p_i)K_2(p_j,p_k)}
{\rho_1(p_1)\rho_1(p_2)\rho_1(p_3)} \ \ \ \ {\rm{and}} \ \ \ \
\stackrel{\sim}{K}_3(p_1,p_2,p_3) =
\frac{K_3(p_1,p_2,p_3)}{\rho_1(p_1)\rho_1(p_2)\rho_1(p_3)}$$
represent the mixed three-boson system in which only two of them are
correlated and
the three-boson correlation. In analogy to $C_2$, one 
defines a correlation function $C_3$ which measures the genuine
three-boson correlation, by subtracting from $R_3$ the term which
contains the two-boson correlation contribution. Thus
\begin{equation}
\label{eq_genuine3}
C_3 \ \equiv \ R_3 - R_{1,2} \ = \ 1 + \stackrel{\sim}{K}_3(p_1,p_2,p_3) \ ,
\end{equation}
which depends only on the genuine three-boson correlation. For the
study of the three-boson correlation one often uses 
the variable $Q_3$ which,
analogous to the variable $Q^2_2$, is defined as
\[ Q^2_3 \ = \ \sum_{(3)}q^2_{i,j} \ = \ M^2_3 - 9\mu^2 \ , \]
where the summation is taken over all the three different $i,j$ 
boson-pairs. Here $M^2_3$ is the invariant mass squared of the three-boson
system and $\mu$ is the mass of the single boson. 
From the definition of this three-boson variable it is clear
that as $Q_3$ approaches zero so do all the three related $q_{i,j}$ 
values
which eventually reach the region where the two-boson BEC enhancements
are observed.
It has been shown \cite{mark_pipi}
that the genuine three-pion correlation function $C_3(Q_3)$ 
can be parametrised by the expression 
\begin{equation}
C_3(Q_3)\ =\ 1\ +\ 2\lambda_3 e^{-Q^2_3 r^2_3}\ ,
\label{eq_c3}  
\end{equation}   
where $\lambda_3$ is the chaoticity parameter which may assume  
a value between zero and one.\\

The method outlined here for the extraction of the three-boson
BEC enhancement can in principle be extended to higher orders, 
however it becomes too cumbersome to be of a practical use 
and on top of it, it requires very high statistics data. 
For these reasons two other approaches have been advocated and utilised
experimentally.
In the first, one measures experimentally the over-all correlation,
as defined by $R_n$ in Eq. (\ref{eq_r}), and then with the help of
various models one tries to extract the higher order genuine BEC parameters
$r_n$ and $\lambda_n$. Such an approach is adopted for example 
in references \cite{biyajima_high,neumeister}. In the second approach
one analyses the data in terms of correlation functions
\cite{car,dewolf} known as
factorial cumulant moments or, in their integrated form are
referred to as the semi-invariant 
cumulants of Thiele which constitutes an important element in
proving the central limit theorem in statistics (see
e.g. \cite{barlow}). 
Having $n$ particles
in a given domain then the  corresponding factorial
cumulant moments will be
different from zero only if a genuine $n$-particle correlation exists.
The shortcoming of the cumulant approach is however the fact 
that its results for a given higher order ($n \geq 3$) pertain to the overall 
genuine correlation present in the data so that the   
partial contribution of the
genuine BEC, parametrised by $r_n$ and $\lambda_n$ 
is not transparent.

\subsection{Bose-Einstein correlation in two and three dimensions}
\label{sec_2d}
\begin{figure}[h]
\centering{\epsfig{file=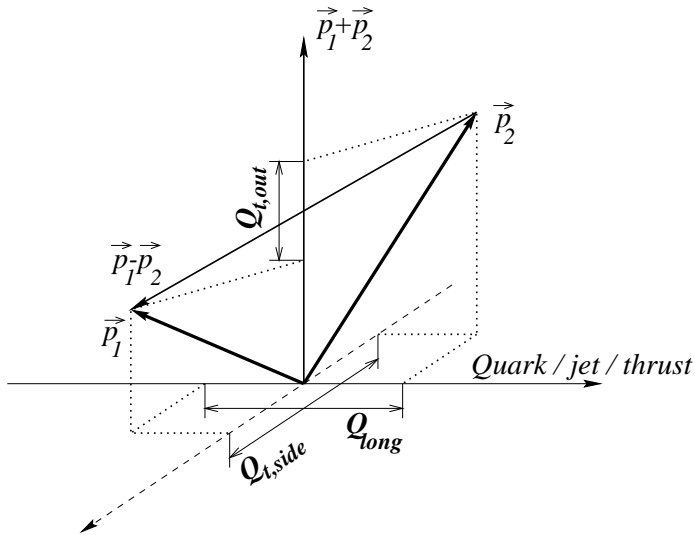,height=7cm}}
\caption{\small The Longitudinal Centre-of-Mass-System coordinates 
(taken from Ref. \cite{delphi_2d}). 
}
\label{fig_lcms}
\end{figure}

In the GGLP interferometry analysis the emitter shape is considered to
be of a spherical shape with a Gaussian density distribution. 
A method to explore the possibility that the time-space extent of the
particle emission region deviates from a sphere, and in fact is
characterised by more than one dimension, has been recently proposed
\cite{csorgo,heinz,geiger}. To this end the BEC analysis is 
carried out in the Longitudinal Centre-of-Mass System (LCMS) shown
schematically in Fig. \ref{fig_lcms}.
This coordinate system is defined for each pair of identical bosons
as the system in which the sum of the boson-pair 3-momenta
$\pbf_1 +\pbf_2$, referred to as the {\it out} - axis, 
is perpendicular
to the {\it thrust} (or {\it jet}) direction of the multi-hadron event
defined as the $z$ - axis
($\equiv$ {\it long} - axis). 
The momentum difference of the pion-pair ${\bf Q}$ is then resolved into
the longitudinal direction
$Q_z\ \equiv \ Q_{||}$ parallel to the thrust axis, to the $Q_{out}$
- axis which is  
collinear with the pair momenta sum and the third axis, $Q_{side}$
which
is perpendicular to both $Q_z$ and $Q_{out}$. In this system the projections
of the total momentum of the particle-pair onto the longitudinal and
side directions are equal to zero. The total $Q^2$
value is given by
\begin{equation}
Q^2=Q^2_z +  Q^2_{side} + Q^2_{out}(1-\beta^2),\ \ \ {\rm{where}}\ \ \  
\beta=\frac{p_{1,out}+p_{2,out}}{E_1 +E_2}\ .
\end{equation}
Here the indices 1 and 2 refer to the first and second boson.
Since $p_{1,z} = -p_{2,z}$ 
one has
 $$Q_z\ =\ Q_{\|}\ =\ p_{1,z} - p_{2,z}\ =\ 2p_{1,z}\ =\ 2p_z\ .$$ 
The difference in the
emission time of the pions, which couples to the energy difference between
them, appears only in the $Q_{out}$ direction.
The 3-dimension correlation function is given by\\
\begin{equation}
C_2(Q_z, Q_{side},Q_{out})\ =\  1\ +\ \lambda_2 e^{-(r^2_zQ^2_z\
+\ r^2_{side}Q^2_{side}\ +\ r^2_{out}Q^2_{out})}\ .
\label{threedim}
\end{equation}

In many cases due to lack of sufficient statistics, one wishes to reduce
the number of parameters to be fitted in the correlation function by
defining the transverse component $r_T$ in the LCMS  
to be $r_T^2=r_{side}^2 + r_{out}^2$ corresponding to
$$Q^2_T\ =\ Q^2_{out}\ +\  Q^2_{side}\ .$$
Thus the correlation function, which is fitted to the data, is of the
form
\begin{equation}
C_2(Q_z, Q_T)\ =\  1\ +\ \lambda_2 e^{-(r^2_zQ^2_z\
+\ r^2_TQ^2_T)}\ ,
\label{twodim}
\end{equation}
where $r_z$, estimated from Eq. (\ref{twodim})
as $Q_z$ approaches zero,
is the longitudinal geometrical radius and $r_T$ is composed of
the transverse radius and the emission time difference.
The experimental findings in heavy ion collisions 
\cite{na44a,na35a},
in hadron-hadron reactions \cite{ehs_na22a,ehs_na22b}
and $e^+ e^-$ annihilations 
\cite{l3a_2d,delphia_2d,opala_2d}, verify the theoretical
expectations of the Lund string 
model \cite{lund_trans1,lund_trans2} 
 that the ratio $r_T/r_z$ is significantly smaller than one
(see Sec. \ref{lund}).\\

The assignment of a well defined physical direction in the LCMS, 
such as thrust axis, allows to study in the framework 
of the BEC analysis an additional meaningful variable namely,
the average transverse mass $m_T$ of the two identical hadrons.
This transverse mass is defined as 
\begin{equation}
m_T\   =\ \frac{1}{2}\left[m_{1,T}\ +\ m_{2,T}~\right]\ =\
\frac{1}{2}\left[\sqrt{m_1^2+p_{1,T}^2}\ 
+\ \sqrt{m_2^2+p_{2,T}^2}~\right]\ ,
\label{mt}
\end{equation}
where the indices 1 and 2 refer to the first and second hadron.

\section{Fermi-Dirac correlation }
\label{sect_fdc}
The application of two-boson correlation to the estimation of an
$r$ dimension 
has recently been extended to identical 
pairs of fermions (baryons) \cite{alex_lipkin}. This extension is 
based on the Fermi-Dirac statistics feature
which prohibit 
the total spin to have the value
${\bf S} = {\bf S}_1 + {\bf S}_2 = 1$ when the two identical fermions
are in an s-wave ($\ell=0$) state. Thus in the so called Fermi-Dirac 
correlation (FDC) method one can,
similarly to the BEC 
analysis, study the contributions of the $S = 0$ and $S= 1$ states to the
di-fermion system as $Q$ approaches zero where, in the absence of
di-baryon resonances,  only the s-wave state survives.\\

The estimation of $r$ from the rate of depletion of the $S=1$ 
population as $Q \to 0$
can be achieved in two ways. The first consists of
a direct measurement of the relative contributions of \SL = 1
and \SL = 0 spin states to the di-baryon system as a 
function of $Q$ (see next section). 
An alternative method consists of a measurement 
of the di-baryon density decrease  
as $Q$ approaches zero, 
assuming its origin to be due to the Pauli exclusion principle. 
This second method is equivalent to that used in the BEC analysis of
two identical bosons. 

\subsection{The spin-spin correlation}
If $F_0(Q)$ and $F_1(Q)$ are respectively
the fraction of the \SL = 0 and the \SL = 1 
contributions, at a given $Q$ value, to the di-baryon system then
one can e.g. study two correlation functions, $C_{S=0}(Q)$ and
$C_{S=1}(Q)$, defined by the ratios:
$$C_0(Q) = \frac{2 F_0(Q)}{F_0(Q) + F_1(Q)/3} 
\ \ \ \ \ \ {\rm{and}} \ \ \ \ \ \
C_1(Q) = \frac{2 F_1(Q)/3}{F_0(Q) + F_1(Q)/3}\ ,$$ 
where $F_1(Q)$ is divided by 3 to
offset the statistical $2S + 1$ spin factor.
\begin{figure}[h]
\centering{\epsfig{file=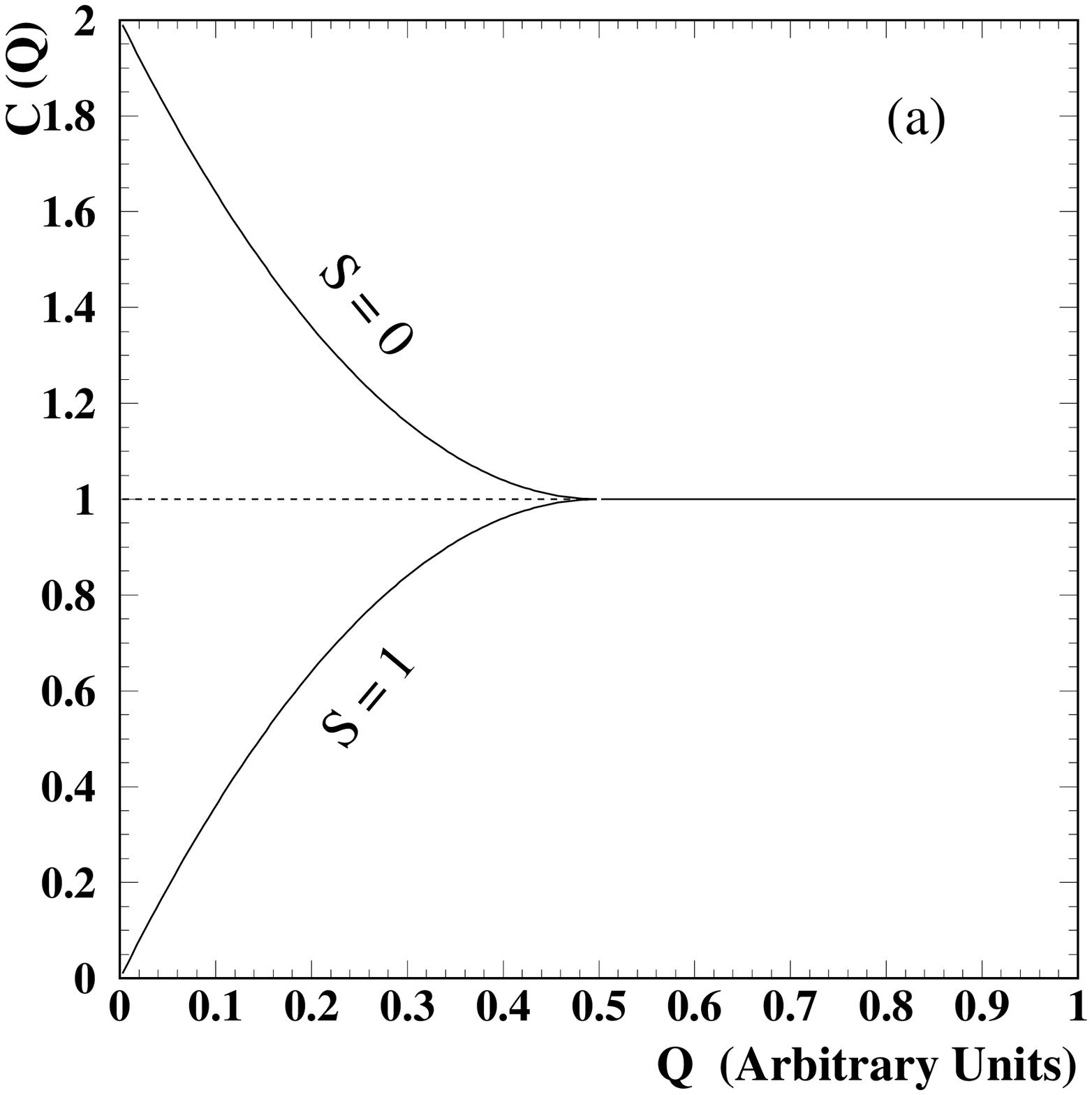,height=7cm,
bbllx=15pt,bblly=145pt,bburx=548pt,bbury=665pt,clip=}\ \ \ \ \
\epsfig{file=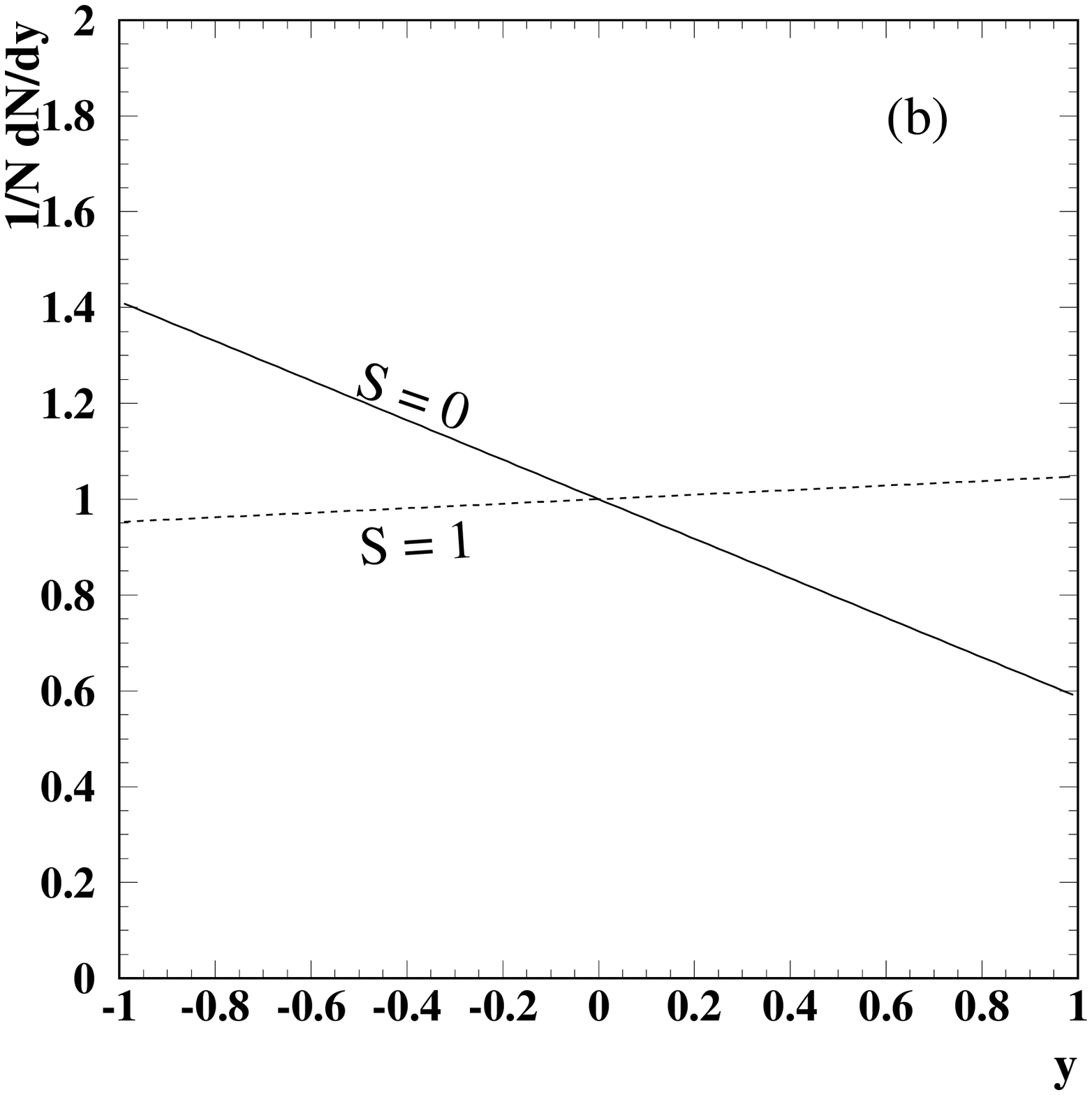,height=7cm,
bbllx=15pt,bblly=145pt,bburx=548pt,bbury=665pt,clip=}}
\caption{\small
(a) Schematic view of the FDC functions $C_0$ and $C_1$ dependence on $Q$,
for the \SL = 0  and  
\SL = 1 states, of an identical spin 1/2 two-baryon system.
The dashed line represents the expectation for a baryon anti-baryon
system in the absence of resonances. (b) 
The $(1/N)dN/dy$ behaviour of the $S=0$ and $S=1$ states of a
two identical baryons sample.} 
\label{fig_fd1}
\end{figure}
At high $Q$ values, where the highest angular momentum $\ell_{max}$ 
is large, one may expect that
$C_0 \cong C_1 \cong 1$ corresponding to a statistical spin
mixture ensemble. On the other hand when $Q = 0$, 
one will observe 
$C_0 \approx 2$ and $C_1 \approx 0$. Thus the $C_0$ behaviour
as a function of $Q$ is similar to the BEC function $C_2(Q)$,
of two identical mesons, which rises as the $\ell_{max}
\rightarrow 0$. As for the di-baryon correlation function $C_1(Q)$
there is no parallel case in the identical charged di-boson system. 
One further assumes the emitter to be a sphere with a Gaussian distribution
so that, analogous to the BEC parametrisation, the di-baryon
correlation functions can be parametrised as \cite{alex_lipkin}:
$$ C_0(Q) = 1 + \lambda e^{-r^2Q^2}\ \ \ \ \ \ {\rm{and}}
\ \ \ \ \ \ C_1(Q) = 1 -  \lambda e^{-r^2Q^2}\ .$$
Here the $\lambda$ parameter, which
can vary between 0 and 1, measures the strength of the effect and is
mainly sensitive to the purity of the data sample. 
A schematic view of the dependence of  $C_0$ and $C_1$ on $Q$ is shown
in Fig. \ref{fig_fd1}a. The study of either $C_0$ or $C_1$ can render 
a value for $r$ however, since the diminishing contribution of the 
\SL = 1 state
is responsible for the variation of the correlation function  
as $Q$ decreases, one usually analyses $C_1$. 
Here it should be noted that 
unlike the BEC effect which can be extended to
many-boson states, the FDC is limited to two identical fermions since
already the third one is in an $\ell > 0$ state 
(Pauli exclusion principle).\\ 
 
A general method for the direct measurement of the total spin composition 
of any two spin 1/2 baryons system
which decay weakly has been outlined in reference \cite{alex_lipkin}.
Here we illustrate this method (later referred as Method I)
in its application to the
$\Lambda \overline{\Lambda}$ and the $\Lambda \Lambda$ systems. 
The simplest measurable case is the one where the
$\Lambda$ decays into $p_1 + \pi^-$ and the other $\Lambda$, 
or $\overline{\Lambda}$, to
$p_2 +\pi^+$ where $p_i$ stands for the decay proton or anti-proton
both of which we will further refer to as protons.
Let us further denote by $y$ the cosine of the angle (in space)
between $p_1$ and $p_2$ and $\langle y\rangle$ as its average,
defined in a system reached after two transformations. The first,
to the CMS of the di-$\Lambda$ pair and the second
where each proton is transformed to the 
CMS of its parent $\Lambda$. To note is that at
$Q = 0$ the second transformation is superfluous.\\ 
 
In the decay of a single\ \sfl , the proton angular 
distribution with respect to the spin direction 
in the centre of mass,
is given by \cite{segre}:
\begin{equation}
dw/d\cos\theta \propto 1 - \alpha_{\Lambda} \cos\theta\ ,
\label{1}
\end{equation}
\begin{flushleft}
where the parity violating  parameter 
$\alpha_{\Lambda}= - \alpha_{\overline{\Lambda}}$ 
was measured experimentally \cite{pdg} to be $0.642 \ \pm \ 0.013$.
Setting $x = \cos\theta$ \ the average \ $\langle x_{\Lambda}\rangle$ \
is given by:
\end{flushleft}
\begin{eqnarray}
\langle x_{\Lambda}\rangle \ = \ \frac{1}{2\pi N}
\int^{2\pi}_0 d\phi \int^{x_{max}}_{x_{min}}
(1 - \alphale x) x dx\ \ \ \ \ {\rm{where}} \ \ \ \ \ 
N \ = \ \int^{x_{max}}_{x_{min}} (1 - \alphale x) dx\ .
\end{eqnarray}
In particular,
the average over
the whole angular range from
$x_{min} = -1 $ to \mbox{$x_{max} = +1 $} yields
$$ \langle x_{\Lambda}\rangle \ = \ -0.214 \pm 0.004 \ \ {\rm{and}} \ \
 \langle x_{\overline{\Lambda}}\rangle \ = \ +0.214 \pm 0.004\ .$$
 
General arguments do not allow $dN/dy$ distribution of
a two spin-1/2 hyperon system at threshold to have a $y$ dependence
higher than its first power.
This means that $dN/dy$ is of the form
$$ dN/ dy = A [ 1 + B_S~y]\ .$$
The factor $B_S$ can then be determined
independent of the total angular momentum $J$ value 
at $Q = 0$
or nearby from the value of $\langle y \rangle$
using the Wigner-Eckart theorem \cite{wigner,eckart}.
Thus for the 
$\Lambda \Lambda$ and $\overline{\Lambda}  \overline{\Lambda}$ pairs 
one obtains
$$B_{S=0} \ = \ - 9 \ \langle x_{\Lambda}\rangle ^2 \ = \ -0.4122
\ \ \ \ \ {\rm{and}}
\ \ \ \ \ B_{S=1} \ = \ + 3 \ \langle x_{\Lambda}\rangle^2 \ = \ +0.1374\ .$$
As a result one then has for the \fl (\aflafl)\ system at, or very near, its
threshold
\begin{equation}
dN / dy|_{_{S = 0}} \ \propto  \ 1 - 0.4122 y
\ \ \  {\rm{and}}\ \ \  
dN / dy|_{_{S = 1}} \ \propto  \ 1 + 0.1374 y\ .
\label{fd_eq1}
\end{equation}
To note is that the $\ell = 0$, $S=1$ state is forbidden by the Pauli
principle.
These two distinctly different distributions,
are shown in Fig. \ref{fig_fd1}b. 
In a similar way one obtains for the $\Lambda \overline{\Lambda}$ system\\
\begin{equation}
dN / dy|_{_{S = 0}} \ \propto  \ 1 + 0.4122 y\
\ \ \ {\rm{and}} \ \ \ \
dN / dy|_{_{S = 1}} \ \propto  \ 1 - 0.1374 y\ .
\label{fd_eq2}
\end{equation}
Even though the Wigner-Eckart theorem is applicable to the di-baryon
system at $Q = 0$, it is shown in Ref. \cite{alex_lipkin} that
Eqs. (\ref{fd_eq1}) and (\ref{fd_eq2}) can also be applied to 
at $Q > 0$
as long as the protons emerging from the $\Lambda$ decays are 
non-relativistic. 
That this method can be extended even to higher $Q$ values has been
pointed out
in Ref. \cite{lednicky}. 
If the parameter $\epsilonl$ is defined as
the fraction of the \SL = 1 contribution to the two-baryon system then
it can be measured as a function of $Q$ by fitting the expression\\
\begin{equation}
dN / dy\ =\ (1\ -\ \epsilonl)\ dN / dy|_{S = 0}\ +\ 
\epsilonl \ dN / dy|_{S = 1}\ ,
\label{fd_eq3}
\end{equation}
to the data.
In the case of a statistical spin mixture, $\epsilonl$ is equal to 
0.75 which yields a constant $dN / dy$ distribution. This
spin analysis  method has the advantage that it directly
measures the \SL =1 depletion as expected from the Pauli principle.
In addition it does not require any reference sample which, as
is well known from the BEC analyses, is the major contributor to the
systematic errors of the measurement (see Sec. \ref{choice}). 
The main disadvantage of the
direct spin measurement is its need for
a rather large data sample. In addition, it is limited to  
spin 1/2 baryons which decay weakly into hadrons and thus is
not applicable, for example, to pairs of protons or neutrons.

\subsection{The phase space density approach}
The probability to observe two particles,
emitted from an interaction with 3-momenta ${\bf k_1}$  and 
${\bf k_2}$, is proportional to the square of its total wave function
$\Psi_{1,2}$ which is equal to the orbital wave function $\psi_{1,2}$ 
times the spin part. In the 
case of identical spin 1/2 baryons, the function $\Psi_{1,2}$ should be 
anti-symmetric under the 
{\bf 1 ${\bf \leftrightarrow}$ 2}
exchange. Assuming plane waves and a
completely
incoherent emitter (i.e. the arbitrary phases can be set to zero) the 
symmetric and the anti-symmetric orbital wave functions are given by

$$ \psi_{1,2}^s\ =\ \frac{1}{\sqrt{2}}\left [e^{i(\kbf_1\rbf_1
+\kbf_2\rbf_2)} + e^{i(\kbf_1\rbf_2+\kbf_2\rbf_1)}\right ]\ \ \ \ \ \ 
{\rm{and}}
\ \ \ \ \ \ 
\psi_{1,2}^a\ =\ \frac{1}{\sqrt{2}}\left [e^{i(\kbf_1\rbf_1
+\kbf_2\rbf_2)} - e^{i(\kbf_1\rbf_2+\kbf_2\rbf_1)}\right ]\ .
$$
From these, as is shown in Sec. \ref{basic},
one obtains for the symmetric orbital wave function, applicable 
to two identical hadrons 
$$ |\psi_{1,2}^s|^2=1 + \cos[(\kbf_1 - \kbf_2)(\rbf_1 -
\rbf_2)] = 1 + \cos(\Delta \kbf \Delta \rbf)\ ,$$
whereas one obtains
$$|\psi_{1,2}^a|^2\ = 1 - \cos[(\kbf_1 - \kbf_2)(\rbf_1 -\rbf_2)]
= 1 - \cos(\Delta \kbf \Delta \rbf)\ ,$$
for the anti-symmetric orbital wave function. 
If we further consider a source with a spherical symmetric Gaussian
density distribution \cite{goldhaber,goldhaber1}
$$f(r) \propto e^{-r^2/(2r_0^2)}\ ,$$
then the rate will be
$$|\psi_{1,2}^s|^2 = 1 + e^{-r_0^2\Delta \kbf^2} \ \ \ \ \ \ {\rm{and}}
\ \ \ \ \ \ \ |\psi_{1,2}^a|^2 = 1 - e^{-r_0^2\Delta \kbf^2}\ .$$ 
Because of the Fermi-Dirac statistics the 
symmetric orbital part of the di-baryon system, 
$|\psi_{1,2}^s|^2$, is coupled to the anti-symmetric spin part, that
is $S=0$, whereas the anti-symmetric orbital part 
$|\psi_{1,2}^a|^2$ is coupled to
the symmetric $S=1$ spin part. If we further assume that at high $Q$
values we face a statistical spin mixture ensemble,
where the probability to find a di-baryon in an S state is
proportional to $2S+1$, we finally obtain for the emission rate,
when properly normalised that
$$|\Psi_{1,2}|^2 = 0.25[1 + e^{-r_0^2\Delta \kbf^2} + 3 
(1 - e^{-r_0^2\Delta \kbf^2})] = 1 - 0.5 e^{-r_0^2\Delta
\kbf^2}\ ,$$ 
or in its Lorentz invariant form 
$$|\Psi_{1,2}(Q)|^2 = 1 - 0.5 e^{-r^2Q^2}\ .$$
Thus one expects at $Q = 0$
a reduction in the correlation function 
to half of its value at the high $Q$ region provided the following conditions 
are satisfied:\\
\vspace{2mm}

\noindent
$\bullet$ At high $Q$ values one has a statistical spin mixture ensemble;\\
$\bullet$ The di-baryon emitter is completely incoherent;\\
$\bullet$ Absence of resonance states at low $Q$ values;\\
$\bullet$ Final states interactions can be neglected.\\
\vspace{2mm}

To accommodate in this FDC analysis method (further referred to as Method II)
the cases where the emitter is not completely incoherent
one introduces, as in the BEC case, a chaoticity factor $\lambda$,
that can vary between 0 and +1,
so that the correlation function assumes the form
\begin{equation}
C(Q) = N(1 - 0.5\lambda e^{-r^2Q^2})\ ,
\label{eq_cqfd}
\end{equation}
where $N$ represents a normalisation factor.
\section{Experimental procedure and data analysis}
\label{sect_exp}
Well suited reactions for a BEC analysis are those  
which lead to multi-hadron final states where the correlations due to
resonances and conservation laws, like energy-momentum 
and charge balance, have 
minor effects. Since in these reactions the fraction of pions is
the highest one, in many of the analyses one assumes that all  
the outgoing particles are pions 
and the contamination from other hadrons are accounted
for by  
a proper correction factor and/or by an increase of the systematic
error. 
The correlation study of kaon and proton pairs require special 
hadron selection criteria
which tend to reduce the data statistics and introduce larger
systematic errors.\\

In the BEC analysis the space 
density of the data hadron pairs dependence on $Q$ is  
compared to a reference sample distribution which serves as a yardstick. 
The correlation  
function, of the form given in
Eq. (\ref{c_two}), is then fitted   
to the ratio of the data density distribution to that of the reference
sample. 
To account in the GGLP 
parametrisation, for the so called 'long range
correlations' due e.g. to energy and momentum conservation, 
Eq. (\ref{c_two}) is often modified
to include linear and quadratic terms in $Q$, namely
\begin{equation}
C(Q) = N(1 + \lambda e^{-r^2Q^2})(1+ \delta Q + \eta Q^2)\ ,
\label{eq_longrange}
\end{equation}
where $\delta$ and $\eta$ are free parameters to be determined 
by the fit to the data.
In assessing the fitting results obtained for the $r$ and $\lambda$ 
parameters  
one has to be aware of the fact that in many cases these are 
not independent and a correlation between the two does exist.
The weak part of the BEC analysis is undoubtedly the fact that it depends 
rather strongly on the chosen reference sample which thus is
the main contributer to the over all systematic errors associated with
the fitted parameters.  The different available 
reference samples and the effect 
of the final state interactions on the BEC analysis results 
are discussed in some details in the following sections.\\

\subsection{Choice of the reference sample}
\label{choice}
As mentioned above the study of the second and higher order particles' 
BEC enhancements requires a yardstick against which they can be detected
and measured as is also evident from the $C_2(\kbf_1\kbf_2)$
definition given by Eq. (\ref{eq_c2}). 
This yardstick is given by the so called reference sample that should
be identical to the analysed data in all its aspects but free from  
Bose-Einstein or Fermi-Dirac statistics effects.
An ideal solution to this requirement really does not exist but one
has a choice of possibilities which satisfy approximately this
requirement. These are divided mainly into two categories.
Reference samples constructed out of the data themselves and those 
supplied by Monte Carlo generated samples which are subject to a full 
simulation of the experimental setup with its 
particle detection capabilities. 

\subsubsection{Reference samples derived from the data} 
The methods where the reference samples are constructed from
the data themselves are often preferred as they are expected to retain
many  of the kinematic and dynamical data correlations such as those 
originating from charge, momentum and energy conservation as well as
correlations arising from hadronic resonances. In the following we will
briefly describe some of the data derived reference samples which, were
and still are, utilised in the BEC analyses of pion-pairs.\\

a) In many studies of the BEC of identical
charged pion-pairs ($\pi^{\pm}\pi^{\pm}$)
the simplest reference sample is used. Namely,
the one which is constructed 
out of the correlation of opposite charged pion-pairs ($\pi^{\pm}\pi^{\mp}$)
present in the same data sample. This choice
however has the following rather severe drawback. Whereas the 
$\pi^{\pm}\pi^{\pm}$ is a so called exotic system void of bosonic
resonances, the origin of the $\pi^{\pm}\pi^{\mp}$ pairs may also be 
resonances where among them the most dominant one is the $\rho(770)$. 
To overcome
this deficiency the expression of the correlation function, like that
given by Eq. (\ref{c_two}), is fitted to the experimental results only in the
$Q$ regions where the data is known to be free of resonances.\\

b) Another frequently used reference sample is the one known under the
name 'mixed-event' sample. In this method one couples two identical pions
each originating from a different data event. In this way one is 
guaranteed that
no BEC effects will exist in the sample but at the price
that all other kinds of correlations, like those arising from
kinematic conservation laws, 
are also eliminated. Furthermore,
as long as the data analysed is produced at low energies where the
the particles emerge to a good approximation isotropically, this
event mixing procedure may be satisfactory. However at higher
energies where the particles emerge in hadron-jets, the mixed event technique
can only be applied to two-jets events where the mixing takes place
between two events with a thrust (sphericity) axes lying very nearby
or alternatively  
after one event is rotated so that  
the two events axes coincide.\\

c) A third method in use for the generation of a reference sample, 
which avoids the need
to rotate the event, 
is constructed by folding 
each  data event along its sphericity or thrust axis so that the
emerging hadrons 
are divided into two hemispheres. The entries to  
the reference sample are then all possible identical pion-pairs 
belonging to different hemispheres. 
This method, as the former one, can only be applied 
to two-hadron jet events.\\ 

Finally to note is that the choice of data derived 
reference samples 
for a BEC analysis of kaon-pairs or for a FDC Method II analysis of 
baryon-pairs is much more
restricted. For example, in the BEC analyses of $K^{\pm}K^{\pm}$
or $K^0_SK^0_S$ pairs the data $K^{\pm}K^{\mp}$ pairs  cannot be 
used due to the strong presence of the $\phi(1020) \to K^{\pm}K^{\mp}$
decay which lies at the very low $Q$ range  
where the BEC interference is near its maximum.
 
\subsubsection{Monte Carlo generated reference samples}
\label{lund}
Modelling of the hadron production in particle reaction plays a
central role in any experimental data analysis in high energy
physics. Its aim is mainly the 
evaluation of the various experimental deficiencies due to the
imperfection of the detection system such as the geometrical
acceptance, angular resolution and separation of tracks 
as well as the limited particle identification capabilities. 
In addition, the hadron production simulation serves as a yardstick
against which various physics phenomena and hypotheses can be 
detected and measured.
In this last capacity modelling of hadron production is also
used extensively in the study of particle correlations.\\

\begin{figure}[ht]
\centering{\epsfig{file=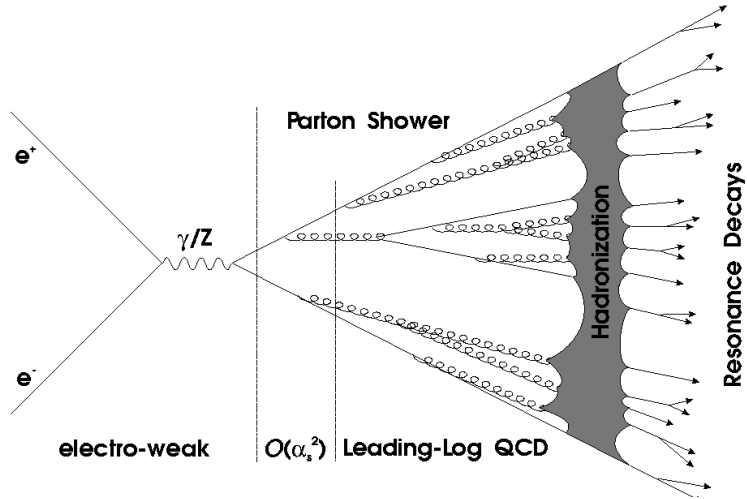,height=10cm,angle=-90,
bbllx=59pt,bblly=85,bburx=467pt,bbury=705pt,clip=}}
\caption{\small
A schematic diagram for the production of hadrons in $e^+e^-$ annihilation. 
}
\label{fig_partsh}
\end{figure}
To illustrate the hadronisation process we show in
Fig. \ref{fig_partsh} a schematic diagram for 
the production of hadrons in the process 
$e^+e^- \to Z^0/\gamma^{\star} \to q \bar q$. 
The $q \bar q$, 
which is not directly detected, results in a relative large number of
hadrons. This final step process is governed by the strong
interactions and can be divided into two parts. Firstly  
a parton cascade develops in which the $q$ and $\bar q$ pair radiates
gluons, 
which in turn may radiate additional gluons and split into 
new $q \bar q$ pairs.
This process is followed by the hadronisation stage which emits the final
observed particles. These two last stages are described in the
standard model of particles and fields in terms of
the Quantum Chromo Dynamics (QCD) theory where the force field between 
partons is the so called colour field.\\

\begin{figure}[ht]
\centering{\epsfig{file=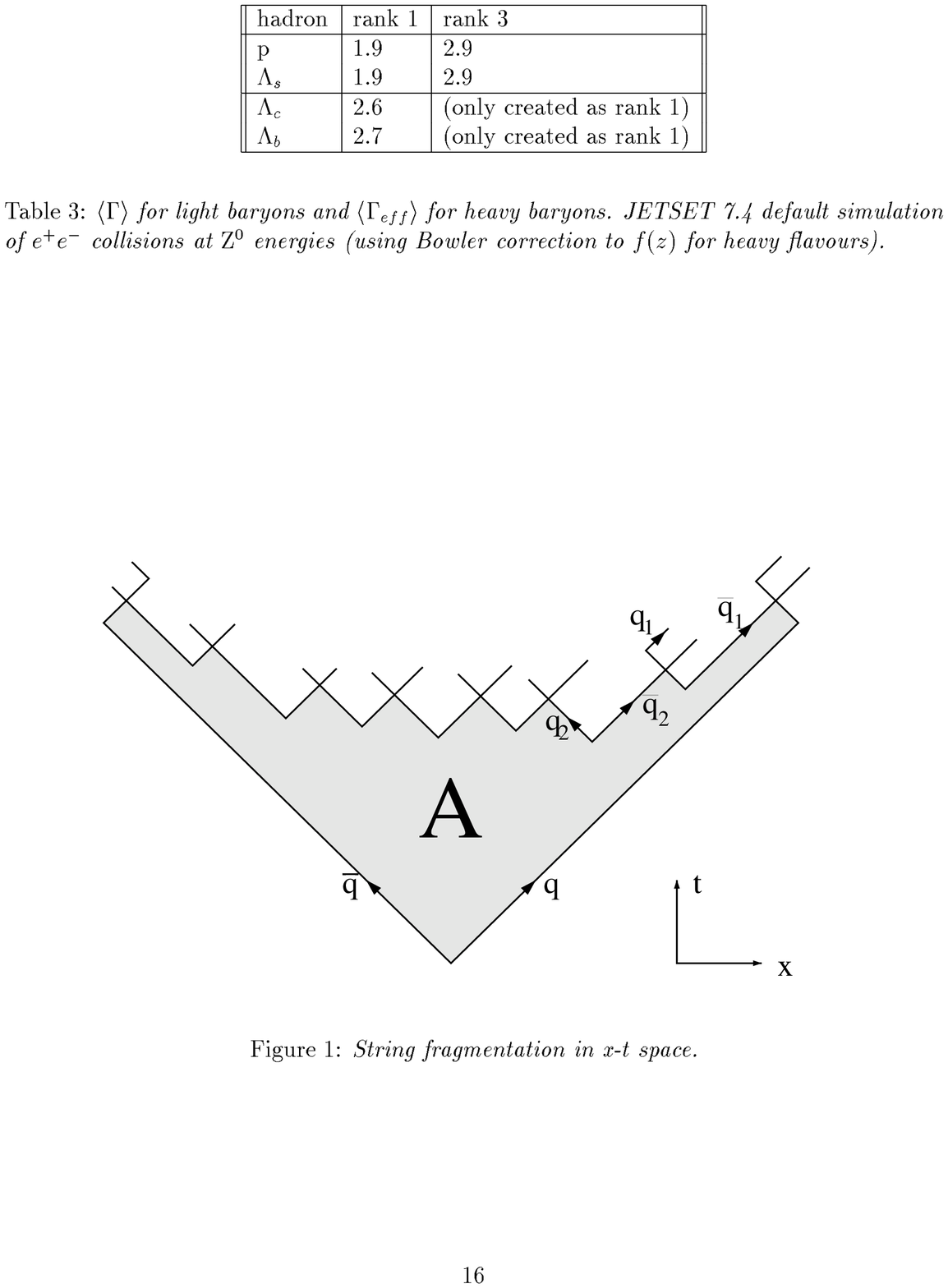,height=7cm,bbllx=85pt,bblly=152,
bburx=510pt,bbury=422pt,clip=}}
\caption{\small
The hadronic decay of a Lund model string spanning the time-space area A. 
}
\label{fig_string}
\end{figure}

At short distances and over short times the quarks and gluons can be 
considered as free particles where the perturbative QCD can be applied.
This is not the case at the  hadronisation stage where there is a need
to rely on non-perturbative QCD models.
One of the leading successful models, which currently is applied 
to many of the high energy
particle reactions leading to hadronic final states, is the
Lund string fragmentation model \cite{lund},
shown
schematically in Fig. \ref{fig_string}. In this Lund model the probability
$|M(q\bar q \to h_1....h_n)|^2$
to emit $n$-hadrons is proportional $e^{-bA}$ where $A$ is the colour
field area 
spanned in time-space by the primary $q \bar q$ pair and $b$ is
a real positive constant.\\ 

In the simplest case where the $q$ and $\bar q$ emerge in opposite
directions, the colour field spanned between them is
approximated by a massless relativistic string with constant energy
density. These quarks oscillate back and forth and if given by
collision enough energy
the string will tear and a new pair of quarks will be produced.
This process repeats itself until only ordinary hadrons will remain.
In this model the produced hadrons are given a
small transverse momentum with respect
to the string axis. To note is that the production of 
quark pairs with $m_q > 0$ costs energy which is taken from the constant
energy density string which means that they cannot be produced at 
the same location but will be separated by some distance proportional to their
mass.
The success of the Lund model stems not only from its ability 
to describe many of the high energy interaction features but
also from the fact that it can be formulated stochastically
as an iterative process and therefore is well suited for computer
Monte Carlo simulation programs like the JETSET 
and its more recent advanced versions.\\

Another approach to the hadronisation has been taken by the cluster
model \cite{webber}. This model is based on the observation that
partons generated in a branching process tend to be arranged in
colour singlet clusters formed from $q \bar q$ pairs 
with limited extension in coordinate and momentum space. These
clusters have masses typically of the energy scale at which the parton
shower terminates. 
Very massive clusters decay first into lighter clusters pairs and then
each cluster decays isotropically into observable
hadron pairs with branching ratio determined by the density of states.
This model is implemented in the HERWIG Monte Carlo program ready to
be used in the experimental analyses.\\

At this point it is important to note that although outgoing particle
tracks generated in the Monte Carlo programs do have a charge ascription 
their Coulomb effects are not included.  
In addition, all generated particles are spinless. 
In the Lund model derived Monte Carlo versions a facility has been 
added so that the Bose-Einstein statistics effects 
present in identical boson system can be incorporated \cite{lund_bec}.
Another deficiency
of the Monte Carlo programs is the need to adjust
quite a few free parameters, like 
the ratio between vector and scalar resonances and their
production cross sections, to the experimental 
findings. As a consequence, 
one should proceed with caution when comparing a particular
effect seen in the data with the Monte Carlo program prediction. Moreover,
the setting of the many free parameters may well interfere with the
possibility to test the validity of the assumptions and underlying 
building elements of the particular hadronisation model.\\ 

\subsection{Final state interactions}
\label{sec_final}
In the BEC and FDC analyses, final state interactions (FSI)
may also play a role. There are two major FSI types.
The first is the Coulomb interaction which affects the charged
hadron systems and the second, the strong interaction final state
which is present in both charged and neutral
hadron systems. So far many of the reported BEC analyses have 
included 
the Coulomb effect whereas the strong final state interactions 
have in general been avoided due to their complexity 
and the realisation that their effect on the $\lambda$ and $r$
parameters is relatively small and can be absorbed in
the overall systematic errors (see Sec. \ref{sub_strong}).

\subsubsection{The Coulomb effect}
\begin{figure}[ht]
\centering{\epsfig{file=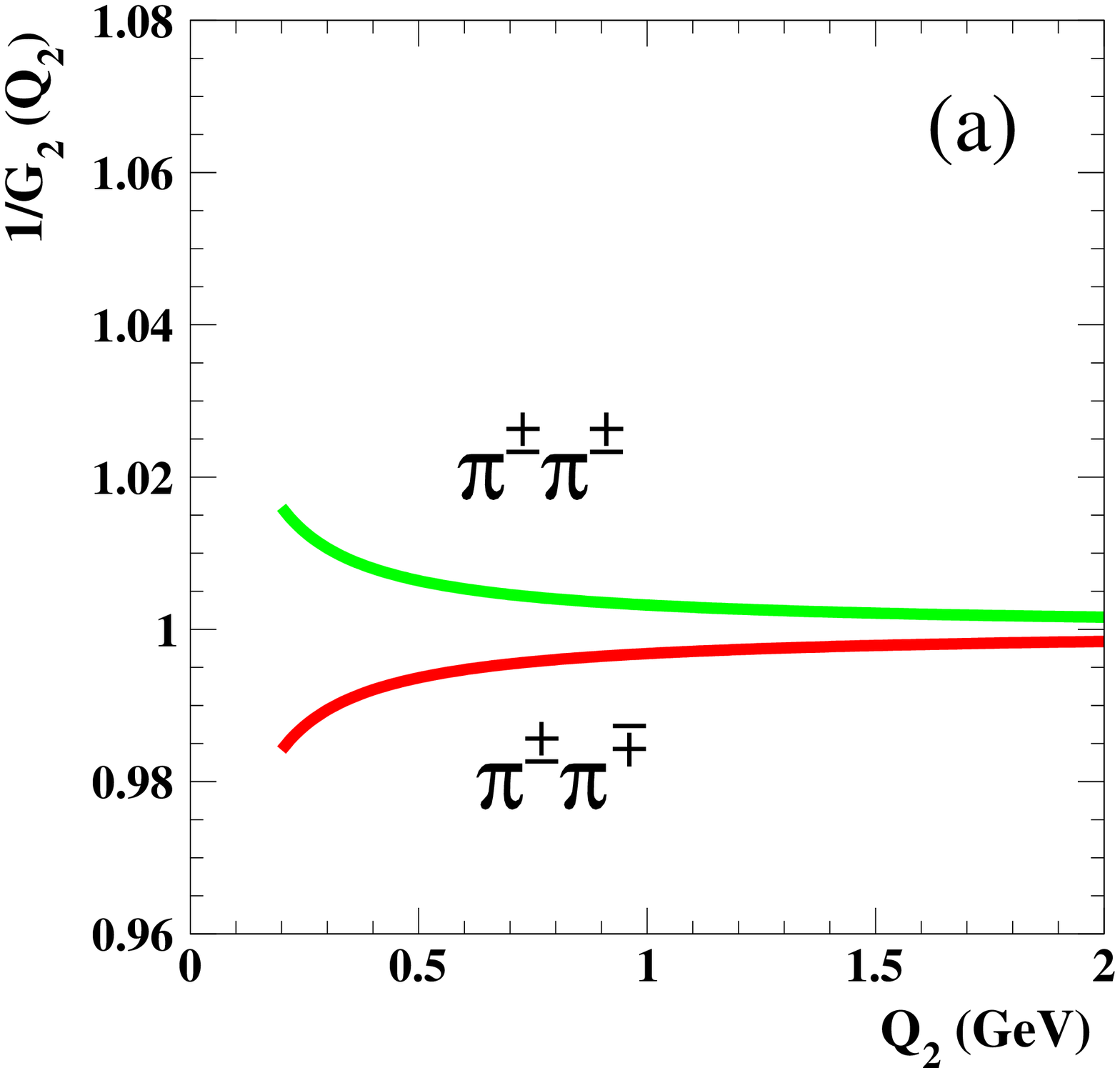,height=7cm}\ \ \ \
\ \epsfig{file=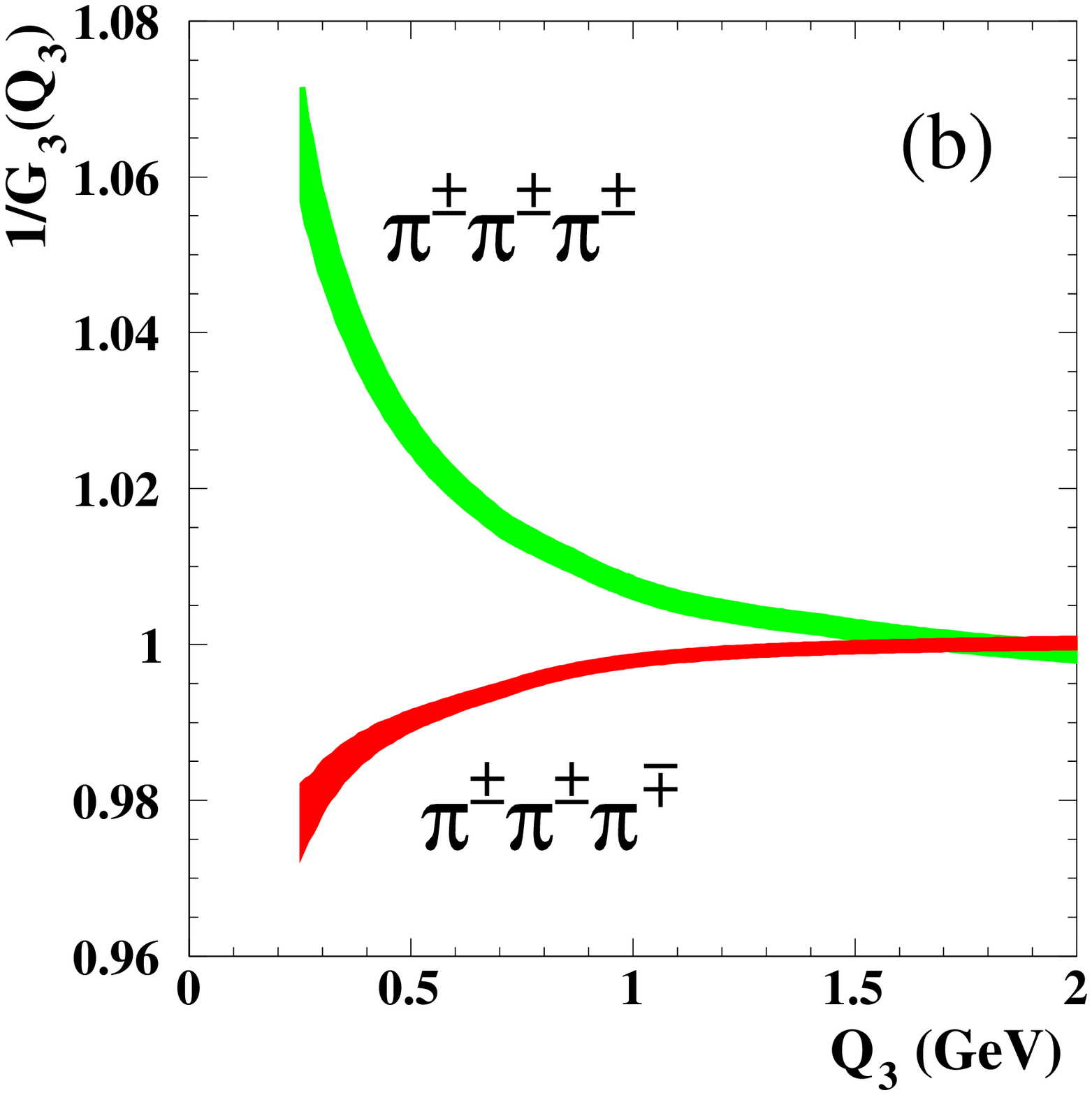,height=7cm}}
\caption{\small
The Coulomb correction to the BEC correlation function \cite{elad}.
(a) For two identical charged-pion systems as a function of $Q_2$ and 
(b) for the three charged pion systems as a function of $Q_3$.
}
\label{fig_coulomb}
\end{figure}

In the Bose-Einstein correlation of identical charged hadrons
the Coulomb repulsive force tends to
reduce the enhancement signal. If $C_2^{meas}(Q)$ denotes the
measured correlation of two hadrons then its relation to the
true correlation $C_2(Q)$ is given by
\begin{equation}
C_2^{meas}(Q)\ =\ C_2(Q) \times G_2(Q)\ ,
\label{eq_coulomb}
\end{equation}
where $G_2(Q)$ is determined through the Gamow penetration factor \cite{gamow}
\begin{equation}
G_2(Q)\ =\ \frac{2\pi\eta_{1,2}}{e^{2\pi\eta_{1,2}}\ -\ 1}\ \ \ \ 
{\rm{where}}\ \ \ \ \eta_{1,2}\ =\ 
\epsilonl_1\epsilonl_2\frac{\alpha~m}{Q}\ .
\label{eq_gamow}
\end{equation}
Here $\epsilonl_i$ are the charges in positron units of the hadrons, 
$m$ their mass and $\alpha$ is the fine-structure constant. 
As seen from Eq. (\ref{eq_gamow}), the Coulomb effect increases 
as $Q$ approaches zero. The measured correlation is corrected 
by the factor $1/G_2(Q)$ which may 
produce an exaggerated BEC signal \cite{bowler2} and therefore 
proper caution has to  be exercised
in its application. A somewhat different   
Coulomb correction method has been advocated in   
\cite{biyajima_cou}. To note is that the Coulomb correction for 
a di-pion system is rather small (see Fig. \ref{fig_coulomb}) 
and even at $Q_2=0.2$ GeV does not
amount to more than 2$\%$ and therefore in many reported BEC studies
this correction was ignored.\\                                      

The Coulomb effect on the BEC analysis of three identical charged hadrons
can be expressed \cite{Liu}, to a good approximation,
in terms of two-hadron Coulomb factors as 
\begin{equation}
G_3(Q_3)\ =\ G_2(Q_{1,2})\times G_2(Q_{1,3})\times G_2(Q_{2,3})\ .
\label{gamow3}
\end{equation} 
Further improvements to Eq. (\ref{gamow3}) 
have been proposed  (see e.g.  
Refs. \cite{cramer,alt1}). 
Unlike the di-pion case 
the Coulomb correction for three equally charged pions 
at $Q_3=0.25 $ GeV is not negligible and, as seen in Fig. \ref{fig_coulomb},
amounts to about 7$\%$.\\

\subsubsection{Strong final state interactions}
\label{sub_strong}
The simultaneous effect of both the Coulomb and the 
strong interaction scattering
of two identical charged pions on the BEC analyses have
been lately worked out and reported in Ref. \cite{final-state} where
references to earlier studies are also included. 
The FSI of the strong type are limited, due to the short range of
strong interaction, to the s-wave alone. For the $\pi^{\pm}\pi^{\pm}$
system the FSI dependence on $Q$ is given by the well measured  
$I_{\pi\pi} = 2$ phase shift
$\delta^{(2)}_0(Q)$ which can be incorporated into the BEC function.\\
  
An example for the inclusion of FSI in a BEC analysis
Ref. \cite{final-state} is illustrated
in Fig. \ref{fig_example}. In this figure the 
data points shown
are the $\pi^{\pm}\pi^{\pm}$ correlation function versus 
$Q$ as measured by the OPAL collaboration \cite{acton} in the hadronic $Z^0$
decays. The solid line represents the fit results to the
BEC function including Coulomb and strong FSI
whereas the dashed line is the outcome of a fit where the FSI were 
ignored. 
In a systematic study of the FSI in several BEC analyses applied to  
$e^+e^- \to pions$ data collected at energies on the 
$Z^0$ mass and below, it was found that in general
the inclusion of the FSI tends to increase $\lambda$ 
and decrease the $r$ value \cite{final-state}.\\

\section{The 1-Dimensional correlation results}
\label{sect_oned}
\subsection{The $\pi\pi$ Bose-Einstein correlation}
\label{pipibec}
\renewcommand{\arraystretch}{1.2}
\begin{table}[h]
\caption{\small The two-pion GGLP emitter dimension $r_2$ and the 
chaoticity parameter $\lambda_2$ deduced from BEC studies of
$e^+e^-$ annihilations using one or two choices for the   
reference sample. In Method I the reference
sample was the $\pi^+ \pi^-$ data sample. In Method II the reference
sample used was either Monte Carlo generated events or a sample
created by the event mixing technique. 
The given errors are the statistical and systematic 
errors added in quadrature. The values given are without a 
Coulomb correction.}
\begin{center}
\begin{tabular}{||lc||c|c|c|c||}
\hline\hline
\multicolumn{2}{||c||}{$\pi^{\pm}\pi^{\pm}$ {\small BEC}}  
&\multicolumn{2}{c|}{Method I}& \multicolumn{2}{c||}{Method II}\cr
\hline
Experiment &$\sqrt{s_{ee}}$ {\small [GeV]}&  $r_2$ [fm]& $\lambda_2$  
& $r_2$ [fm]& $\lambda_2$ \cr
\hline \hline
{\small MARK II} \cite{mark_pipi}& 29 &0.75 $\pm$ 0.05 &0.28
$\pm$0.04 &0.97 $\pm$ 0.11 &0.27 $\pm$ 0.04 \cr
{\small TPC} \cite{tpc_pipi}&29  &$-$ &$-$ & 0.65 $\pm$ 0.06  &0.50 $\pm$ 0.04 \cr
{\small TASSO} \cite{tasso_pipi}&34 &0.82 $\pm$ 0.07 &0.35 $\pm$ 0.03 
&$-$ &$-$ \cr
{\small AMY} \cite{amy_pipi}&58 &0.73 $\pm$ 0.21 &0.47 $\pm$ 0.07 &0.58
$\pm$ 0.06 &0.39 $\pm$ 0.05 \cr
{\small ALEPH} \cite{aleph_pipi}& 91 & 0.82 $\pm$ 0.04  &0.48 $\pm$
0.03 &0.52 $\pm$ 0.02 &0.30 $\pm$ 0.01 \cr
{\small DELPHI} \cite{delphi_pipi}& 91 &$0.83 \pm 0.03 $ &0.31 $\pm$
0.02 &0.47 $\pm$ 0.03 &0.24 $\pm$ 0.02 \cr
{\small L3} \cite{l3_pipi} & 91 &$-$ &$-$ &0.46 $\pm$ 0.02  &0.29 $\pm$ 0.03 \cr
{\small OPAL} \cite{opal_pipi}& 91   &0.96 $\pm$ 0.02  & 0.67 $\pm$
0.03&0.79 $\pm$ 0.02  &0.58 $\pm$ 0.01 \cr
\hline\hline
\ \ \ \ \ \ \ \ \ \ \ \ \ \ \ \ \ $\pi^0\pi^0$ {\small BEC} & & $r_2$ [fm]&
$\lambda_2$  
& $r_2$ [fm]& $\lambda_2$ \cr
\hline
{\small L3} \cite{l3_pipi,sanders} & 91 &$-$ &$-$ &0.31
$\pm$ 0.10  &0.16 $\pm$ 0.09 \cr
{\small OPAL} \cite{opal_pi0pi0} & 91 &$-$ &$-$ &0.59
$\pm$ 0.11  &0.55 $\pm$ 0.15 \cr
\hline\hline 
\end{tabular}
\end{center}
\label{tab_2pions}
\end{table}
\renewcommand{\arraystretch}{1.0}
\begin{figure}[ht]
\centering{\epsfig{file=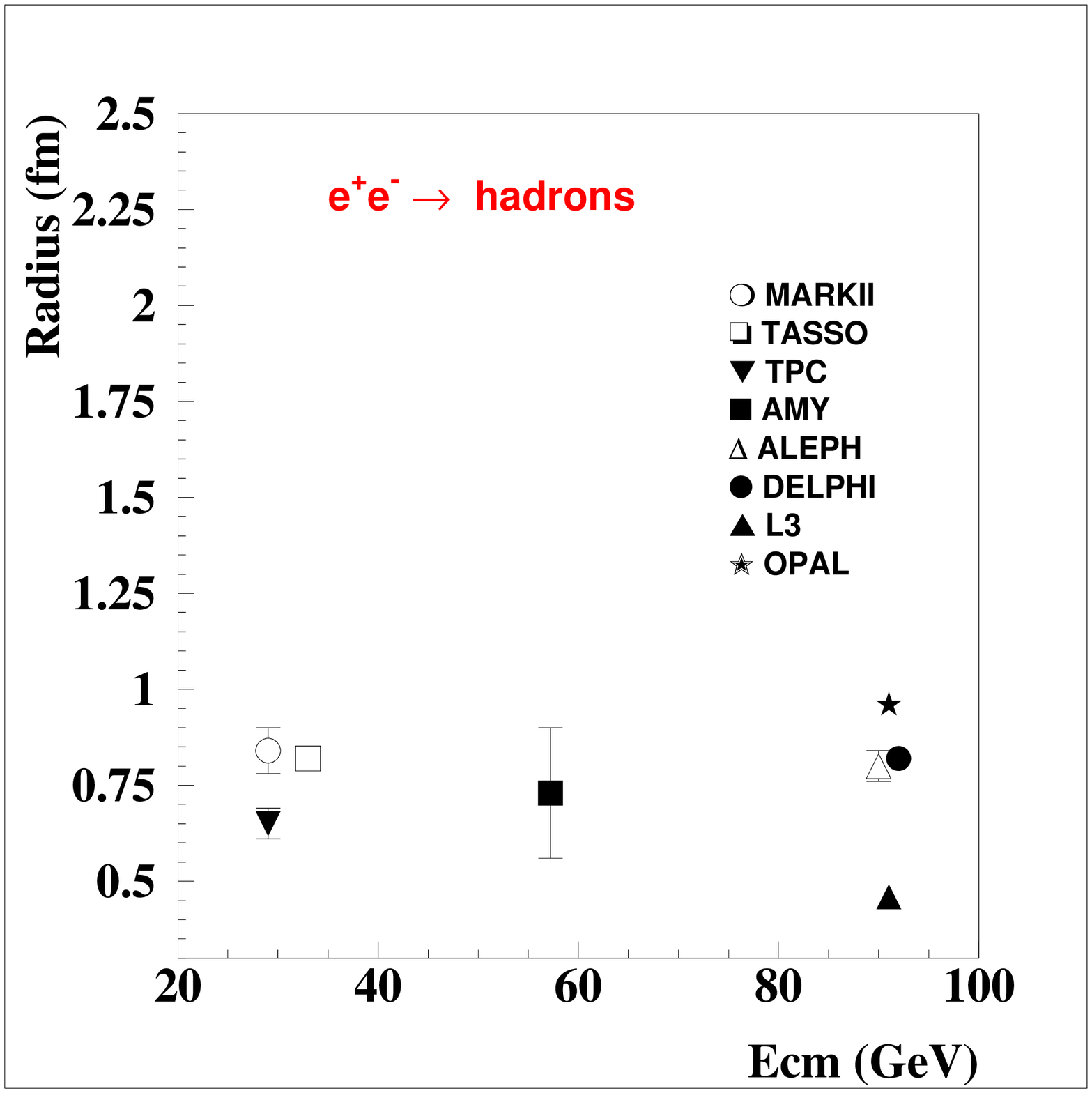,height=9cm,
bbllx=26pt,bblly=150pt,bburx=541pt,bbury=696pt,clip=}}
\caption{\small The emitter dimension $r_{\pi^{\pm}\pi^{\pm}}$ 
obtained from BEC
analyses of $e^+e^-$ annihilation into hadrons versus the
centre of mass energy of the $e^+e^-$ system. 
The values were taken from references \cite{mark_pipi,tpc_pipi,
tasso_pipi,amy_pipi,aleph_pipi,delphi_pipi,opal_pipi,l3_pipi}
where all but TPC and L3 used a data reference sample 
(see also Table \ref{tab_2pions}).}
\label{fig_comp}
\end{figure}
\renewcommand{\arraystretch}{1.2}
\begin{table}[h]
\caption{\small A representative sample of the
two-pion emitter dimension $r$ and the chaoticity 
parameter $\lambda_2$ deduced from BEC studies of 
hadron-hadron, lepton-hadron and $\gamma \gamma$ reactions
leading to multi-hadron final states. The data selection and reference
samples, as well as the application of the Coulomb correction, differ 
from experiment to experiment. In most cases the errors given are only
the statistical ones. The systematic errors are typically of the
order of 10 to 20$\%$ of the measured parameter values. 
}
\begin{center}
\begin{tabular}{||lc|c||c|c||}
\hline\hline
\multicolumn{3}{||c||}{$\pi^{\pm}\pi^{\pm}$ BEC analyses}
&\multicolumn{2}{c||}{Parameter
}\cr
\cline{1-5}
Reaction&  & E$_{CM}$~[GeV] & $r_2$ [fm]& $\lambda_2$ \cr
\hline \hline
$\gamma\gamma \to h$ &\cite{mark_pipi}& $<5>$
& 1.05 $\pm$ 0.08 &1.20 $\pm$ 0.13 \cr
$\gamma\gamma \to 6\pi^{\pm}$ &\cite{asher}&1.6 $-$ 7.5 &
0.54 $\pm$ 0.22 &0.59 $\pm$ 0.20 \cr
\hline
$\nu (\bar \nu) N \to h$ &\cite{nubebc}&8 $-$ 64  &
0.64 $\pm$ 0.16 &0.46 $\pm$ 0.16 \cr
\hline
$\mu p \to h$ &\cite{arneodo}&23& 0.65 $\pm$ 0.03& 0.80 $\pm$
0.07 \cr
\hline
$\pi^+ p \to h$ &\cite{adamus}&21.7& 0.83 $\pm$ 0.06& 0.33 $\pm$
0.02 \cr
\hline
$p p \to h$ &\cite{bailly}&26& 1.02 $\pm$ 0.20& 0.32 $\pm$
0.08 \cr
$p p \to h$ &\cite{aguillar}&27.4&1.20 $\pm$ 0.03& 0.44 $\pm$
0.01 \cr
$p p \to h$ &\cite{akesson}&63&0.82 $\pm$ 0.05& 0.40 $\pm$
0.03 \cr
\hline
$\overline{p} p \to h$ &\cite{cplear}&1.88& 1.04 $\pm$ 0.01& 1.96 $\pm$
0.03 \cr
$\overline{p} p \to h$ &\cite{albajar}&200 - 900 & 0.73 $\pm$
0.03 & 0.25 $\pm$ 0.02 \cr
\hline\hline
$e~p \to e~h$&\cite{h1_pipi}&$6<Q^2_{\gamma}<100$ [GeV$^2$]& 0.68 $\pm$ 
0.06 & 0.52 $\pm$ 0.20 \cr
$e~p \to e~h$&\cite{zeus_pipi}&$110<Q^2_{\gamma}$\ \  [GeV$^2]$& 0.67 $\pm$ 
0.04 & 0.43 $\pm$ 0.09 \cr
\hline\hline
\end{tabular}
\end{center}
\label{tab_2pionsh}
\end{table}
\renewcommand{\arraystretch}{1.0}

The major information concerning the BEC features 
is coming from the $\pi^{\pm}\pi^{\pm}$
system which has been analysed in a large variety of particle reactions and
over a wide range of centre of mass energies. In Table \ref{tab_2pions}
results obtained from the 1-dimensional BEC analyses of two 
identical pions produced in
$e^+ e^-$ annihilation in the energy range from 29 to 91 GeV are listed.
The values for $r_2$ and $\lambda_2$ are divided into two groups
according to the reference sample type that was used. In Method I the
reference sample chosen was the data $\pi^{\pm}\pi^{\mp}$ system whereas
in Method II the reference sample used was either a generated  
Monte Carlo sample plus a full detector simulation 
or a sample obtained by the event mixing technique.\\

From the $\pi^{\pm}\pi^{\pm}$ BEC analyses results shown in 
Table \ref{tab_2pions} one can make the following observations.
Firstly the values of $r_2$ are in general found to be higher in
the Method I analysis than in the Method II analysis whereas the
$\lambda_2$ values are 
roughly the same. Secondly there is no clear evidence for an
$r$ dependence on the centre of mass energy. To note is also the
fact that 
the values of $r_2$ are smaller than 1 fm and their average in Method I
is in the vicinity of 0.8 fm. In Method II the $r$ values are
less stable and fluctuate from experiment to experiment.
It is generally believed that these
fluctuations of the $r$ values are just a reflection
of the fact that different experiments used different
BEC analysis procedures, namely in their data selection criteria and
in their 
choice of the reference sample. This situation is also clearly
seen in Fig. \ref{fig_comp} where $r_2$ is plotted against 
the $e^+e^-$ centre of mass energy, $E_{cm}$. 
In particular to note
are the three $r_2$ values obtained in Method I for 
the $\pi^{\pm}\pi^{\pm}$ pairs
emitted from the $Z^0$ gauge boson where the OPAL value differs by
several standard deviations from those reported by ALEPH and
DELPHI. The L3 collaboration has not given an $r$ value in the 
framework of the Method I analysis.    
Unlike the abundant results on the $\pi^{\pm}\pi^{\pm}$ correlations
the information on the BEC analyses of $\pi^0\pi^0$ pairs is
very limited due to the
experimental difficulties in identifying the two neutral bosons.
At the same time there exists an interest in the two neutral pion system
both in the framework of the string model and in the 
so called generalised BEC where isospin invariance is 
incorporated into the analysis.  
The two $r$ values, obtained from the BEC analysis of the  $\pi^0\pi^0$
system, present in the hadronic $Z^0$ decay and listed in 
Table \ref{tab_2pions}, do not allow presently  
a meaningful comparison to the string model and isospin invariance 
predictions.\\

A compilation of $\pi^{\pm}\pi^{\pm}$ BEC parameters 
obtained from analyses carried out in a
variety of particle reactions, other than
$e^+e^-$ annihilation, is given in Table \ref{tab_2pionsh}.
Again no clear evidence can been seen for an $r_2$ dependence
on the type of reaction and/or energy. The $r$ values 
reported are, like in $e^+ e^-$ annihilations, 
again near, or somewhat lower, than 1 fm.\\ 

The BEC of the $\pi^{\pm}\pi^{\pm}$ pairs produced in 
the reaction $e^+ e^- \to W^+ W^- \to hadrons$, where the pions are
emerging from different $W$ bosons, have been recently investigated in
view of the concern that they might affect the $W$ gauge boson mass 
estimation \cite{wwlep}.
Whereas the preliminary results of DELPHI found some evidence
for a BEC enhancement \cite{delphi_ww}, the L3 and
ALEPH collaborations have not seen any effect of this kind
\cite{l3_ww,aleph_ww}. 
\subsection{The $K^{\pm}K^{\pm}$ and $K^0_SK^0_S$ systems}
\label{sec_kk}
The results from BEC analyses of the $K^{\pm}K^{\pm}$ system are 
meager in comparison to those reported from
the $\pi\pi$ BEC analyses. This is mainly due to the fact that
the rate of the charged kaons is much smaller than that of the 
charged pions which
even in the hadronic $Z^0$ decays amounts only to \mbox{$\simeq$ 2.4/17}. 
An additional reason for the relatively poor 
kaon-pairs statistics stems from the need to remove from their 
sample contributions from protons, anti-protons and charged pions.
This is achieved by  applying hadron identification criteria which
in most cases are effective only in a limited range of
momentum. In the $e^+ e^-$ annihilations on the $Z^0$ mass energy
two $K^{\pm}K^{\pm}$ BEC analyses have been reported
\cite{delphi_kpkp,opal_kpkp}. Their extracted $r_{K^{\pm}K^{\pm}}$ 
values are listed in Table \ref{rofm}.\\ 

Unlike the $K^{\pm}K^{\pm}$ pairs, the origin of the $K^0_SK^0_S$ system
is not only fed from strangeness $\pm$2 states but also from the
boson-antiboson $K^0\overline{K}^0$
(strangeness = 0) system which in turn may be fed from
the decay of resonances. 
Neglecting CP violation\footnote{The inclusion of CP violation effect
in the following discussion
is straightforward but with negligible consequences at the
current experimental precision level.}
we show in the following that
a Bose-Einstein like threshold enhancement is nevertheless expected
in the $Q$($K^0_S,K^0_S$) distribution even when
its origin is the $K^0 \overline{K}^0$ system 
\cite{lipkink1,lipkink3,lipkink2,lipkina}.\\

In general a boson-antiboson, $B \overline{B}$,
pair is an eigenstate of the charge conjugation
operator \cal{C}. In the absence of outside constraints, 
one can write the probability amplitude
for a given charge conjugation eigenvalue $C_n$ as follows:
\begin{equation}
\frac{1}{\sqrt{2}} \ \left| B ;\overline{B} \right>_{C_n
= \pm 1} = \frac{1}{2} \
\left| B(\pbf);
\overline{B}(-\pbf) \right>
\ \pm \
\frac{1}{2} \
\left| \overline{B}(\pbf);
B(-\pbf) \right>\ ,
\label{bb1}
\end{equation}
where $\pbf$ is the three momentum vector
defined in the $B \overline{B}$ centre of mass system.
In the $Q = 0$ limit,
where the BEC effect should be maximal, $\pbf$ is equal to zero
and Eq. (\ref{bb1}) reads
\begin{equation}
\frac{1}{\sqrt{2}} \ \left|B;\overline{B}\right>_{C_n =
\pm 1} = \frac{1}{2} \
\left| B(0);
\overline{B}(0) \right>
\ \pm \
\frac{1}{2} \
\left| \overline{B}(0); B(0) \right>\ .
\label{bb2}
\end{equation}
This means that,
at $Q=0$, the probability amplitude for the $C_n = -1$ state
(odd $\ell$ values)
is zero whereas the $C_n = +1$ state (even $\ell$ values)
is maximal.
Here it is important to note that
these equations do hold for any spinless boson-antiboson pair
such the $K^0 \overline{K}^0$, the $K^+ K^-$ and the
$\pi^+ \pi^-$ systems. 
What does distinguish the $K^0 \overline{K}^0$ from the other
spinless boson-antiboson pairs is
the simplicity by which one is able
to project out the $C_n = +1$ and the $C_n = -1$
parts of the probability amplitude.\\
 
As is well known, the
$K^0$ and the $\overline{K}^0$ mesons
are described in terms of the two \cal{CP} eigenstates,
$K^0_S$
with $CP_n = +1$ and $K^0_L$ with $CP_n=-1$.
From this follows that, when the
$K^0 \overline{K}^0$
pair is detected through
their $K^0_S$ and $K^0_L$ decays,
a definite eigenvalue $C_n$ of the
$K^0 \overline{K}^0$ system can be selected.
Thus, as $Q$ approaches zero, an enhancement
should be observed in the probability
to detect $K^0_S~K^0_S$ pairs and/or
$K^0_L K^0_L$ pairs ($C_n = +1$),
whereas a decrease should be seen in the probability
to find $K^0_S K^0_L$ pairs \mbox{($C_n = -1$).} 
The BEC can thus be analysed by forming for
the eigenstate $C_n=+1$ and $C_n=-1$ respectively the correlation functions:

\begin{equation}
C_+(Q)\ =\ 
\frac{2P(K^0_S K^0_S)\ + \ 2P(K^0_L K^0_L)}{P(K^0~\overline{K}^0)}
\ \ \ \ {\rm{and}} \ \ \ \ 
C_-(Q)\ =\ 
\frac{2P(K^0_S K^0_L)\ + \ 2P(K^0_L K^0_S)}{P(K^0~\overline{K}^0)}\ ,
\label{cqkk}
\end{equation}
where $P$ stands for the production rate of a given two-boson state
as a function of $Q$.\\ 

\begin{figure}[h]
\centering{\epsfig{file=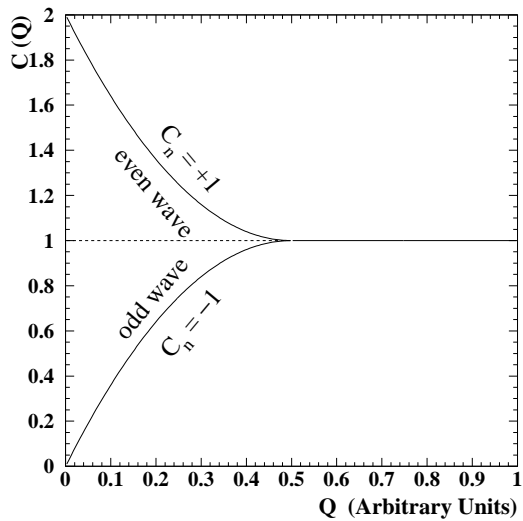,height=7cm,
bbllx=15pt,bblly=145pt,bburx=548pt,bbury=665pt,clip=}}
\caption{\small
Schematic behaviour of the correlation functions $C(Q)$ of the
$K^0~\overline{K}^0$ system given by Eq. (\ref{cqkk}). 
The $C_n = +1$ probability
amplitude 
reaches the value 2 at the limit of $Q$ = 0 whereas the $C_n = -1$
probability amplitude reaches the value 0. The sum of $C_n = +1$ and $C_n =
-1$ remains constant down to $Q$ = 0.
}
\label{fig_kzero}
\end{figure}

These $C_+(Q)$ and $C_-(Q)$
dependence on $Q$, for $\lambda$ = 1, are
drawn schematically in Fig. \ref{fig_kzero}.
As seen, when $Q$
approaches zero, $C(Q)$ splits into two branches.
The first rises up to the value two and the other decreases
to zero in such a way that their sum remains
constant and equal to one at all $Q$ values.
This means, that if
all the decay modes of the $K^0 \overline{K}^0$
pairs are detected and used simultaneously
in the same correlation analysis then,
according to Eqs. (\ref{bb1}) and (\ref{bb2}), no BEC
effect will be observed at $Q=0$. This however should not come as
a surprise if one recalls that
the $K^0 \overline{K}^0$ system is at the very end not composed of
two identical bosons.\\

The interpretation of a BEC enhancement in the $K^0_SK^0_S$ system 
is not straightforward as its origin 
may also be the decay of the scalar $f_0(980)$ resonance.
This resonance, which is mainly observed in its decay to pion-pair 
in the decay of the $Z^0$ gauge boson \cite{delphif}, 
lies within its width at the $K^0\overline{K}^0$ threshold of 995.3
MeV. Thus the low $Q$ enhancement seen in the 
$K^0_S~K^0_S$ correlation at $Q \sim 0$ may be formed from the
Bose-Einstein effect and/or from the $f_0(980)$ resonance decay.
A possible method to disentangle the two contributions 
is given by isospin considerations \cite{alex_iso} which is
outlined in the following section. 
Values for $\lambda$ and $r$ obtained for the $K^0_S~K^0_S$ system,
produced in $e^+e^-$ annihilation on the $Z^0$ mass, 
are given in Table \ref{tab:comp_lep}.

\subsection{Isospin invariance and generalised Bose-Einstein correlation}
\label{sec_isospin}
In the sector of strong
interacting hadrons where BEC and FDC analyses are performed, 
the isospin is a conserved quantity. 
In analogue to the so called
generalised Pauli principle, which is the extension of that principle
from two identical nucleons, the proton-proton and 
neutron-neutron  pairs, to the proton-neutron system,
one can generalise the Bose-Einstein statistics 
by the inclusion of the isospin invariance. This extension 
may relate for example, 
BEC results of the $\pi^{\pm}\pi^{\pm}$ 
pairs to those obtained for the $\pi^+\pi^0$ system. 
This extension was pointed out by several 
authors (see e.g. references
\cite{biyajima_iso,suzuki,bowler1,weiner}) and recently was worked out
in details \cite{alex_iso} to produce concrete relations between 
the BEC effects of hadron systems which are connected 
by isospin and are proposed to be tested
experimentally.
So far however the concept of the
generalised BEC has not been verified  
as it involves pion systems
which include one or more neutral pions which experimentally
are hard to detect and identify. However if the generalised BEC
assumption is confirmed then 
its effect on particles correlations is    
not negligible. In the following we briefly discuss some aspects
of the generalised BEC features and consequences.\\
 
First to note is that
in many cases boson pairs, like the $KK$ and $\pi\pi$ systems, 
are produced together with other hadrons, here  denoted by $X$, from an
initial state which is 
isoscalar to a very good approximation. These include 
e.g. pairs produced from an initial multi-gluon state, 
pairs from hadronic decays of 
$J/\psi$ and $\Upsilon$, or pairs produced by $Z^0$ decays. 
In some of these
cases the initial state is however not pure I=0 but has also 
some contamination of an
I=1 component which is mixed in like in those processes  
where the $J/\psi$ and $\Upsilon$  
decay to hadrons via one photon annihilation.
According to the specific case methods can be applied to reduce, or even
eliminate, this contamination. For example, the subsample of the
$C_n = -1$ quarkonia which decays into an odd number of pions assures, 
due to G-parity, that the hadronic final state is in an I=0 state. 
Forming such a subsample from the hadronic $Z^0$ decays is not useful,
but on the other hand multi-hadron final states which originate 
from $Z^0$ decays into $s \bar s,\ c \bar c $ and 
$b \bar b$ quark pairs, are in an I=0 state.\\  

Several specific relations in the framework of the generalised
Bose-Einstein statistics between charge and neutral 
K-mesons have been proposed in reference \cite{alex_iso}.
Among others, of a particular 
interest are those relations which have a bearing on the analysis of the
enigmatic scalar $f_0(980)$ resonance, whose  mass and width lie in the
KK threshold region, and which has a long history
regarding its nature and decay modes \cite{pdg,fzero,fzero1,fzero2}.
For details concerning these relations the reader is advised to turn to
reference \cite{alex_iso}.\\

Next we turn to the effect of the generalised BEC assumption on the
$\pi\pi$ system. 
In this system there are more states and more 
isospin amplitudes than the two, I=0 and I=1 of the KK system. 
In the low $Q$ region where only the 
s-wave contributes, the $\pi^+\pi^+$, $\pi^+\pi^-$ and $\pi^0 \pi^0$ 
amplitudes 
depend upon only two isospin amplitudes, I=0 and I=2. Their intensities in this
region satisfy a triangular inequality \cite{alex_iso}
\begin{eqnarray}  
\sum_X\left |~\sqrt{
(2/3)\cdot P[i_o \rightarrow (\pi^0\pi^0) X]} - 
\sqrt{ (1/3)\cdot P[i_o \rightarrow (\pi^+\pi^-)_e X]}~\right | \leq  
\cr 
\leq                            
\sum_X\sqrt{ P[i_o \rightarrow (\pi^\pm\pi^\pm) X]} =
\sum_X\sqrt{ P[i_o \rightarrow (\pi^\pm\pi^0)_e X]} \leq \hspace*{1cm}
\cr 
\leq                            
\sum_X\left |~\sqrt{(2/3)\cdot P[i_o \rightarrow (\pi^0\pi^0) X]} + 
\sqrt{ (1/3)\cdot P[i_o \rightarrow (\pi^+\pi^-)_e X]}~\right |\ , 
\end{eqnarray}
where the notation $(\pi\pi)_e$ is used to indicate that only even
partial waves for the $(\pi\pi)$ final state are included in the sum.
Here $i_o$ denotes an initial I=0 state and $X$ stands for the hadrons
produced in association with the two earmarked pion-pair.\\
 
In this last relation the middle line is of a particular interest as
it states that at low $Q$ values, where only the s-wave survives,
the BEC enhancement in the $\pi^{\pm}\pi^0$ system 
should be equal to that present in the $\pi^{\pm}\pi^{\pm}$ pairs.
Finally the BEC features of the $\pi^0\pi^0$ system 
should not be far from that of the $\pi^+\pi^+$ state as
long as the contribution of the $\pi^{\pm}\pi^{\mp}$ 
to the triangle inequality relation remains small.

\subsection{Observation of higher order Bose-Einstein correlations}
\begin{figure}[ht]
\centering{\epsfig{file=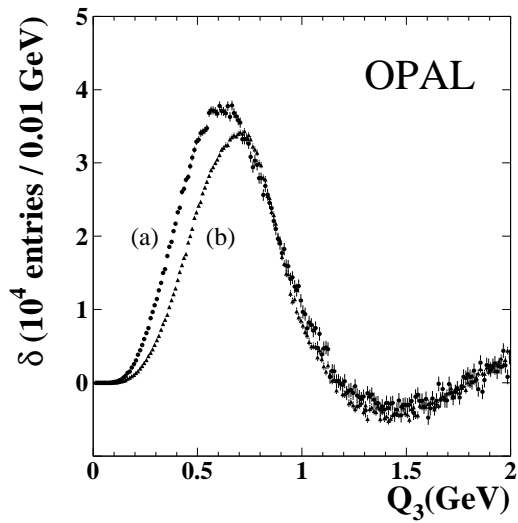,height=7.5cm}}
\caption{\small BEC results from an OPAL 
study of three identical charged pions detected in
the hadronic $Z^0$ decays \cite{opal_pipipi}. The over-all normalised
BEC enhancement is described by the (a)
distribution whereas distribution (b) corresponds to the three-pion BEC
enhancement generated from the lower order BEC enhancement of
the two-pion system.}
\label{fig_opalc3}
\end{figure}

As mentioned in Sec. \ref{higher}, the directly measured higher 
order BEC are essentially limited to the case of three identical 
charged pions. 
In these analyses, the contribution from the two-hadron BEC was
subtracted in each experiment with somewhat different method. The genuine 
$\pi^{\pm}\pi^{\pm}\pi^{\pm}$ BEC were also measured 
in the reaction $e^+ e^-\to Z^0$ followed by an hadronic
decay \cite{delphi_pipipi,opal_pipipi,l3_pipipi}. 
The experimental
normalised distribution, which is marked by (a) in Fig. \ref{fig_opalc3}, 
is that found by OPAL \cite{opal_pipipi} 
to describe the
contribution to $R_3(Q_3)$, as defined in
Eq. (\ref{eq_r3}), from the over-all three identical charged
pion BEC enhancement. In the same figure 
the three-pion distribution, denoted by  (b), corresponds to 
the BEC enhancement due to the lower order BEC correlations i.e. the
two-pion system. A clear evidence is seen for 
the genuine three-pion BEC enhancement which can be extracted by
subtracting the distribution (b) from that of (a) which then allows the
evaluation of $r_3$.\\

Results for the three-boson emitter dimension are listed 
in Table \ref{tab_3pionsh} 
where the values of $r_3$ and the ratio $r_3/r_2$ are given.
As can be seen, it does not seem to exist a significant dependence of the
$r_3$ and $r_3/r_2$ values on the type and/or energy 
of the particles reaction.\\

\renewcommand{\arraystretch}{1.2}
\begin{table}[ht]
\caption{\small Results for $r_3$  and $r_3/r_2$ values obtained from
BEC analyses of two and 
three identical charged pions. Values for genuine three-pion 
BEC are marked by (g) otherwise they are marked by (n). 
}
\begin{center}
\begin{tabular}{||lc|c||c|c||}
\hline\hline
\multicolumn{3}{||c||}{2$\pi$ and 3$\pi$ BEC analyses}  &
\multicolumn{2}{c||}{Parameter}\cr
\cline{1-5}
Reaction& & E$_{CM}$~[GeV] & $r_3$ [fm]& $r_3/r_2$ \cr
\hline \hline
$\pi^+p,K^+p \to h$ &\cite{agababian}& 22
& 0.51 $\pm$ 0.01\ (g) &0.61 $\pm$ 0.12 \cr
$pp \to h$  &\cite{na22_bec}&26&
0.58 $\pm$ 0.07\ (n) &0.58 $\pm$ 0.25 \cr
$pp \to h$  &\cite{aguillar}&27.4& 0.54 $\pm$ 0.01\ (n) &0.45 $\pm$ 0.07  \cr
$pp \to h$ &\cite{akesson}& 63  &
0.41 $\pm$ 0.02\ (n) &0.50 $\pm$ 0.13 \cr
\hline
$e^+ e^- \to h$ &\cite{mark_pipi}&3.1& 0.53 $\pm$ 0.03\ (n)& 0.65 $\pm$
0.17 \cr
$e^+ e^- \to \gamma \gamma$ &\cite{mark_pipi}&5& 0.55 $\pm$ 0.03\ 
(n)& 0.65 $\pm$ 0.20 \cr
$e^+ e^- \to h$ &\cite{mark_pipi}&4$-$7& 0.45 $\pm$ 0.04\ (n)& 0.63 $\pm$
0.18 \cr
$e^+ e^- \to h$ &\cite{mark_pipi}&29&0.64 $\pm$ 0.06\ (n)& 0.77 $\pm$
0.25 \cr
$e^+ e^- \to h$ &\cite{tasso_pipi}&29$-$37& 0.52 $\pm$ 0.07\ (n)& 0.59 $\pm$
0.21 \cr
$e^+ e^- \to h$ &\cite{delphi_pipipi}&91& 0.66 $\pm$ 0.05\ (g)
& 0.80 $\pm$ 0.16 \cr
$e^+ e^- \to h$ &\cite{opal_pipipi}&91& 0.58 $\pm$
0.05\ (g) & 0.73 $\pm$ 0.13 \cr
$e^+ e^- \to h$ &\cite{l3_pipipi}&91& 0.65 $\pm$ 0.07\ (g)
 &1.00 $\pm$ 0.12 \cr
\hline\hline
\end{tabular}
\end{center}
\label{tab_3pionsh}
\end{table}
\renewcommand{\arraystretch}{1.0}
In Ref. \cite{mark_pipi}
a relation is derived
between the two-pion emitter radius $r_2$ and the 
three-pion emitter radius $r_3$ as extracted from a fit of $R_3(Q_3)$ 
to the over-all (non-genuine) three-boson correlation distribution.
This relation, which is based on the Fourier transform 
of a Gaussian shape hadron source, is given by the inequality
\begin{equation}
r_2/\sqrt{3} \ \leq \ r_3 \ \leq \ r_2/\sqrt{2} \ .
\label{inequality}
\end{equation}
so that $r_3/r_2$,
lies in the range of 0.55 to 0.71.
The inequality relation (\ref{inequality}) is  reduced to the equality 
\begin{equation}
\label{eq_r3e}
r_3 \ = \ r_2/\sqrt{2} \ ,
\end{equation}
when 
the emitter radius $r_3$ is determined from the genuine BEC distribution
parametrised by $C_3(Q_3)$ so that 
$r_3/r_2$ should exactly be equal to 0.71.
The experimental
results, shown in Table \ref{tab_3pionsh}, are seen to fulfil the
expected relations between $r_2$ and $r_3$ and in particular the
results from the genuine three-pion correlation obtained from the 
hadronic $Z^0$ decay are consistent within 
one to two standard deviations with the 
expectation of Eq. (\ref{eq_r3e}).\\

\subsection{Experimental results from Fermi-Dirac correlation  analyses}
\begin{figure}[h]
\centering{\epsfig{file=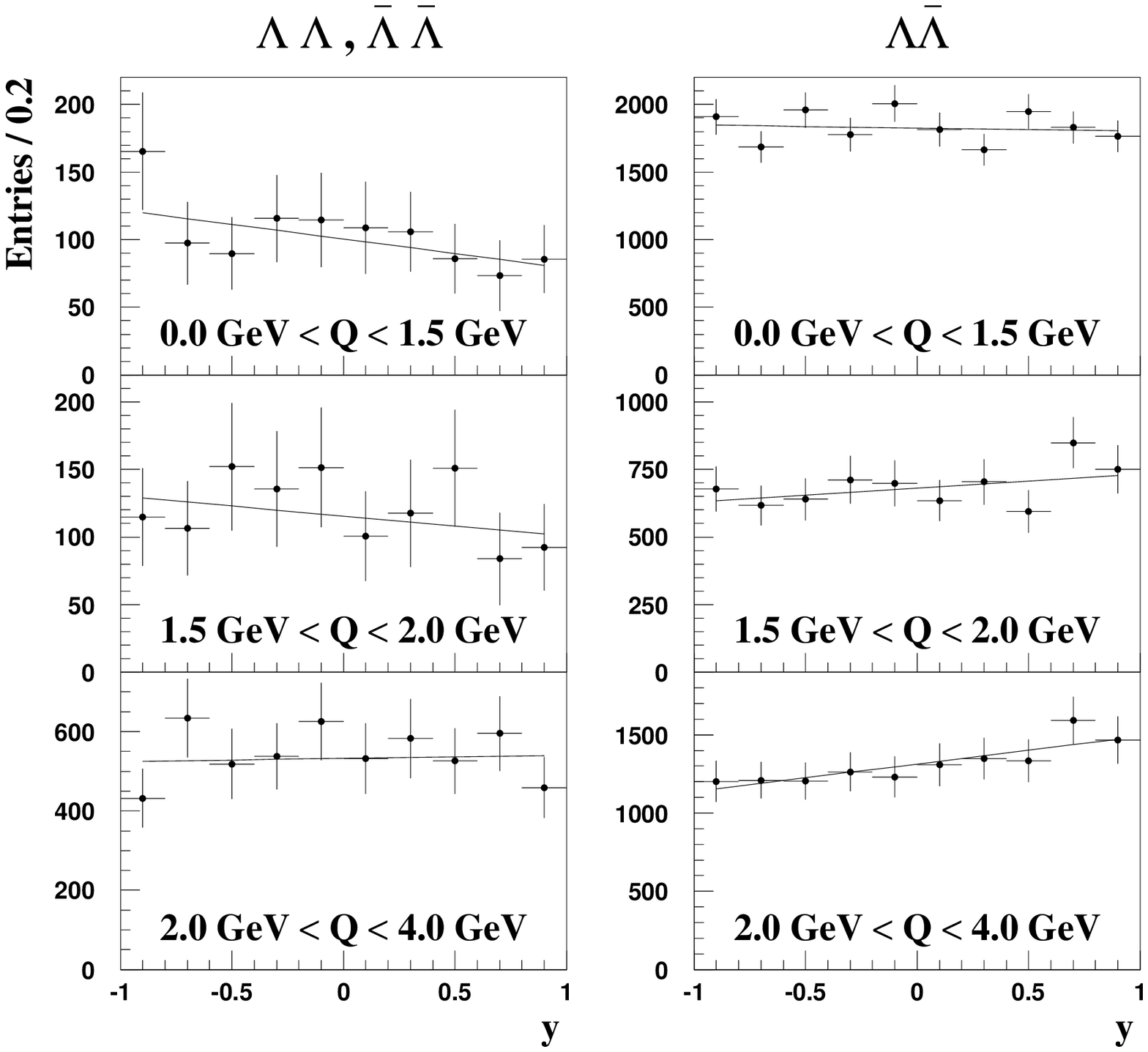,height=10.3cm,
bbllx=29pt,bblly=28pt,bburx=533pt,bbury=528pt,clip=}}
\caption{\small The ALEPH results from the spin-spin
FDC analysis of the $\Lambda \Lambda$ and $\Lambda
\overline{\Lambda}$ pairs observed in the hadronic $Z^0$
decays \cite{aleph_ll} given at several $Q$ regions 
and presented in terms of $dN/dy$ where $y$ is the cosine
angle between the two decay protons. The lines represent
the best fit of Eq. (\ref{fd_eq3}) to the data.} 
\label{fig_fd2}
\end{figure}

\begin{figure}[h] 
\centering{\epsfig{file=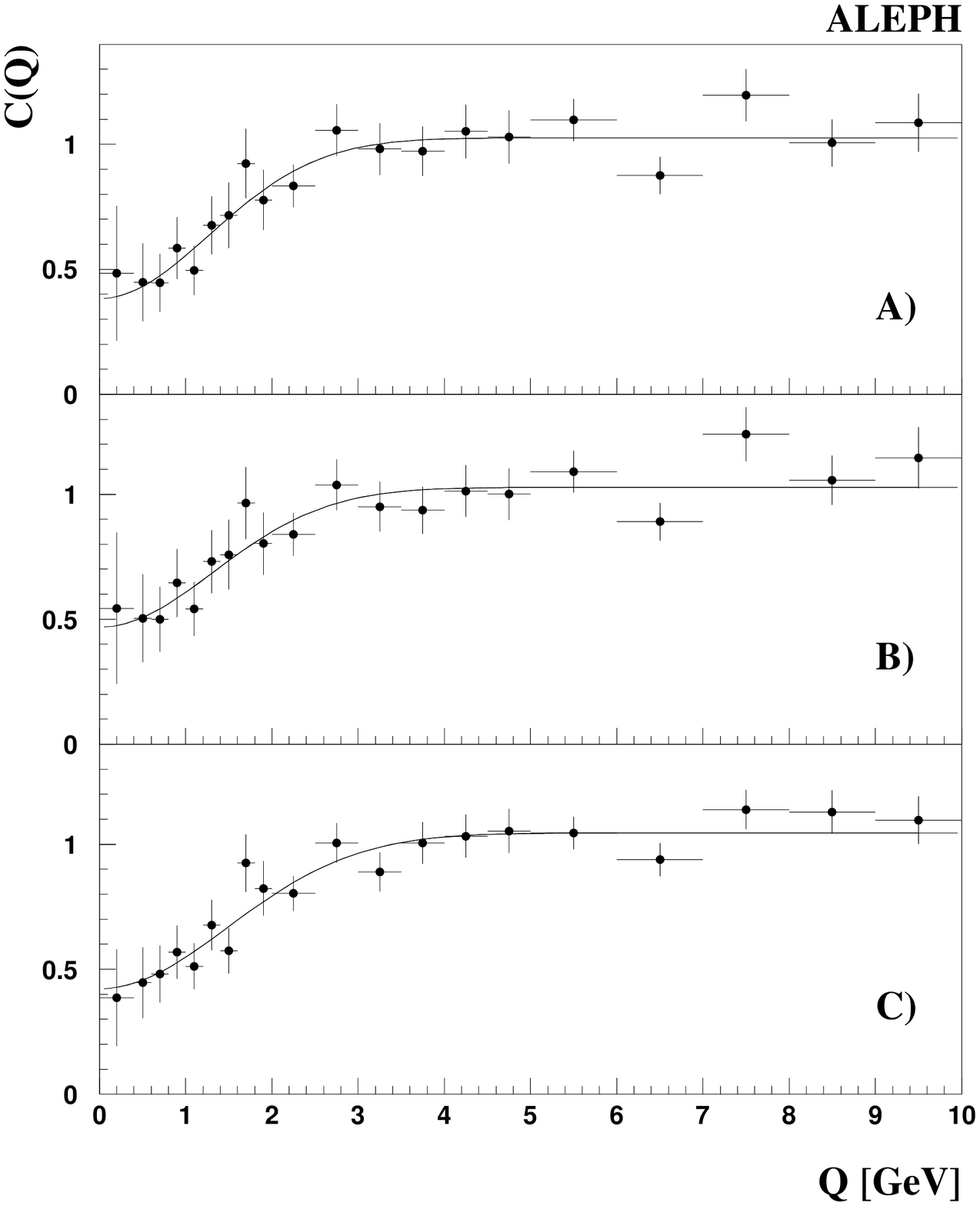,height=5.45cm}
\ \ \ \ 
\epsfig{file=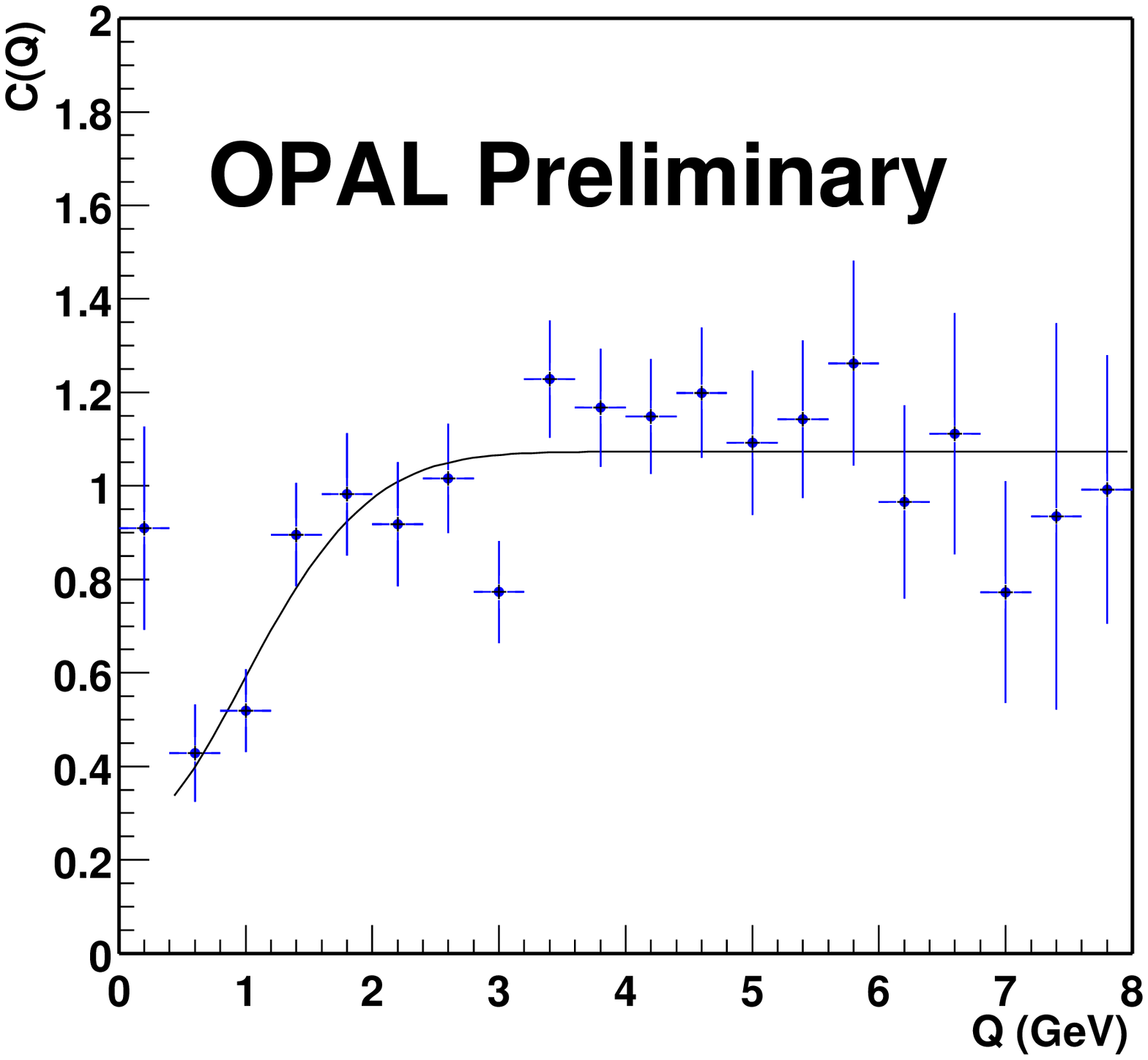,height=7cm}}
\caption{\small  
Left: The correlation function $C(Q)$ obtained by the ALEPH collaboration
\cite{aleph_ll} using
Method II for the $\Lambda \Lambda$ pairs for three different reference
samples. A) Monte Carlo, B) and C) two, 
somewhat differently treated, mixed events samples.
Right: The OPAL $C(Q)$ distribution of the
$\overline{p}~\overline{p}$ system taken from Ref. \cite{opal_pp}.
The continuous lines
represent the best fit of Eq. (\ref{eq_cqfd}) to the $C(Q)$ results.} 
\label{fig_fd3}
\end{figure}

Until recently FDC analyses of fermion-pairs were prohibited due 
to the low production rate of identical di-baryons in particle reactions
at low and intermediate energies.
This situation changed with the commission of the
LEP collider, operating on the $Z^0$ mass energy,
where each of its four experiments have accumulated some 3 to 4 million 
hadronic $Z^0$ decays. 
The inclusive $Z^0$ decay branching ratio of $\sim$0.34 $\Lambda$ per
event was sufficiently high to allow a FDC analysis 
of the $\Lambda ~\Lambda (\overline{\Lambda} ~\overline{\Lambda})$
and $\Lambda ~\overline{\Lambda}$ pairs which are free of Coulomb
effects and are relatively easy identified. 
Results from these analyses 
have been reported by the OPAL \cite{opal_ll},
DELPHI \cite{delphi_ll} and ALEPH \cite{aleph_ll} collaborations
using the spin-spin correlation method
(see Fig. \ref{fig_fd2}). The ALEPH
collaboration has further repeated its analysis with Method II, 
i.e. with the phase space density approach 
and found the results of both methods to be consistent within errors
(see Fig. \ref{fig_fd3}). In the same figure is also shown the 
OPAL \cite{opal_pp}
recent 
$\overline{p}~\overline{p}$ FDC measurement carried out  
in the hadronic $Z^0$ decays.\\ 
\renewcommand{\arraystretch}{1.2}
\begin{table}[h]
\caption{\small Experimental results for $r$ and $\lambda$ extracted from 
Bose-Einstein and Fermi-Dirac correlations of boson and baryon-pairs
present in the hadronic $Z^0$ decays produced in $e^+ e^-$ 
annihilations in the LEP1 collider.}
\label{tab:comp_lep}
 \begin{center}
\begin{tabular}{||c||c|l|lr|l}
      \hline\hline
h~h     & {\large  $\lambda_2$ }  & ~~~~~~~~{\large r$_2$}~[fm]
&~~~~Experiment 
& \\
  \hline\hline 
 {\large $\pi^{\pm}~\pi^{\pm}$}  & $-$    & {$ 0.78 \pm
   0.01 \pm 0.16$} & 
{\small LEP}1 Average &\cite{acl}\cr
\hline\hline
${ K^{\pm}~K^{\pm}}$ & 0.82 $\pm$ 0.11 $\pm$ 0.25 & 0.48 $\pm$ 0.04 $\pm$
  0.07  & {\small DELPHI}& \cite{delphi_kpkp} \\  
  & 0.82 $\pm$ 0.22 ~~$^{+0.17}_{-0.12}$ & 0.56 $\pm$ 0.08 ~~$^{+0.07}_{-0.06}$  &
OPAL &\cite{opal_kpkp}  \\
\hline
  ${K^0_S ~K^0_S}$ & ${1.14 \pm 0.23 \pm 0.32}$ & ${
    0.76 \pm 0.10 \pm 0.11}$ & 
{\small OPAL}& \cite{opal_kzkz}\\
       & ${0.96 \pm 0.21 \pm 0.40}$ & ${0.65 \pm 0.07 \pm
         0.15}$ & {\small ALEPH}& \cite{aleph_kzkz}\\
       &  ${0.61 \pm 0.16 \pm 0.16}$ &  ${0.55 \pm 0.08 \pm 0.12}$ &
       {\small DELPHI} &\cite{delphi_kpkp}  \\
\hline
 ${\overline{p}~\overline{p}}$& $-$ &${0.15 \pm 0.02 \pm
 0.04}$& {\small OPAL} &\cite{opal_pp} \\ 
\hline
$\Lambda~\Lambda$  &  $-$    & ${0.11 \pm 0.02
  \pm 0.01 }$ & {\small ALEPH}& \cite{aleph_ll} \\ 
\hline\hline
 $\Lambda ~\Lambda$ & Spin Analysis &$0.19 ~~^{+0.37}_{-0.07} \pm 0.02
 $ & {\small OPAL}& \cite{opal_ll}  \\
       &   ``    & $0.11 ~~^{+0.05}_{-0.03} \pm 0.01 $ & {\small
         DELPHI} &
\cite{delphi_ll} \\
       &   ``    & $0.17 \pm 0.13 \pm 0.04$ &  {\small ALEPH}& \cite{aleph_ll} \\
\hline\hline
\end{tabular}
\end{center}
\label{rofm}
\end{table}
\renewcommand{\arraystretch}{1.0}

A summary of the measured di-baryon $r$ values is given in Table
\ref{tab:comp_lep} together with the average LEP1 value for the
$\pi^{\pm}~\pi^{\pm}$ pairs and values for the $K^{\pm}~K^{\pm}$ and
${K^0_S ~K^0_S}$ systems
obtained from BEC analyses.
To note is the fact, as already discussed in
Sec. \ref{sec_kk}, that  the ${K^0_S ~K^0_S}$ pairs 
may also be the decay product of resonances
like the $f_0(980)$, whereas the other hadron pairs
listed in the Table form, so called, exotic states, that is they are
in isospin 2 or strangeness $\pm 2$ states or they belong to an 
identical di-baryon state.
The most striking feature that emerges from the measured data
listed in Table \ref{rofm}
is the very small $r$ values obtained for
the identical di-baryon systems which lie in the vicinity of
$\sim$0.15 fm, way below the values obtained for the mesons 
and also much smaller than the proton charge and nuclear radius 
\cite{selex} 
of $\sim 0.83 \pm 0.02$ fm. 
Finally the fact that the data
in Table \ref{tab:comp_lep} utilised the same reaction at the same centre
of mass energy, affords a unique opportunity to
study the emitter dimension $r$ as a function of the hadron mass. This
feature is dealt with in the following sections. 
\begin{figure}[h]
\centering{\epsfig{file=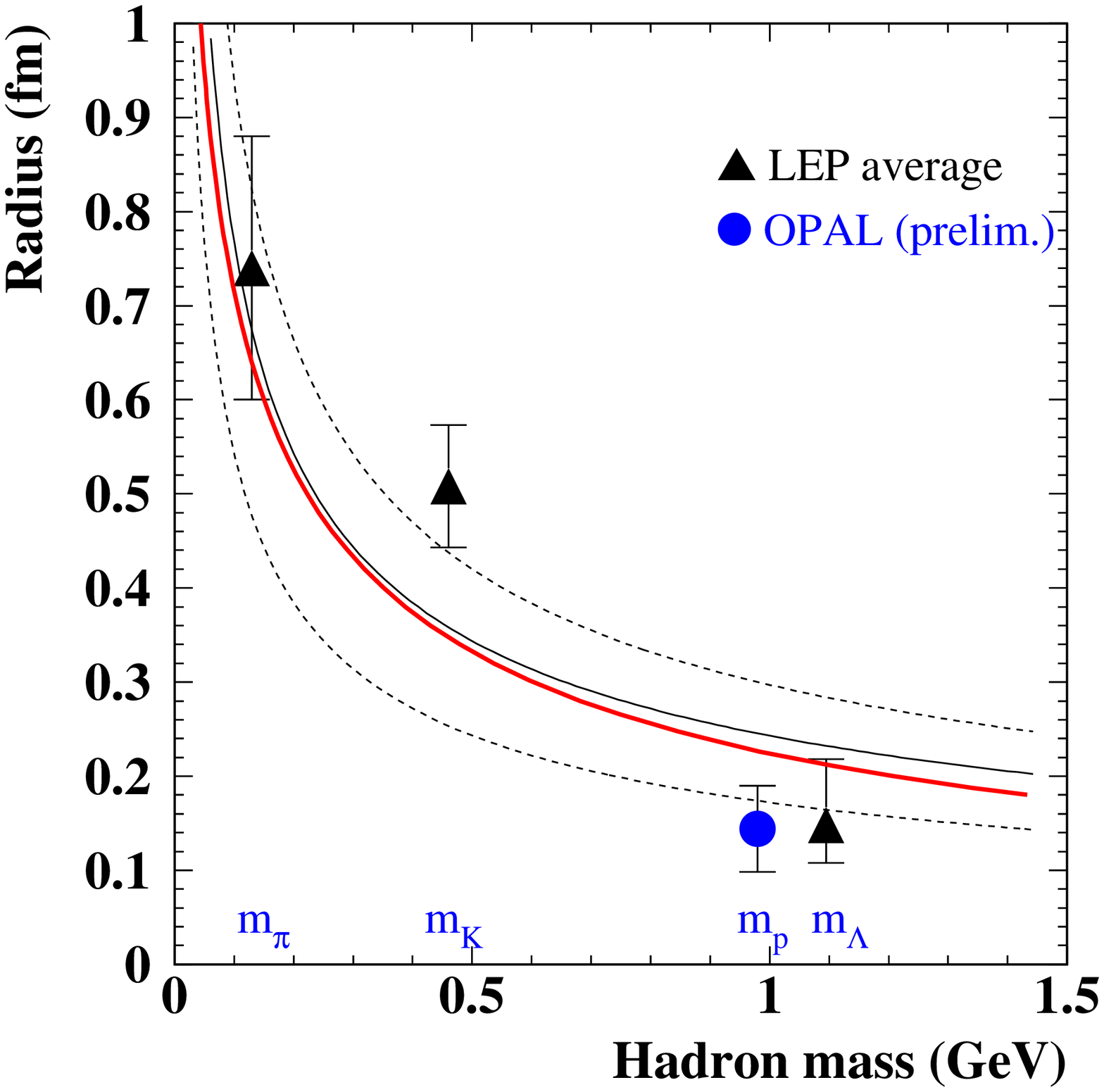,height=8.5cm}}
\caption{\small  
The dependence of $r_2$ on the hadron
mass obtained from BEC and FDC analyses of identical hadron pairs 
emerging from the $Z^0$ decay. 
For clarity the data points are plotted at slightly displaced mass values.
The error bars correspond to the statistical and systematic
errors added in quadrature. The thin solid
line represents the expectation from the Heisenberg uncertainty  
relations
setting $\Delta t = 10^{-24}$ seconds \cite{acl}. The upper and the
lower
dashed lines correspond to $\Delta t = 1.5\times 10^{-24}$
and $0.5\times 10^{-24}$ seconds respectively. The thick solid line
represents the dependence of $r_2$ on the hadron mass as expected from a QCD 
potential given by Eq. (\ref{VT5}).
} 
\label{fig_fd4}
\end{figure} 

\section{The r dependence on the hadron mass}  
\label{sect_rdep}
The $r_2$ dependence on the hadron mass, $r(m)$, shown in
Fig. \ref{fig_fd4}, exhibits a relatively strong decrease in $r$ as the
hadron mass increases, a feature which however currently rests mainly on the 
results obtained for the baryon-pairs. 
The possibility that the origin of this behaviour arises from 
purely kinematic considerations has been explored in Ref. \cite{smith}
with the conclusion that this by itself cannot account for the rather
sharp decrease of $r_2$ with $m$.
Thus this $r_2$ dependence on the hadron mass 
has to be faced by all non-perturbative QCD
models which attempt to describe the hadron production process. 
In particular $r(m)$ poses a challenge to the string approach 
which also constitutes a basis
for the Lund model of hadronisation described in details in
\cite{lund}, which expects that $\partial r/\partial m > 0$
in its rudimental form.
Whereas the somewhat smaller value of $r(m_K)$ as compared to 
that of $r(m_{\pi})$,
may still be acceptable, 
the very small $r_2$ values extracted from the $\Lambda$ and proton pairs
cannot apparently be accommodated within the Lund string 
approach \cite{andersson}.  

\subsection{The r(m) description in terms of the Heisenberg relations} 
\label{heis}
The maximum of the BEC and FDC effects, from which the dimension $r$
is deduced, occurs when the $Q$ value of the two identical
hadrons of mass $m$ approaches zero 
i.e., the hadrons are almost at rest in their CMS
which also means that
the three vector momentum difference $\Delta p$ approaches zero.
This motivated the attempt to link 
the $r(m)$ behaviour  
to the Heisenberg uncertainty 
principle \cite{acl}. 
From these uncertainty relations one has that
\begin{equation}
\Delta p\,\Delta r \,\, = \,\,2\,\mu\,\emph{v}\,r\ = \ m\,\emph{v}\,r\
= \  \hbar\,c \ ,
\label{heis1}
\end{equation}
\noindent
where $m$ and $\emph{v}$ are the hadron mass and its velocity.
Here $\mu$ is
the reduced mass of the di-hadron system and $r$ is the 
geometrical distance between them.
The momentum difference $\Delta p$ is measured in GeV and 
$r \ \equiv \ \Delta r$ is given in Fermi units
while $\hbar c \ = \ 0.197$ GeV fm.
From Eq. ({\ref{heis1}) follows that
\begin{equation}
r\ = \ \frac{\hbar c}{m \emph{v}}\ = \ \frac{\hbar c}{p}\ .
\label{r2}
\end{equation}
\noindent
Simultaneously one also utilises the uncertainty
relation expressed in terms of time and energy
\begin{equation}
\Delta E \Delta t \,\, =\,\,  \frac{ p^2}{m}\, \Delta t \,\, =\,\,
\hbar \ ,
\label{heis2}
\end{equation}

\noindent
where the $\Delta E$ is given in GeV and $\Delta t$ in seconds.
Thus one has
\begin{equation}
p^2 = \hbar\,m/\Delta t \ \ \ \ {\rm{so \ that}} \ \ \ \
p  \,\, = \,\, \sqrt{\hbar\,\,m/\Delta t}\ .
\label{eqv}
\end{equation}
\noindent
Inserting this last expression for $p$ in Eq. (\ref{r2})
one finally obtains
\begin{equation}
r(m) \ = \ \frac{\hbar c/ \sqrt{\hbar /\Delta t}}{\sqrt{m}}\ = \
\frac{c \sqrt{\hbar \Delta t}}{\sqrt{m}}\ .
\label{final}
\end{equation}
If one further assumes that $\Delta t$ is independent of the hadron
mass and its identity and just represents the time scale of
strong interactions, of the order of $\Delta t = 10^{-24}$ seconds,  
then $r(m) \ = \ A/\sqrt{m}$ \ with
$A$\ = \ 0.243 fm GeV$^{1/2}$. This $r$ dependence on $m$
is shown in Fig. \ref{fig_fd4} by the thin solid line. 
The sensitivity of
$r(m)$ to the value of $\Delta t$ is represented by the dashed lines in
the same figure. The upper and the lower dashed lines correspond
respectively to the
values of $\Delta t = 1.5\times 10^{-24}$ and  $0.5\times
10^{-24}$ seconds. A fit of Eq. (\ref{final}) to the
data points plotted in Fig. \ref{fig_fd4} yields
$\Delta t \ = \ (1.2 \pm 0.3)\times 10^{-24}$ seconds where 0.3
represents the statistical error. 
As seen in the figure, the dependence of $r$ on the mass $m$,
as determined from the Heisenberg relations, follows   
within errors  
the measured values   
and thus offers a natural explanation for the experimental found 
hierarchy 
$r_{\pi \pi} \ >\  r_{KK}\  > \ r_{pp} \ \sim  \  r_{\Lambda \Lambda}$.\\

\subsection{QCD description of r(m) via the virial theorem} 
In the previous section it has been shown that the observed hierarchy 
$\partial r/\partial m < 0$
has a natural and a rather general
explanation. Here we would like to demonstrate that such a dependence
could shed  light on the character of the ``soft" interaction or, in other
words, on the non-perturbative QCD which is responsible for the 
interaction at long distances.  In that we follow \cite{acl} 
and use the semi-classical 
approximation as in Sec. \ref{heis}, which means that the angular
momentum is given by
\beq \label{VT1}
\ell \,\,=\,\,|
\,\vec{p}_1\,\,-\,\,\vec{p}_2\,|\,b_t\,\,=\,\,2
\,p\,b_t\,\,\approx\,\,\hbar\,c\,\,,
\eeq
where $b_t$ is  the impact parameter. To estimate the value of $p$ one 
uses the virial theorem \cite{VT} which provides a general connection
between the average values of the kinetic and the potential energies
(T and V), namely
\beq \label{VT2}
2\,\langle \,T_{t}\,\rangle \,\,=\,\,\langle \,\,\vec{b}_t
\,\cdot\,\vec{\nabla}_t\,V(r)\,\,\rangle \,\,,
\eeq
where $t$ denotes the transverse
direction. Since the motion in the transverse direction is always
finite, one can safely apply the virial theorem.
Substituting in \eq{VT2} the kinetic
energy by its relation to the momentum, $T_t\,\,=\,\,p^2_t/ m$, we obtain
\beq \label{VT3}
\langle \ p^2_t\,\rangle \,\,= \,\,m\,\langle \,\,\vec{r}_t
\,\cdot\,\vec{\nabla}_t\,V(r)\,\,\rangle\,\,,
\eeq
where $r$ stands for the distance between the two particles with $r_t $
being equal to $2 b_t$.\\

From \eq{VT3} follows that the relation between
$\langle p^2\,\rangle$ and $r$
depends crucially on the $r$ dependence of the potential energy $V(r)$.
Here we take  $V(r)$ to be independent 
of the interacting particles (quarks) mass. 
This assumption is based on the Local Parton Hadron Duality 
\cite{DUAL,DUAL1} 
concept, which states
that one can consider the production of hadron as interaction of partons, 
i.e. gluons and quarks, in spite of the unknown mechanism of hadronisation. 
This approach is supported experimentally by the single and 
double inclusive
hadron production in $e^{+ }\,e^{-}$ annihilation \cite{DUAL,DUAL1}. 
Since the interaction
between partons is independent of the produced hadron mass it follows that 
$\partial V(r)/\partial m = 0$.
Based on this assumption and using
\eq{VT1} and \eq{VT2},
one obtains an equation for the typical size of the emission volume, namely
\beq \label{VT4}
r^2\,\,\langle\,\,\vec{r}
\,\cdot\,\vec{\nabla}\,V(r)\,\rangle \,\,\,\approx\,\,\, \frac{(\,\hbar
\,c\,)^2}{m}\ . 
\eeq
Further one applies to \eq{VT4} the general QCD potential
\beq \label{VT5}
V(r)\,\,\,=\,\,\,\kappa\,\,r\,\,\,-\,\,\,\frac{4}{3}\,\frac{\alpha_S
\hbar c}{r}\,\,,
\eeq
widely used to derive the wave functions and decay constants
of hadrons \cite{RCP,RCP1}. 
Setting the potential parameters values 
to $\kappa = 0.14$ \ GeV$^2$ = 0.70 GeV/fm\ , 
$\alpha_s \,\,= 2 \pi /9 \,\ln(\delta + \gamma/r )$ with $\delta = 2 $ and
$ \gamma = 1.87$ \ GeV$^{-1} = 0.37$ fm 
as derived from hadron wave 
functions and decay constants \cite{RCP},
the resulting dependence of $r$ on the hadron mass,
given by the solid thick line
in Fig. \ref{fig_fd4}, is seen to follow rather well the 
the data.  
Moreover, this QCD expectation essentially coincides with the prediction 
derived from the Heisenberg uncertainty relations (thin solid line in
the same figure). 

\section{Results from multi-dimensional Bose-Einstein correlation  analyses}
\label{sect_multi}
The study of multi-dimensional BEC has gained 
considerable interest in recent
years, in particular in heavy-ion collisions sector, but also in particle
reactions.
In the framework of the Lund hadronisation model it is expected that
the emitter shape will deviate significantly from a sphere and that
an elongation' of the source should occur. Three
LEP experiments, in their study of  the process
$e^+e^- \to \gamma^{\star}/Z^0 \to hadrons$, have  
analysed their data in the Longitudinal Centre-of-Mass System with the
aim of extracting values for the longitudinal $r_z\ (=\ r_{\ell})$,
transverse $r_T$ and side $r_{side}$ dimensions
(see Sec. \ref{sec_2d}). These are summarised in Table \ref{comp_2d} 
together with the
results from the ZEUS experiment at HERA $ep$ collisions
which confirm that the ratios
$r_{side}/r_z$ and $r_T/r_z$ are significantly smaller than
one. Missing however is a control experiment that will show 
that in the case of a genuine spherical emitter the multi-dimensional
analysis procedure will in fact yield the results that
$r_{side}/r_z\ \approx\  r_T/r_z \approx\ 1$.\\

A somewhat surprising result which came out of the BEC
multi-dimensional analysis of
pion-pairs is the dependence of $r_z$, as well as $r_{side}$
and $r_T$, on the transverse mass $m_T$. An example of this 
$r_z$ dependence on $m_T$ is the one 
reported by the DELPHI collaboration
\cite{delphi_tran1} which is shown in Fig. \ref{fig_rl}. This 
$r_z(m_T)$ behaviour
is very similar to that seen in heavy ion collisions where it
is often attributed to nuclei collective effects which should be 
absent in particle reactions and in particular
in $e^+e^-$ annihilations. A theoretical understanding in terms of the
Lund model of 
this effect in particle reactions is apparently still 
lacking even if only for the
reason that the inclusion of the BEC effect into the model is 
a highly non-trivial task and still seems to need further improvements
\cite{smirnova,todorova}.   

\renewcommand{\arraystretch}{1.2}
\begin{table}[h]
\caption{\small Evidence for the elongation of the source from
experimental results for the ratios
$r_{side}/r_z$ and $r_T/r_z$
obtained from multi-dimensional BEC
analyses of identical charged pions. The first and the second errors
are respectively the statistical and systematic ones.}
 \begin{center}
    \begin{tabular}{||l|c|ll||l|l||}
      \hline\hline
Reaction & $\sqrt{s}$ [GeV] &Experiment&  & $r_{side}/r_z$ & $r_T/r_z$\cr
  \hline\hline 
$e^+e^-\to Z \to$h & 91 & L3 &\cite{l3a_2d} & 
$0.80\pm0.02^{+0.03}_{-0.18}$ & \cr
$e^+e^-\to Z \to$h & 91 & OPAL& \cite{opala_2d} &
$0.82\pm0.02^{+0.01}_{-0.05}$ & \cr
$e^+e^-\to Z \to$h & 91 & DELPHI& \cite{delphia_2d} &
&$0.62\pm0.02\pm0.05$ \cr
\hline
$e~p \to e~h$ & $110<Q^2_{\gamma}$ [GeV$^2$] & ZEUS& \cite{zeus_pipi}&
&$0.67\pm0.08$ \cr
\hline\hline
\end{tabular}
\end{center}
\label{comp_2d}
\end{table}
\renewcommand{\arraystretch}{1.0}
\begin{figure}[h]
\centering{\epsfig{file=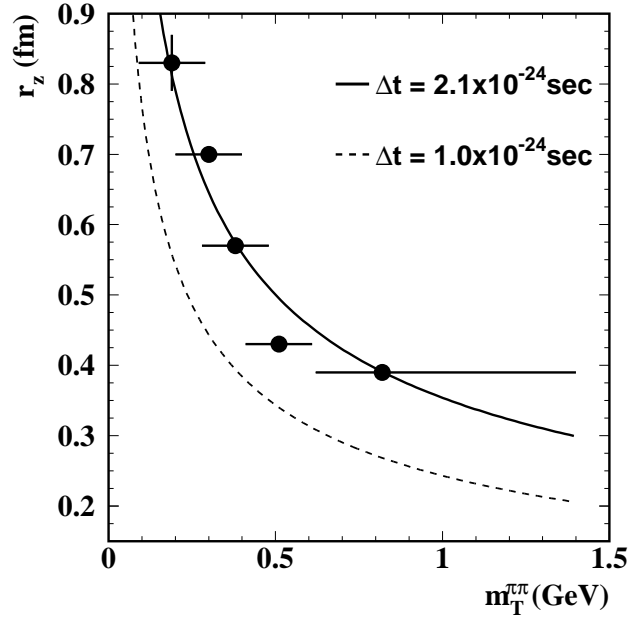,height=9cm}}
\caption{\small Preliminary results of DELPHI \cite{delphi_tran1}
for the dependence of the longitudinal emitter dimension $r_z(\pi\pi)$
on the transverse mass  $m_T(\pi\pi)$
in hadronic $Z^0$ decays.
The data are compared with the expression for $r_z$ given in  
Eq. (\ref{finalmt}) setting $\Delta t$ to the best fitted value of  
2.1 $\times 10^{-24}$ sec (continuous line) and the expectation for
1.0  $\times 10^{-24}$ sec (dashed line).}
\label{fig_rl}
\end{figure}

\subsection{\rzbf\ description in terms of the Heisenberg relations}
We have seen that there is a a great similarity between
the dependence of the 1-dimensional $r_2$ on the hadron mass $m$ and the
dependence of the $r_z$ on the transverse mass $m_T$.
This is not surprising once one realises that 
the longitudinal distance $\Delta r_z$,
and $\Delta p_z$, 
the difference in the longitudinal momentum
of the two hadrons in the LCMS, are also conjugate observables which obey
the uncertainty relation \cite{alex_be}

\begin{equation}
\Delta r_z \Delta p_z\ =\ \hbar c\ .
\label{drdp}
\end{equation}

\noindent
Here $\Delta p_z$ is
measured in GeV, $\Delta r_z\ \equiv\ r_z$ is given in Fermi units and 
$\hbar c\ =\ 0.197$ GeV fm. In the LCMS one has 
$$\Delta p_z\ =\ 2\mu v_z\ =\ p_z\ ,$$  
where $\mu\ =\ m/2$ is the reduced mass of the two identical hadrons of mass
$m$ and the longitudinal velocity $v_z$ of these hadrons. 
Thus
\begin{equation}
r_z\ =\ \frac{\hbar c}{p_z}\ . 
\label{rz}
\end{equation}

\noindent
Simultaneously one considers the
uncertainty relation expressed in terms of energy and time
\begin{equation}
\Delta E \Delta t\ =\ \hbar\ ,
\label{dedt}
\end{equation}
where $\Delta E$ is given in GeV and $\Delta t$ in seconds. 
In as much as the total energy $E$ of the two-hadron
system is given by their mass and their kinetic energy,
i.e. the potential energy can be neglected, one has
\begin{equation}
E\ = \sum_{i=1}^2 \sqrt{m^2 + p^2_{i,x} +  p^2_{i,y} +  p_z^2}\ =\ 
\sum_{i=1}^2 \sqrt{m^2_{i,T} +  p_z^2}\ ,
\label{deltae}
\end{equation}
where $m_{1,T}$ and $m_{2,T}$ are the transverse mass of the first and
second hadron.
As $Q_z$ decreases the longitudinal momentum $p_z$ vanishes 
so that one can, once  $p_z^2 <  m^2_{i,T}$,
expand the hadron energy $E$ in terms of $p_z^2/m^2_{i,T}$ and 
retain only the two first terms,

\begin{equation}
E\ = \sum_{i=1}^2 m_{i,T} \sqrt{1\ +\ \frac{p_z^2}{m^2_{i,T}}}\ 
\approx\ \sum_{i=1}^2 m_{i,T} + 
\sum_{i=1}^2 \frac{1}{2}\frac{p_z^2}{m_{i,T}}\ .
\label{expand}
\end{equation}
\noindent
Next one can order the identical bosons so that 
$m_{1,T} \geq m_{2,T}$ and define\\
\begin{equation}
\delta m_T = \frac{m_{1,T} - m_{2,T}}{m^2_T -(\delta m_T)^2}
\ \ \ \ {\rm{while}}\ \ \ \ m_T =\frac{m_{1,T} + m_{2,T}}{2}\ .
\end{equation}
Inserting these relations into Eq. (\ref{expand}) one gets, after a few
algebraic steps that\\
\begin{equation}
E\ =\ 2 m_T\ +\ \frac{m_T p^2_z}{m^2_T - (\delta m_T)^2}\ . 
\end{equation}
Thus as long as $m^2_T \gg (\delta m_T)^2$ the total energy of the
di-boson system is approximately equal to
\begin{equation}
E\ \approx\ 2 m_T\ +\ \frac{p^2_z}{m_T}\ . 
\label{finale}
\end{equation}

\noindent
In the 2-dimensional analysis the $r_z$ estimation is deduced from
the BEC enhancement in the longitudinal direction, keeping $m_T$
fixed and
letting $Q_z$ and therefore also $p_z$, approach zero.  
Thus as long as the uncertainty in 2$m_T$ is much smaller than
that of $p_z^2/m_T$ one has
\begin{equation}
\Delta E \Delta t\ =\ \frac{p_z^2}{m_T}\Delta t\ =\ \hbar\ . 
\label{deltaf}
\end{equation}
Combining Eqs. (\ref{rz}) and (\ref{deltaf})
one finally has

\begin{equation}
r_z(m_T)\ \approx\ \frac{c\sqrt{\hbar \Delta t}}{\sqrt{m_T}}\ .  
\label{finalmt} 
\end{equation}

This last equation is identical to the one given in Eq. (\ref{final}) 
for the dependence of the emitter dimension on the boson mass  
when $r$ and $m$ are replaced by $r_z$ and $m_T$. A fit of
Eq. (\ref{finalmt}) to the data shown in Fig. \ref{fig_rl} 
yields $\Delta t\ =\ (2.1 \pm 0.4)\times 10^{-24}$ sec
so that $r_z(m_T)\ =\  0.354/\sqrt{m_T[GeV]}$ fm.
This $\Delta t$ value is compatible with the corresponding value of 
$(1.2 \pm 0.3)\times 10^{-24}$ sec obtained in
\cite{acl} from a fit to $r(m)$ in view of 
the considerable 
spread of the $r$ values obtained from the 1-dimensional BEC analyses
carried out with the LEP1 data. 
In heavy-ion collisions the longitudinal range $r_z$ was also observed
to be inversely proportional to the square root of $m_T$ 
\cite{heinz_bec} namely, $r(m_T) \approx 2/\sqrt{m_T[GeV]}$ fm where the
difference between the proportionality factor of 0.354 and 2.0 can
be accounted for by the difference in the extent of
the heavy ion targets as compared to that existing in the $e^+ e^-$  
annihilation.\\    

The dependence of the transverse dimension $r_T$ on the 
two-pion transverse
mass $m_T$ has also been measured in the hadronic $Z^0$ decays
\cite{L3_twod}. 
Here again the transverse range was found to decrease
as $m_T$ increases. However 
unlike $r_z$ which is a geometrical quantity, $r_T$ is a mixture of the 
transverse radius and the emission time difference between the two
hadrons so that a straightforward
application of the uncertainty relations is improper.

\subsection{Application of the Bjorken-Gottfried relation to r(m$_T$)} 
\begin{figure}[h]
\centering{\epsfig{file=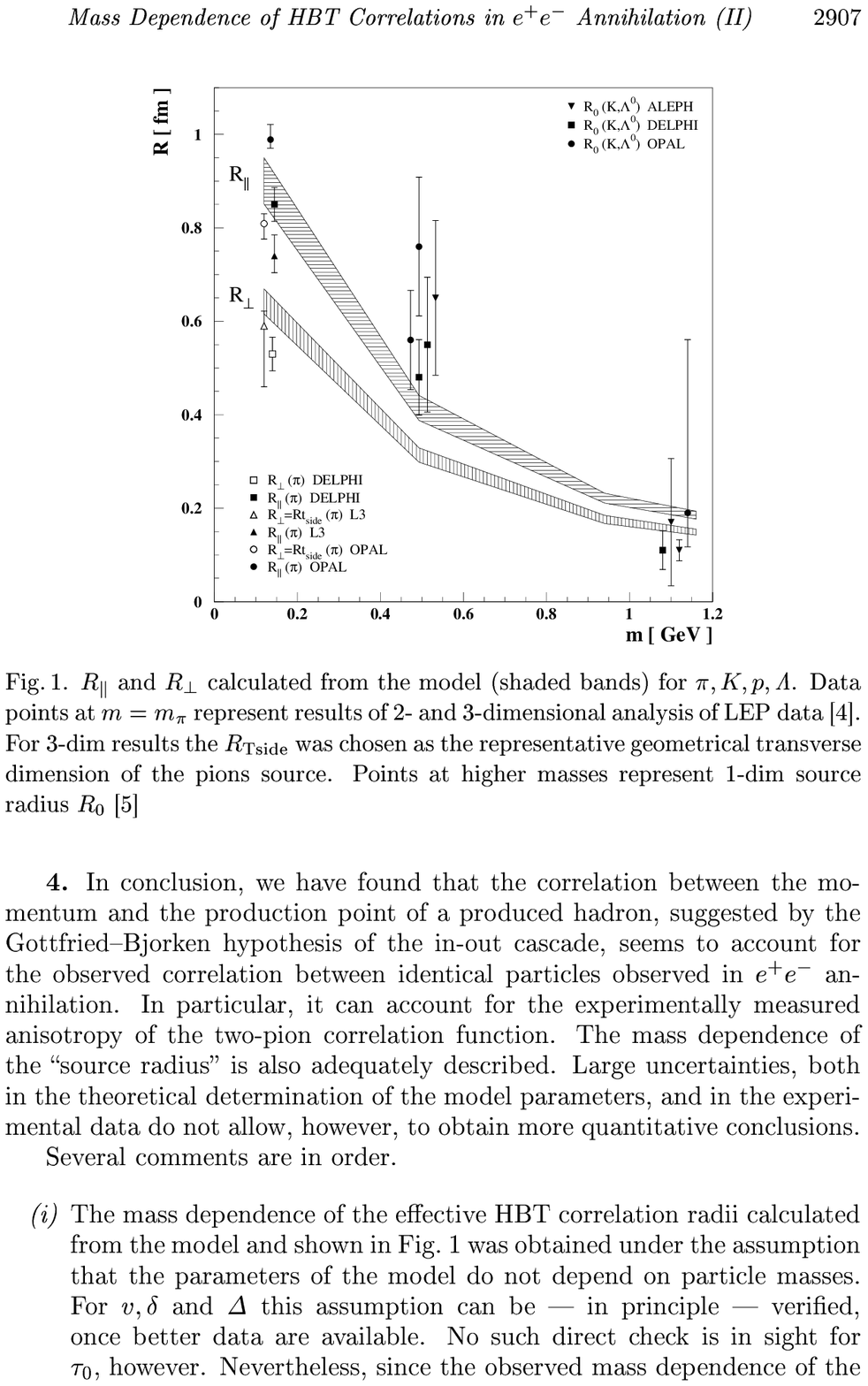,height=7cm,
bbllx=168pt,bblly=447pt,bburx=436pt,bbury=694pt,clip=}}
\caption{\small $r_z$ (=${\small R}_{||}$) and $r_T$ (=${\small R}_{\perp}$) 
as a function of the 
transverse mass $m_T$ taken from Refs. \cite{bialas1,bialas2}. 
The dashed bands 
were calculated from a model based on the 
Bjorken-Gottfried relation. The data points are taken from 
BEC analyses of hadronic $Z^0$ decays.
}
\label{fig_bialas}
\end{figure}
Another approach to account for the dependence of $r_Z$ and $r_T$ on the mass
of the hadron has been taken in Refs. \cite{bialas1,bialas2}. In that
approach
the parameters characterising the particle emission region are
mass independent and the observed change in radii is solely
the consequence of
the momentum-position correlation as expressed in the
Bjorken-Gottfried condition
\cite{gottfried1,gottfried2,bjorken1,bjorken2}. In this hypothesis
there exists a linear relation between the 4-momentum of the produced
particle and the time-space position at which it is produced, namely
$$q_{\mu}\ =\ \lambda x_{\mu}\ \ \ \ \ {\rm{which\ implies}}\ \ \ \ \ 
\lambda\ =\ m_T/\tau\ ,$$
where $m_T$ is the transverse mass and $\tau = \sqrt{t^2-z^2}$ is the
longitudinal proper time after the collision. 
In the framework of this hypothesis relations between   
the longitudinal and transverse radii and the transverse mass can 
be derived. These are given in Fig. \ref{fig_bialas} by the 
dashed bands where they are seen to adequately describe
the results obtained from the 1-dimensional BEC analyses
carried out with the LEP1 data.

\subsection{r(m) and the interatomic separation in Bose condensates}   
The observation that the dimensions $r$ and $r_z$ derived from 
BEC analyses are seen respectively to be
inversely proportional to $\sqrt{m}$ and $\sqrt{m_T}$, 
naturally arouses the curiosity
whether similar behaviour does also exist in other
phenomena associated to the Bose-Einstein statistics. 
To this end we
turn here to the 
Bose condensate states where we will show   
that the average interatomic separation is 
inversely proportional to the square-root of the 
atomic mass. In this presentation   
we follow closely reference \cite{alex_be} and first 
summarise briefly, for the benefit of the non-expert reader, the main 
features of the Bose condensates and focus on their
interatomic separation.\\

When bosonic atoms are cooled down below a critical temperature $T_B$, 
the atomic wave-packets overlap and the 
equal identity of the particles becomes significant. 
At this temperature, these atoms undergo a 
quantum mechanical phase transition and form a Bose
condensate, a coherent cloud of atoms all occupying 
the same quantum mechanical state.
This phenomenon, first surmised by
A. Einstein in 1924/5, is a consequence of 
quantum statistics \cite{einstein,einstein1}.
Detailed theoretical aspects of the Bose-Einstein condensation
can be found in Ref. \cite{theory} and its  
recent experimental situation is described in Ref. \cite{experiment}.
Concise summaries, aimed in particular to the non-expert, 
both of the experimental aspects and the theoretical
background, can be found in Refs. \cite{ketterle,burnett}. 
To form Bose condensates, extremely dilute gases
have to be cooled down below the critical temperature
$T_B$, so that the formation time of molecules and
clusters in three-body collisions is slowed down to seconds or
even minutes to prevent the creation of the more familiar transitions into
liquid or even solid states.\\

The existence of Bose-Einstein  condensation was first 
demonstrated in 1995 by three groups
\cite{discovery,discovery1,discovery2} in cooling down rubidium, sodium and lithium.
Typical temperatures where Bose condensates occur are in the
range of 500 nK  to 2 $\mu$K with atom densities between
$10^{14}$ and $10^{15}$ cm$^{-3}$. The largest sodium condensate 
has about 20 million atoms whereas hydrogen 
condensate can reach even one billion atoms.\\

Let us now
consider a dilute homogeneous ideal gas of $N$ identical bosonic atoms
of spin zero
confined in a volume $V$. These atoms occupy $\epsilon_i$
energy levels, handled 
here as a
continuous variable, which are distributed according to the  
Bose-Einstein statistics. 
We further set the ground state to be
$\epsilon_0 = 0$. If $N_0$ is the number of atoms in this ground state
and $N_{ex}$ is the number of atoms in the excited states then
$N\ =\ N_0\ +\ N_{ex}$. For a 
homogeneous ideal gas of identical bosonic atoms it can be shown
(see e.g. Ref. \cite{book}) that at a low temperature $T$ one has
\begin{equation}
N_{ex}=2.612\, V\left(\frac{2\pi mkT}{h^2}\right )^{3/2}\ ,
\label{exvn}
\end{equation}
where $V$ is the volume occupied by the atoms of mass $m$. 
Since $T_B$ is defined as
the temperature where almost all bosons are still in excited states,
we can, to a good approximation, equate $N$ with $N_{ex}$. That is
\begin{equation}
N = 2.612\, V \left(\frac{2\pi mkT_B}{h^2}\right )^{3/2}\ .
\label{exnn}
\end{equation}
For $T< T_B$ one obtains from Eqs. (\ref{exvn}) and (\ref{exnn})  
that the number of atoms $N_0$ which are in the condensate state is, 
\begin{equation}
N_0= N - N_{ex} = N\left[1- \left(\frac{T}{T_B}\right )^{3/2}\right ]\ ,
\label{nzeron}
\end{equation} 
where $T$ is the temperature of the atoms lying at the excited energy
states
above the condensate energy level $\epsilon_0 =0$. 
The atomic density of the Bose gas at very low temperatures,
$T/T_B << 1$ where $N \approx N_0$, is then given by
\begin{equation}
\rho=\
\frac{N}{V}=2.612\left(\frac{2\pi mkT}{h^2}\right )^{3/2}\ ,
\label{exdn}
\end{equation}
\noindent
where $k$ is the Boltzmann constant and
$\rho$, the atomic density, has the dimension of L$^{-3}$.
From this follows that $\rho^{-1/3}$ is the average interatomic 
separation in the Bose condensate. At the same time the thermal 
de Broglie wave length is equal to
\begin{equation}
\lambda_{dB}=\left(\frac{h^2}{2\pi mkT}\right)^{1/2}\ . 
\label{broglie}
\end{equation}
Combining Eqs. (\ref{exdn}) and (\ref{broglie}) one has
for the state of a Bose condensate the relation
\begin{equation}
\rho \lambda ^3_{dB} \approx 2.612\ .
\label{condition}
\end{equation}
Thus the average interatomic distance in a Bose condensate,
$d_{BE}$, is equal to
\begin{equation}
d_{BE} \equiv \rho^{-1/3} \approx \lambda_{dB}/1.378\ .  
\end{equation}

Next one can consider two different bosonic gases, having atoms with masses
$m_1$ and $m_2$, which are cooled down to the same very low
temperature $T_0$, below the critical temperature 
$T_B$ of both of them. In this case one produces two Bose condensates 
with interatomic distances
\begin{equation}
d_{BE}(m_i)\ \approx 
\ \frac{\sqrt{2\pi}}{1.378}\left(\frac{\hbar^2}{m_ikT_0}\right)^{1/2}\
;\ \ \ i\ =\ 1,2\ .
\label{dbfinal}
\end{equation}  
From this follows that when two condensates are at the same
fixed temperature $T_0$ one has
\begin{equation}
\frac{d_{BE}(m_1)}{d_{BE}(m_2)} = \sqrt{\frac{m_2}{m_1}}\ ,
\label{mmm}
\end{equation}
which is also the expectation of Eq. (\ref{final}) 
for the $r$
dependence on the mass of the hadron produced in high energy 
reactions provided $\Delta t$ is fixed. Finally it is
interesting to note that in as much as one 
is justified, since almost all the atoms occupy the same 
quantum mechanical state, to replace in Eq. (\ref{dbfinal}), at very low
temperatures, 
$kT_0$ by 
$\Delta E$ and use the uncertainty relation 
$\Delta E = \hbar /\Delta t$ one derives the expression for $r(m)$ 
as given by Eq. (\ref{final}) multiplied by the factor 
$\sqrt{2\pi}/1.378$. Thus in Bose-Einstein condensates $r(m)$ measures the
interatomic separation and not the dimension of the condensate
ensemble,
a deduction which will be further referred to in Sec. \ref{inter}.\\
 
In relating the condensates to the production 
of hadrons in high energy reactions one should    
keep in mind that the interatomic separation proportionality
to $1/\sqrt{m_{atom}}$ does not necessarily imply
that it should also apply to 
hadrons produced in high energy reactions. 
Common to both systems is their
bosonic nature which allows all hadrons (atoms) to occupy
the same lowest energy state. In addition the condensates are 
taken to be in
a thermal equilibrium state. Among the various models
proposed for the production of hadrons in high energy $e^+ e^-$
and nucleon-nucleon reactions
some attempts have also been made to explore the application of
a statistical thermal-like model \cite{becattini}. 
Whether this approach will eventually prevail
is at present questionable. Finally the
condensates are taken to be in a  coherent state. In the case of 
hadrons one is able to measure $r$ via BEC only if the chaoticity
factor $\lambda$ in Eq. (\ref{c_two}) is different from zero 
i.e., only if the source
is not 100$\%$ coherent. However so far there is no evidence from BEC
studies in hadron production for
a dependence of $r$ on $\lambda$ apart from that introduced by
the correlated errors between $\Delta r$ and $\Delta \lambda$ 
produced by fitting Eq. (\ref{c_two}) to the data.

\section{On the relation between r and the emitter size}
\label{inter}
As it has been pointed out in the introduction, the HBT interferometry 
in astronomy and its equivalent, the GGLP correlations 
in its application to heavy ion collisions, have been
rather successful in estimating respectively the dimensions 
and configurations of stellar objects and the geometrical extent 
of nuclei. In the elementary particle sector 
it looks as if the $r$ dimension is very little if at all
dependent on the type and energy of the reaction. Furthermore 
$e^+ e^-$ annihilations, there seem to be several 
features which pose a challenge to the general belief 
that $r$ measures a quantity which represents 
the emitter volume and corresponds to its radius in the spherical
approximation.\\

In the Lund string model one expects that 
the extension of the emitter of hadrons in
elementary particle reactions will increase with multiplicity
and hence with energy  
and certainly in $e^+e^-$ annihilation at 91 GeV will 
spread over more than several fm. 
From the Bose-Einstein correlation analyses
there is no evidence for such a behaviour.
In fact, 
the $r$ values obtained from $\pi^{\pm}\pi^{\pm}$ pairs produced in
$e^+ e^-$ annihilation seem to be rather independent
of energy having values in the vicinity of 0.75 fm 
(see Table \ref{tab_2pions}).\\ 

Moreover, recent experimental results show that 
the dimension $r$, measured in $e^+e^-$ annihilations on the $Z^0$ mass energy,  
does depend on the hadron mass. 
As shown in Sec. \ref{sect_rdep}, the $r$ values at the proton
and $\Lambda$ masses decrease to the very low value of $\sim$0.15 fm
as compared to approximately 0.75 fm for pions and kaons.
This behaviour of $r(m)$ is very similar to  
the interatomic separation dependence on the atomic mass
in equal temperature Bose condensates. 
In as much as one is justified to infer from the 
Bose condensates features to the hadron
production sector, $r(m)$ apparently does not measure the 
emitter radius but rather the distance between two neighbouring identical
hadrons.\\ 

In any case, the experimental results found for the dependence of $r$
on the hadron mass are not reconcilable with the notion 
of a unique emitter radius for all outgoing hadrons.   
However the possibility that baryons are emitted from 
a much smaller source than pions and kaons
has its own difficulty \cite{alex_buda}
if one considers
the source energy density $\epsilo$. 
To illustrate this problem
let us assume that the emitter source is a sphere of
radius $r$, as determined from the correlation analysis, filled 
homogeneously with at least the energy which corresponds to
the sum of the rest masses of the two outgoing identical hadrons. 
Then, for
a given hadron mass $m_h$ and its emitter radius
$r_h$, the energy density of the source is given by
\begin{equation}
{\epsilo_{measured}}\ =\ \frac{3}{4\pi}\frac{2m_h}{r^3_h}
\label{dense_meas}
\end{equation}
which is plotted in Fig. \ref{fig_density} for the pion, kaon,
proton and $\Lambda$ pairs. 
The dashed lines in the figure are the predicted dependence of
the energy density on the hadron mass 
computed from Eq. (\ref{dense_model}), which is based on  
the expression of $r(m)$ given in Eq. \ref{final}, namely
\begin{equation}
{\epsilo_{model}}\ =\ \frac{3}{2\pi}\frac{m^{5/2}_h}
{c^3(\hbar\Delta t)^{3/2}}\ .
\label{dense_model}
\end{equation}
Whereas the energy density of the pion and kaon pairs are
still lying within an acceptable range, the energy density
of the proton and $\Lambda$ emitter sources reach 
the very high value of about 100 GeV/fm$^3$.\\ 

\begin{figure}[h]
\centering{\epsfig{file=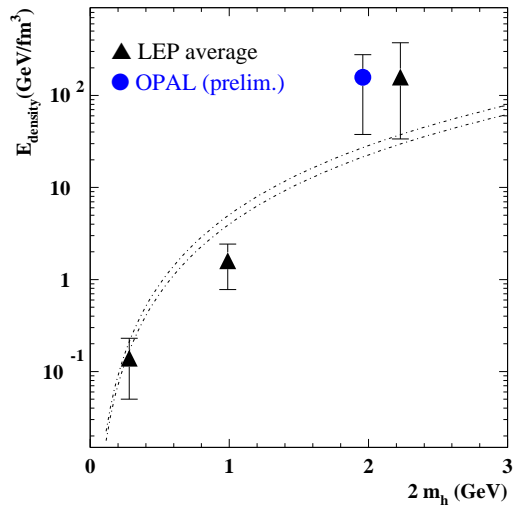,height=7.5cm}}
\caption{\small Energy density of the hadron emitter, 
calculated from Eq. (\ref{dense_meas}), 
as a function of the hadron mass assuming a spherical emitter having 
the radius $r$ which was extracted from the BEC analysis. 
The dashed lines are the expectations from Eq. 
(\ref{dense_model}) 
setting $\Delta t$ to plus and minus three standard deviations 
away from
its median value of $1.2\times 10^{-24}$ seconds \cite{alex_buda}.}
\label{fig_density}
\end{figure}

In trying to cope with these observations, it may be helpful to try 
and identify the underlying variable or variables, apart from the
hadron mass, 
which $r$ depends on. 
A partial answer to this quest \cite{gaes}, which is
briefly
outlined in the following section, is provided by the
relation between the BEC analysis 
and the genuine particle correlation method which utilises
the factorial cumulant moments (see Sec. \ref{higher}).
  
\subsection{The $r$ dependence on the number of emitting sources} 

Let us consider a simplified description of heavy ions (AA)
interactions at
high energies which is given in terms of several distinct, 
say $S$, primary nucleon-nucleon (nn) collisions leading to hadrons.
In this approach the hadrons produced in a single
primary nucleon-nucleon reaction  
are emitted from an $S_{nn}$ source. 
The hadron produced
in heavy ion (AA) collisions are then coming from $S_{AA}$ sources 
so that $S_{AA}$/$S_{nn}$=S. Since here the AA collisions are
described in terms of the single primary nucleon-nucleon reactions one can 
define, without loss of generality, $S_{nn}$ as a unit source 
so that $S_{AA}=S$.\\

Next the $\pi^{\pm}\pi^{\pm}$ pairs emerging from a single primary
nucleon-nucleon reaction and those emerging from the
heavy ion collisions are subject to BEC analyses
and the correlation functions
$$C^{S=1}_2(Q)\ =\ 1\ +\ \lambda_{S=1} e^{-Q^2r_{S=1}^2}
\ \ \ \ {\rm{and}}\ \ \ \  C^S_2(Q)\ =\ 1\ +\ \lambda_S
e^{-Q^2r_S^2}$$
are fitted, each to its corresponding data distribution. 
In general for every $S$ value the resulting $r_S$
can be extracted from the fitted correlation functions
by integration ~$C_2 - 1$ ~over $Q$ from zero to infinity so that
\begin{equation}
\int^\infty_0 \left [C^S_2(Q) -1\right ] dQ\ =\ 
\int^\infty_0 \lambda_S e^{-Q^2r_S^2}dQ\ =
{\displaystyle \frac {1}{2}} \,{\displaystyle \frac {\lambda_S 
\sqrt{\pi }}{{r_S}}}\ ,
\label{integration}
\end{equation}
where $\lambda_S$ is, as customary in the BEC analyses,
taken to be independent of $Q$.\\

For the two-particle correlation there exists
a relation (see e.g. Ref. \cite{cumulant_bec})
between the cumulant $K_2$ and the BEC function
$C_2$, namely that 
\begin{equation} 
C_2(Q)\ =\ 1\ +\ \lambda e^{-r^2Q^2}\ =\ 1\ +\ K_2\ .
\label{cs}
\end{equation} 
In addition it has been
shown \cite{lipa,bushbeck,alex_sark} 
that for two particle correlation the cumulant $K_2^S$ of
$S$ sources
is, to a very good approximation, related to the value of the 
corresponding cumulant of one source, $K_2^{S=1}$,
by 
$$K_2^S\ =\ K_2^{S=1}/S\ .$$
Thus the BEC function $C_2^S$ is equal to
\begin{equation}
C_2^S(Q)\ =\ 1\ +\ \lambda_S e^{-r^2_SQ^2}\ =\ 
1\ +\ \frac{K_2^{S=1}}{S}\
=\ 1\ +\ \frac{\lambda_{S=1}e^{- r^2_{S=1}Q^2}}{S}\ .
\label{smain}
\end{equation}
By integrating Eq. (\ref{smain}) over $Q$ from zero to
infinity one finally obtains that
\begin{equation}
r_S\ =\ S \frac{\lambda_S}{\lambda_{S=1}} r_{S=1}\ ,
\label{rsr1}
\end{equation}
where $S$ like the $\lambda$ parameters is taken to be independent of $Q$.
Consequently
if the increase of an emitter volume is coupled to 
an increase in the number of boson sources, then the BEC extracted
$r_S$ will grow as well. 
Further to note is that if  
in heavy-ion collisions $S\lambda_S/\lambda_{S=1}$ is
proportional to $A^{1/3}$ then $r$ will be proportional to
the nucleus radius. As for the HBT analysis applied in astronomy, 
the proportionality between the stellar object dimension
and the number of photon sources is self-evident.\\ 

From the discussion above it is obvious that the choice
of the unit source determines the predicted dependence of $r$
on the various features, such as type and energy, of the hadron 
(or photon) emitting reaction.      
In the case of 
heavy ion collisions the choice of the unit source to be a 
single primary nucleon-nucleon
collision is a natural one. In the case  
of $e^+ e^-$ annihilation one may consider the possibility that $S_{ee}$
itself is the unit source.
Under this assumption the number of sources does not
change with energy and $r$ will remain constant.
This prediction seems to be consistent with the
measured $r$ values allowing for the fact that
the various experiments used  
different analysis methods and reference samples (see Sec. 
\ref{pipibec}). 
On the other hand an increase of $r$ 
as a function of the hadron multiplicity has been 
observed in the BEC analysis \cite{opal_pipi}
of equally non-zero charged pion-pairs emerging from the $Z^0$ decay.
Furthermore the same experiment also found
that the extracted $r$ value from
the $Z^0$ decays into three hadron jets was by about 10$\%$ higher than
that obtained from the $Z^0$ decay events having only two hadron jets. 
Similar rise of $r$ 
with the average hadron multiplicity has also been 
reported by several hadron-hadron collision  
experiments carried out at centre of mass energies above 30 GeV
\cite{albajar,akesson1,akesson2,breakstone}.\\

These last findings indicate that apparently for the $e^+e^-$
annihilations 
a reasonable choice for the unit source is the single hadron jet.
The shortcoming of such a choice is however the fact that the
charge multiplicity of a hadron jet 
depends on its energy and on its origin, i.e. if it is
a quark or a gluon \cite{gary1,gary2,madjid}, which currently
impedes a detailed quantitative prediction for the behaviour of 
$r$ as a function of the centre of mass energy and
the multiplicity. Regardless of these hindrances,
the adaptation of the hadron jet as a unit source
provides nevertheless 
a qualitative explanation for the increase of $r$ with
multiplicity and also expects in $e^+e^-$ annihilations a moderate
dependence of $r$ on the $e^+e^-$ centre of mass energy.\\ 
 
\section{Summary and conclusions}
\label{sect_conclusion}
The 1-dimensional Bose-Einstein 
correlation 
enhancement measured in identical pion-pairs close in time-phase space,
is found to be a universal phenomenon present in a large variety of  
particle reactions and over a wide range of energies. 
Nevertheless, there are still missing systematic BEC studies 
concentrating on 
the same reaction at different energies, and/or different reactions
at the same energy, which will allow a comprehensive
investigation of the
dependence of $r$ on the variables that
characterise the particle interactions.
Notwithstanding the fact that these systematic studies are still missing
and that
BEC analyses cannot currently be considered as precision measurements, 
the extracted $r$ values, which are taken to represent the
hadron emitter source, seem to be 
very little, if at all, dependent on the reaction type or on its
energy
and they are seen to fluctuate in the range of about 0.6 to 1.0 fm.\\

The essentially constant $r$ value seems to be in contradiction 
with the results obtained from the BEC studies of heavy-ion collisions and  
from the HBT analyses of stellar objects. An insight into the origin of  
this difference is offered by the relation 
between the BEC and the cumulant moments
analysis where the introduction of the so called
hadron source plays a central role. From this relation follows that
the value of $r$ is proportional to the number of active hadron
sources which participate in the reaction. 
In particle reactions the proposal to equate the
number of sources with the number of hadron jets offers
an understanding of the $r$ dependence on multiplicity. 
At the same time however the difficulty in defining a
standard hadron jet prohibits presently any quantitative results.\\

The genuine higher order BEC analyses are presently restricted to
three identical charged pions. Assuming a Gaussian
hadron emitter, the expected relation  
$r_2/\sqrt{3} \leq  r_3  \leq  r_2/\sqrt{2}$ between
the two and three-pion BEC dimension parameters, 
is found to be consistent within errors with the 
measurements carried out in $e^+ e^-$ annihilation and hadron collisions.
The direct measurement of the
genuine correlations of four and more pion systems has not been
realised so far. On the other hand
the extracted $r$ values from the , so to speak,
non-genuine correlations, are model dependent and as such
less reliable.\\

Recently first results from 2 and 3-dimensional BEC analyses of
pion-pairs have been reported. These verified the Lund string model
expectation that the hadron emitter
source deviates from a sphere and is elongated in the 
thrust (or sphericity) direction of the multi-hadron final state.
In the approximation that the emitter
has an elongated prolate ellipsoid shape, where the long axis $r_{\ell}$ is 
in the thrust direction and $r_T$ is the axis perpendicular to it,
the ratio $r_T/r_{\ell}$ is found be in the range of
$\sim$0.7 to $\sim$0.8.\\ 
  
The extension of the BEC analysis to the fermion sector was 
shown to be feasible via 
the Pauli exclusion principle. First results from
the so called Fermi-Dirac correlation analysis of the 
$\Lambda\Lambda$ and proton$-$proton pairs, emerging from the 
hadronic $Z^0$ decays,
have demonstrated that the dimension $r$
is strongly dependent on the mass of the 
identical hadrons. It has been further shown that
this dependence can be accounted for by the Heisenberg
uncertainty relations, or alternatively by QCD derived potential,
to yield the relation $r(m) \propto 1/\sqrt{m}$. 
Similar relation is seen to be valid also in the 2-dimensional
BEC analysis when $r$ is replaced by the $r_{\ell}$  
and $m$ by the transverse mass $m_t$. It is interesting further
to note that the interatomic separation in the Bose
condensates having the same low temperature, is proportional
to $1/\sqrt{m_{atom}}$\ .\\ 

So far the FDC analyses were
restricted to the $Z^0$ hadronic decays, therefore  
it is essential
that they will be repeated in other particle reactions to 
verify that indeed $r$ of the $\Lambda$ and the proton pairs is in the
vicinity of 0.15 fm. 
This small $r$ value, 
derived from the FDC analyses of baryon-pairs, 
poses a challenge to the concept that $r$
represents the hadron emitter radius. 
If one does insist on keeping the radius emitter interpretation for
$r$ then one encounters the situation of
an unreasonable high 
energy density value, of some 100 GeV/fm$^3$, of the 
baryons source.
Thus the very small $r_{\Lambda\Lambda}$ and $r_{pp}$ values 
indicate that our level of understanding the
production of hadrons, and in particular baryons, is apparently
far from being satisfactory.\\

\subsection*{\small Acknowledgements}
I would like to express my thanks to my colleagues G. Bella, I. Cohen, 
E.K. Grinbaum-Sarkisyan, P. S\"{o}ding, and E. Tsur 
for many helpful discussions and comments. I am 
in particular indebted to B. Ilan for going 
diligently over the manuscript to ensure that it is free of
errors 
and omissions.\\
\bibliography{befd}

\begin{thebibliography}{100}

\bibitem{hbt1}
R.~Hanbury-Brown and R.~Q. Twiss.
\newblock {\em Phil. Mag.}, 45:663, 1954.

\bibitem{hbt2}
R.~Hanbury-Brown and R.~Q. Twiss.
\newblock {\em Nature}, 177:27, 1956.

\bibitem{hbt3}
R.~Hanbury-Brown and R.~Q. Twiss.
\newblock {\em Nature}, 178:1046, 1956.

\bibitem{boal}
D.~H. Boal, C.~K. Gelbke, and B.~K. Jenning.
\newblock {\em Rev. Mod. Phys.}, 62:553, 1990.

\bibitem{baym}
G.~Baym.
\newblock {\em Acta Phys. Polon.}, B29:1839, 1998.

\bibitem{Wiedemann}
U.~A. Wiedemann and U.~Heinz.
\newblock {\em Phys. Rept.}, 319:145, 1999.

\bibitem{goldhaber}
G.~Goldhaber et~al.
\newblock {\em Phys. Rev. Lett.}, 3:181, 1959.

\bibitem{goldhaber1}
G.~Goldhaber et~al.
\newblock {\em Phys. Rev. Lett.}, 120:300, 1960.

\bibitem{alex_lipkin}
G.~Alexander and H.~J. Lipkin.
\newblock {\em Phys. Lett.}, B352:162, 1995.

\bibitem{chacon1}
A.~D. Chacon et~al.
\newblock {\em Phys. Rev.}, C43:2670, 1991{,} and references therein.

\bibitem{heinz}
U.~Heinz.
\newblock {H}anbury-{B}rown {T}wiss interferometry for relativistic heavy ion
  collisions: theoretical aspects.
\newblock In {\em Proc. of Int. Summer School on Correlations and Clustering
  Phenomena in Subatomic Physics}, page 137, Dronten, The Netherlands, 1996.

\bibitem{heinz_bec}
U.~Heinz and B.V. Jacak.
\newblock {\em Ann. Rev. Nucl. Part. Sci.}, 49:529, 1999{,} and references
  therein.

\bibitem{weiner}
R.~M. Weiner.
\newblock {\em Phys. Rept.}, 327:249, 2000.

\bibitem{ringner}
M.~Ringn{\' e}r.
\newblock {\em Correlations in multiparticle production}.
\newblock PhD thesis, Lund University, Lund, Sweden, 1998.

\bibitem{acton}
{OPAL}~collaboration{,} P.~D.~Acton, et~al.
\newblock {\em Phys. Lett.}, B267:143, 1991.

\bibitem{zeus_pipi}
{ZEUS}~collaboration{,} M.~Derrick.
\newblock {\em Acta Phys. Polon.}, B33:3281, 2002.

\bibitem{final-state}
T.~Osada, S.~Sano, and M.~Biyajima.
\newblock {\em Z. Phys.}, C72:285, 1996.

\bibitem{kopylov1}
G.~I. Kopylov and M.~I. Podgoretskii.
\newblock {\em Sov. J. Nucl. Phys.}, 15:219, 1972.

\bibitem{kopylov2}
G.~I. Kopylov and M.~I. Podgoretskii.
\newblock {\em Sov. J. Nucl. Phys.}, 18:336, 1973.

\bibitem{kopylov3}
G.~I. Kopylov.
\newblock {\em Phys. Lett.}, B50:472, 1974.

\bibitem{hofmann}
W~Hofmann.
\newblock A fresh look at {B}ose-{E}instein correlations.
\newblock Research Note LBL-23108, Lawrence Berkeley Laboratory, UC, 1987.

\bibitem{opal_pipi}
{OPAL}~collaboration{,} G.~Alexander, et~al.
\newblock {\em Z. Phys.}, C72:389, 1996.

\bibitem{dremin}
E.~A. De~Wolf, I.~M. Dremin, and W.~Kittel.
\newblock {\em Phys. Rept.}, 270:1, 1996.

\bibitem{mark_pipi}
{MARK} {II}~collaboration{,} I.~Juricic, et~al.
\newblock {\em Phys. Rev.}, D39:1, 1989.

\bibitem{biyajima_high}
M.~Biyajima et~al.
\newblock {\em Prog. Theor. Phys.}, 84:931, 1990.

\bibitem{neumeister}
{UA1-MINIMUM BIAS}~collaboration{,} N.~Neumeister, et~al.
\newblock {\em Phys. Lett.}, B275:186, 1992.

\bibitem{car}
P.~Carruthers and I.~Sarcevic.
\newblock {\em Phys. Rev. Lett.}, 63:1562, 1989.

\bibitem{dewolf}
E.~A.~De Wolf.
\newblock {\em Acta Phys. Pol.}, B21:611, 1990.

\bibitem{barlow}
R.~J. Barlow.
\newblock {\em Statistics}.
\newblock John Wiley $\&$ and Sons., 1989.

\bibitem{csorgo}
T.~Cs{\"o}rg{\"o} and S.~Pratt.
\newblock Structure of the peak in {B}ose-{E}instein correlation.
\newblock In T.~Cs{\"o}rg{\"o} et~al., editors, {\em Proc. of the Workshop on
  relativistic heavy ion physics}, page~75, Budapest, Hungary, 17-21 June 1991.

\bibitem{geiger}
K.~Geiger et~al.
\newblock {\em Phys. Rev.}, D61:054002, 2000.

\bibitem{delphi_2d}
{DELPHI}~collaboration{,} P.~Abreu, et~al.
\newblock {\em Phys. Lett.}, B471:460, 2000.

\bibitem{na44a}
{NA}44~collaboration{,} H.~Beker, et~al.
\newblock {\em Phys. Rev. Lett}, 74:3340, 1995.

\bibitem{na35a}
{NA}35~collaboration{,} T.~Alber, et~al.
\newblock {\em Z. Phys.}, C66:77, 1995.

\bibitem{ehs_na22a}
{EHS}{/}{NA22}~collaboration{,} N.~M.~Agababyan, et~al.
\newblock {\em Z. Phys.}, C66:409, 1995.

\bibitem{ehs_na22b}
{EHS}{/}{NA}22~collaboration{,} N.~M.~Agababyan, et~al.
\newblock {\em Z. Phys.}, C71:405, 1996.

\bibitem{l3a_2d}
{L}3~collaboration{,} M.~Acciarri, et~al.
\newblock {\em Phys. Lett.}, B458:517, 1999.

\bibitem{delphia_2d}
{DELPHI}~collaboration{,} P.~Abreu, et~al.
\newblock {\em Phys. Lett.}, B471:460, 2000.

\bibitem{opala_2d}
{OPAL}~collaboration{,} G.~Abbiendi, et~al.
\newblock {\em Eur. Phys. J.}, C16:423, 2000.

\bibitem{lund_trans1}
B.~Andersson and M.~Ringn{\' e}r.
\newblock {\em Nucl. Phys.}, B513:627, 1998.

\bibitem{lund_trans2}
B.~Andersson and M.~Ringn{\' e}r.
\newblock {\em Phys. Lett.}, B421:283, 1998.

\bibitem{segre}
E.~Segr{$\grave{\rm{e}}$}.
\newblock {\em Nuclei and Particles}.
\newblock Benjamin/Cummings Publishing Co., 2nd edition, 1982.

\bibitem{pdg}
{Particle Data Group}{,} K.~Hagiwara, et~al.
\newblock {\em Phys. Rev.}, D66:1, 2002.

\bibitem{wigner}
E.~P. Wigner.
\newblock {\em Z. Phys.}, 43:624, 1927.

\bibitem{eckart}
C.~Eckart.
\newblock {\em Rev. Mod. Phys.}, 2:305, 1930.

\bibitem{lednicky}
R.~Lednicky.
\newblock On correlation and spin composition techniques.
\newblock Research Note {PhE-10}, {MPI}, 1999.

\bibitem{lund}
B.~Andersson et~al.
\newblock {\em Phys. Rep.}, 97:31, 1983.

\bibitem{webber}
B.~R Webber.
\newblock {\em Nucl. Phys.}, B238:492, 1984.

\bibitem{lund_bec}
B.~Andersson and M.~Ringn{\' e}r.
\newblock {\em Nucl. Phys.}, B513:627, 1998.

\bibitem{gamow}
M.~Gyulassy et~al.
\newblock {\em Phys. Rev.}, C20:2267, 1979.

\bibitem{bowler2}
M.~G. Bowler.
\newblock {\em Phys. Lett.}, B270:69, 1991.

\bibitem{biyajima_cou}
M.~Biyajima et~al.
\newblock {\em Phys. Lett.}, B353:340, 1995.

\bibitem{elad}
E.~Tsur.
\newblock Bose-{E}instein correlations of three charged pions in hadronic
  {$Z^0$} decays.
\newblock Master's thesis, Tel-Aviv University, Tel-Aviv, Israel, 1997.

\bibitem{Liu}
Y.~M. Liu et~al.
\newblock {\em Phys. Rev.}, C34:1667, 1986.

\bibitem{cramer}
J.~G. Cramer.
\newblock {\em Phys. Rev.}, C43:2798, 1991.

\bibitem{alt1}
E.~O. Alt et~al.
\newblock {\em Eur. Phys. J.}, C13:663, 2000.

\bibitem{tpc_pipi}
{TPC}~collaboration{,} H.~Aihara, et~al.
\newblock {\em Phys. Rev.}, D31:996, 1985.

\bibitem{tasso_pipi}
{TASSO}~collaboration{,} M.~Althoff, et~al.
\newblock {\em Z. Phys.}, C30:355, 1986.

\bibitem{amy_pipi}
{AMY}~collaboration{,} S.~K.~Choi, et~al.
\newblock {\em Phys. Lett.}, D355:406, 1995.

\bibitem{aleph_pipi}
{ALEPH}~collaboration{,} D.~Decamp, et~al.
\newblock {\em Z. Phys.}, C54:75, 1992.

\bibitem{delphi_pipi}
{DELPHI}~collaboration{,} P.~Abreu, et~al.
\newblock {\em Phys. Lett.}, B286:201, 1992.

\bibitem{l3_pipi}
{L}3~collaboration{,} P.~Achard, et~al.
\newblock {\em Phys. Lett.}, B524:55, 2002.

\bibitem{sanders}
M.~Sanders.
\newblock {\em Pion (non-) correlations in hadronic events at the {Z}
  resonance}.
\newblock PhD thesis, The Nijmegen Catholic University, The Netherlands, 2002.

\bibitem{opal_pi0pi0}
OPAL Collaboration, Physics Note PN-513, 2002. Submitted to the XXXII Int.
  Symp. on Multiparticle Dynamics (ISMD), 7-13 September 2002, Alushta,
  Ukraine.

\bibitem{asher}
{CELLO}~collaboration{,} H.~J.~Behrend, et~al.
\newblock {\em Phys. Lett.}, B245:298, 1990.

\bibitem{nubebc}
{B}ig {B}ubble {C}hamber {N}eutrino~collaboration{,} V.~A.~Korotkov, et~al.
\newblock {\em Z. Phys.}, C60:37, 1993.

\bibitem{arneodo}
{EMC}~collaboration{,} M.~Arneodo~et al.
\newblock {\em Z. Phys.}, C32:1, 1986.

\bibitem{adamus}
{NA}22~collaboration{,} M.~Adamus, et~a.
\newblock {\em Z. Phys.}, C37:347, 1988.

\bibitem{bailly}
{NA}23~collaboration{,} J.~L.~Bailly, et~al.
\newblock {\em Z. Phys.}, C43:431, 1989.

\bibitem{aguillar}
{NA}27~collaboration{,} M.~Aguilar-Benitez, et~al.
\newblock {\em Z. Phys.}, C54:21, 1992.

\bibitem{akesson}
{AFS}~collaboration{,} T.~{\AA}kesson, et~al.
\newblock {\em Z. Phys.}, C36:517, 1987.

\bibitem{cplear}
{CPLEAR}~collaboration{,} R.~Adler, et~al.
\newblock {\em Z. Phys.}, C63:541, 1994.

\bibitem{albajar}
UA1~collaboration{,} C.~Albajar, et~al.
\newblock {\em Phys. Lett.}, B226:410, 1989.

\bibitem{h1_pipi}
{H}1~collaboration{,} C.~Adloff, et~al.
\newblock {\em Z. Phys.}, C75:437, 1997.

\bibitem{wwlep}
The~LEP collaborations, ALEPH, DELPHI, L3, OPAL, and the LEP WW~working group.
\newblock CERN EP preprint, 2003.

\bibitem{delphi_ww}
{DELPHI}~collaboration{,} N.~van Remortel, et~al.
\newblock In {\em Proc. of Int. Symp. on Multiparticle Dynamics}, Alushta,
  Ukraine, 7-13 September 2002.

\bibitem{l3_ww}
{L}3~collaboration{,} P.~Achard, et~al.
\newblock {\em Phys. Lett.}, B547:139, 2002.

\bibitem{aleph_ww}
{ALEPH}~collaboration{,} R.~Barate, et~al.
\newblock {\em Phys. Lett.}, B478:50, 2000.

\bibitem{delphi_kpkp}
{DELPHI}~collaboration{,} P.~Abreu, et~al.
\newblock {\em Phys. Lett.}, B379:330, 1996.

\bibitem{opal_kpkp}
{OPAL}~collaboration{,} G.~Abbiendi, et~al.
\newblock {\em Euro. Phys. J.}, C21:23, 2001.

\bibitem{lipkink1}
H.~J. Lipkin.
\newblock {\em Phys. Lett.}, B219:474, 1989.

\bibitem{lipkink3}
H. J. Lipkin, Argonne National Laboratory priprint ANL-HEP-PR-88-66, 1988.

\bibitem{lipkink2}
H.~J. Lipkin.
\newblock {\em Phys. Rev. Lett.}, 69:3700, 1992.

\bibitem{lipkina}
G.~Alexander.
\newblock Bose-{E}instein effect in the ${K^0}~\overline{K}^0$ in
  boson-antiboson system.
\newblock Research Note 93/001, LNF, 1993.

\bibitem{delphif}
{DELPHI}~collaboration{,} P.~Abreu, et~al.
\newblock {\em Phys. Lett.}, B298:236, 1993.

\bibitem{alex_iso}
G.~Alexander and H.~J. Lipkin.
\newblock {\em Phys. Lett.}, B456:270, 1999.

\bibitem{biyajima_iso}
M.~Biyajima et~al.
\newblock {\em Phys. Rev.}, C58:2316, 1998.

\bibitem{suzuki}
M.~Suzuki.
\newblock {\em Phys. Rev.}, D35:3359, 1987.

\bibitem{bowler1}
M.~G. Bowler.
\newblock {\em Phys. Lett.}, B197:443, 1987.

\bibitem{fzero}
D.~Morgan and M.~R. Pennington.
\newblock {\em Phys. Lett.}, B258:44, 1991.

\bibitem{fzero1}
D.~Morgan and M.~R. Pennington.
\newblock {\em Phys. Rev.}, D48:1185, 1993.

\bibitem{fzero2}
F.~E. Close et~al.
\newblock {\em Phys. Lett.}, B319:291, 1993.

\bibitem{delphi_pipipi}
{DELPHI}~collaboration{,} P.~Abreu, et~al.
\newblock {\em Phys. Lett.}, B355:415, 1995.

\bibitem{opal_pipipi}
{OPAL}~collaboration{,} K.~Ackerstaff, et~al.
\newblock {\em Eur. Phys. J.}, C5:239, 1998.

\bibitem{l3_pipipi}
{L3}~collaboration{,} P.~Achard, et~al.
\newblock {\em Phys. Lett.}, B540:185, 2002.

\bibitem{agababian}
{EHS/NA22}~collaboration{,} N.~M.~Agababian, et~al.
\newblock {\em Z. Phys.}, C68:229, 1995.

\bibitem{na22_bec}
{NA}23~collaboration{,} J.~L.~Bailly, et~al.
\newblock {\em Z. Physik}, C43:341, 1989.

\bibitem{opal_ll}
{OPAL}~collaboration{,} G.~Alexander, et~al.
\newblock {\em Phys. Lett.}, B384:377, 1996.

\bibitem{delphi_ll}
{DELPHI}~collaboration{,} {T. Lesiak and H. Palka}.
\newblock Determination of the spin composition of
  ${\Lambda}\overline{\Lambda}$ and ${\Lambda}{\Lambda}~
  (\overline{\Lambda}\overline{\Lambda})$ pairs in hadronic {Z} decays
  ({DELPHI} 98-114 {CONF} 176).
\newblock In {\em Proc. of the XXIX Int. Conf. on High Energy Physics},
  Vancouver, Canada, 22-29 July 1998.

\bibitem{aleph_ll}
{ALEPH}~collaboration{,} R.~Barate, et~al.
\newblock {\em Phys. Lett.}, B475:395, 2000.

\bibitem{opal_pp}
OPAL Collaboration, Physics Note PN486, 2002. Submitted to the Int. Conf. on
  High Energy Physics ({HEP} 2001) 12-18 July 2001, Budapest, Hungary.

\bibitem{acl}
G.~Alexander, I.~Cohen, and E.~Levin.
\newblock {\em Phys. Lett.}, B452:159, 1999.

\bibitem{opal_kzkz}
{OPAL}~collaboration{,} R.~Akers, et~al.
\newblock {\em Z. Phys.}, C67:389, 1995.

\bibitem{aleph_kzkz}
{ALEPH}~collaboration{,} D.~Buskulic, et~al.
\newblock {\em Z. Phys.}, C64:361, 1994.

\bibitem{selex}
{SELEX}~collaboration{,} I.~Eschrich, et~al.
\newblock {\em Phys. Lett.}, B522:233, 2001.

\bibitem{smith}
M.~Smith.
\newblock {\em Phys. Lett.}, B477:141, 2000.

\bibitem{andersson}
B.~Andersson.
\newblock Some theory remarks on {B}ose-{E}instein correlation.
\newblock In {\em Proc. of the XXVth Les Rencontres de Moriond}, Les Arcs,
  France, 18-20 March 2000.

\bibitem{VT}
C.~Cohen-Tannoudji, B.~Diu, and F.~Lalo{\"e}.
\newblock {\em Quantum Mechanics}.
\newblock John Wiley and Sons publishing Co., 1990.

\bibitem{DUAL}
Ya. Asimov et~al.
\newblock {\em Z. Phys.}, C31:213, 1986{,} and references therein.

\bibitem{DUAL1}
V.~A. Khoze and W.~Ochs.
\newblock {\em Int. J. Mod. Phys.}, A12:2949, 1997{,} and references therein.

\bibitem{RCP}
M.~R. Ahmady, R.~R. Mendel, and J.~D. Talman.
\newblock {\em Phys. Rev.}, D55:419, 1997{,} and references therein.

\bibitem{RCP1}
B.~Blok and M.~Lublinsky.
\newblock {\em Phys. Rev.}, D57:2676, 1998{,} and references therein.

\bibitem{delphi_tran1}
B.~L{\"o}rstad and O.G. Smirnova.
\newblock Transverse mass dependence of {B}ose-{E}instein correlation radii in
  $e^+ e^-$ annihilation at {L}{E}{P} energies.
\newblock In {\em Proc. 7th Int. Workshop on Multiparticle Production:
  Correlations and Fluctuations}, page~42, Nijmegen, The Netherland, 1997.
  World Scientific Publishing Co.

\bibitem{smirnova}
O.~Smirnova.
\newblock {\em Nucl. Phys. (Proc. Suppl.)}, B92:301, 2001.

\bibitem{todorova}
S.~Todorova-Nova.
\newblock Bose-{E}instein correlations at {LEP} and {HERA}.
\newblock In {\em Proc. of the XXXI Int. Conf. on High Energy Physics},
  Amsterdam, 24-31 July 2002.

\bibitem{alex_be}
G.~Alexander.
\newblock {\em Phys. Lett.}, B506:45, 2001.

\bibitem{L3_twod}
L3~collaboration{,}.
\newblock Study of {B}ose-{E}instein correlations in {Z} decays at {L}{E}{P}.
\newblock In {\em Proc. XXIX Int. Conf. on HEP}, Vancouver, 1998.

\bibitem{bialas1}
A.~Bialas et~al.
\newblock {\em Phys. Rev.}, D62:114007, 2000.

\bibitem{bialas2}
A.~Bialas et~al.
\newblock {\em Acta Phys. Polon.}, B32:2901, 2001.

\bibitem{gottfried1}
K.~Gottfried.
\newblock {\em Phys. Rev. Lett.}, 32:957, 1974.

\bibitem{gottfried2}
F.~E. Low and K.~Gottfried.
\newblock {\em Phys. Rev.}, D17:2487, 1978.

\bibitem{bjorken1}
J.~D. Bjorken.
\newblock {\em Phys. Rev.}, D27:140, 1983.

\bibitem{bjorken2}
J.~D. Bjorken.
\newblock {\em Phys. Rev.}, D7:282, 1973.

\bibitem{einstein}
A.~Einstein.
\newblock Sitszber. Kgl. Preuss. Akad. Wiss. rep. 261, 1924.

\bibitem{einstein1}
A.~Einstein.
\newblock Sitszber. Kgl. Preuss. Akad. Wiss. rep. 3, 1925.

\bibitem{theory}
F.~Dalfovo et~al.
\newblock {\em Rev. Mod. Phys.}, 71:463, 1999{,} and references therein.

\bibitem{experiment}
See e.g. the Georgia Southern University BEC bibliography on the web at
  \mbox{http:// amo.phy.gasou.edu/bec.html/bibliography.html}.

\bibitem{ketterle}
W.~Ketterle.
\newblock {\em Physics Today}, 52(12):30, 1999.

\bibitem{burnett}
K.~Burnett, M.~Edwards, and C.~Clark.
\newblock {\em Physics Today}, 52(12):37, 1999.

\bibitem{discovery}
M.~H. Anderson et~al.
\newblock {\em Science}, 269:198, 1995.

\bibitem{discovery1}
K.~B. Davis et~al.
\newblock {\em Phys. Rev. Lett.}, 75:3969, 1995.

\bibitem{discovery2}
C.~C. Bradley et~al.
\newblock {\em Phys. Rev. Lett.}, 75:1687, 1995.

\bibitem{book}
D.~H. Trevena.
\newblock {\em Statistical Mechanics $-$ An Introduction}.
\newblock {Prentice-Hall/Ellis Horwood}, 1993.

\bibitem{becattini}
F.~Becattini, L.~Bellucci, and G.~Passaleva.
\newblock Transverse momentum spectra of identified particles in high energy
  collisions with statistical hadronisation model.
\newblock In A.~Giovannini and R.~Ugoccioni, editors, {\em Proc. IX Int.
  Workshop on Multiparticle Production}, page 363, Torino, Italy, {12-17} June
  2000{,} and references therein. Nucl. Phys. B (Suppl.) 92 (2001) 363.

\bibitem{alex_buda}
G. Alexander, {\it Emitter size as a function of mass and transverse mass},
  hep-ph/0108194, In Proceedings of the EPS International Conference on High
  Energy Physics, Budapest, 2001 (D. Horvath, P. Levai, A. Patkos, eds.), JHEP
  (http://jhep.sissa.it/), Proceedings Section, PrHEP-hep2001/027.

\bibitem{gaes}
G.~Alexander and E.~K.~G. Sarkisyan.
\newblock To be published.

\bibitem{cumulant_bec}
J.~G. Cramer and K.~Kadija.
\newblock {\em Phys. Rev.}, C53:908, 1996.

\bibitem{lipa}
P.~Lipa and B.~Bushbeck.
\newblock {\em Phys. Lett.}, B223:465, 1989.

\bibitem{bushbeck}
B.~Bushbeck, H.~C. Eggers, and P.~Lipa.
\newblock {\em Phys. Lett.}, B481:187, 2000.

\bibitem{alex_sark}
G.~Alexander and E.~K.~G. Sarkisyan.
\newblock {\em Phys. Lett.}, B487:215, 2000.

\bibitem{akesson1}
{AFS}~collaboration{,} T.~{\AA}kesson, et~al.
\newblock {\em Phys. Lett.}, B129:269, 1983.

\bibitem{akesson2}
{AFS}~collaboration{,} T.~{\AA}kesson, et~al.
\newblock {\em Phys. Lett.}, B187:420, 1987.

\bibitem{breakstone}
{SFM}~collaboration{,} A.~Breakstone, et~al.
\newblock {\em Z. Phys.}, C33:333, 1987.

\bibitem{gary1}
I.~M. Dremin and J.~W. Gary.
\newblock {\em Phys. Lett.}, B459:341, 1999.

\bibitem{gary2}
A.~Capella et~al.
\newblock {\em Phys. Rev.}, D61:074009, 2000.

\bibitem{madjid}
M.~Boutemeur.
\newblock {\em Fortschr. Phys.}, 50:1001, 2002.

\end{thebibliography}
\end{document}